\documentclass[a4paper,fleqn,usenatbib]{mnras}

\usepackage{newtxtext,newtxmath}

\usepackage[T1]{fontenc}
\usepackage{ae,aecompl}

\usepackage{graphicx}	
\usepackage{amsmath}	
\usepackage{amssymb}	
\usepackage{lipsum}
\usepackage{subfig}
\usepackage{grffile}
\usepackage{lscape}
\usepackage{booktabs}
\usepackage{multirow}
\usepackage[normalem]{ulem}
\useunder{\uline}{\ul}{}

\title[Ionization conditions in Type II QSOs]{Carbon-loud SDSS BOSS QSO2s at z$>$2: High density gas or secondary production of Carbon?}

\author[Silva, M. et al.]{
M. Silva$^{1,2}$\thanks{E-mail: marckelson.silva@astro.up.pt},
A. Humphrey$^{1}$\thanks{E-mail: andrew.humphrey.mexico@gmail.com},
P. Lagos$^{1,3}$
and S.G. Morais$^{1,2}$
\\
$^{1}$Institute of Astrophysics and Space Sciences, Universidade do Porto, CAUP, Rua das Estrelas, 4150-762 Porto, Portugal.\\
$^{2}$Departamento de F\'isica e Astronomia, Faculdade de Ci\^encias, Universidade do Porto, R. Campo Alegre 687, 4169-007 Porto, Portugal\\
$^{3}$Centre for Space Research, North West University, Potchefstroom 2520, South Africa\\
}

\date{Accepted 2020 May 18. Received 2020 May 18; in original form 2019 August 4}

\pubyear{2020}

\begin{document}
\label{firstpage}
\pagerange{\pageref{firstpage}--\pageref{lastpage}}
\maketitle

\begin{abstract}
	We study the ultraviolet (UV) emission-line ratios of a sample of 145 type II quasars (QSO2s) from Sloan Digital Sky Survey III Baryon Oscillation Spectroscopic Survey, and compare against a grid of active galactic nucleus (AGN) photoionization models with a range in gas density, gas chemical abundances, and ionization parameter. Most of the quasars are "carbon-loud", with \ion{C}{IV}/\ion{He}{II} ratios that are unusually high for the narrow-line region, implying higher than expected gas density ($>$10$^{6}$ cm$^{-3}$) and/or significantly super-Solar relative carbon abundance. We also find that solar or supersolar nitrogen abundance and metallicity are required in the majority of our sample, with potentially significant variation between objects. Compared to radio galaxies at similar redshifts (HzRGs; z $>$ 2), the QSO2s are offset to higher \ion{N}{V}/\ion{He}{II}, \ion{C}{IV}/\ion{He}{II} and \ion{C}{III]}/\ion{He}{II}, suggesting systematically higher gas density and/or systematically higher C and N abundances. We find no evidence for a systematic difference in the N/C abundance ratio between the two types of objects. Scatter in the \ion{N}{IV]}/\ion{C}{IV} ratio implies a significant scatter in the N/C abundance ratio among the QSO2s and HzRGs, consistent with differences in the chemical enrichment histories between objects. Interestingly, we find that adopting secondary behaviour for both N and C alleviates the long-standing "\ion{N}{IV]} problem". A subset of the QSO2s and HzRGs also appear to be "silicon-loud", with \ion{Si}{III]} relative fluxes suggesting Si/C and Si/O are an order of magnitude above their Solar values. Finally, we propose new UV-line criteria to select genuine QSO2s with low-density narrow-line regions.
\end{abstract}




\section{Introduction}

High-z quasars (QSOs, z $>$ 2) or more generally active galactic nuclei (AGNs) are among the byproducts of galaxy formation. They are powered by the accretion of material onto a central supermassive black hole (SMBH), and their study provides powerful tools for understanding SMBH growth \citep[e.g.][]{mortlock2011,wu2015}, the re-ionization process \citep[e.g.][]{fan2006a,fan2006b}, and chemical enrichment history \citep[e.g.][]{hamman2002,dietrich2003,baldwin2003,nagao2006a,nagao2006b,xu2018} during a poorly understood early phase of galaxy assembly.

Among the notable observational characteristic of AGNs is the presence of emission lines with widths upwards of 1000 km s$^{-1}$ \citep[e.g.][]{villar-martin1999b,VM1,Hu2,VM2,Hu4,Hu6,sandy2017,marckelson2018,marck2018} and far in excess of any other known class of galaxy. These broad lines are thought to arise in gas clouds of high density (i.e., $n_H\ga$10$^{6}$ cm$^{-3}$) situated close to the SMBHs (broad line region or BLR), which are photoionized by the radiation field from the accretion disk of the SMBH. On the other hand, when the BLR is obscured by the putative dusty torus, the observer is able to see narrow lines that arise in low density gas clouds (i.e., $n_H\la$10$^{6}$ cm$^{-3}$) located several parsecs or more from the AGN (narrow line region or NLR). The ultraviolet and optical emission lines from these regions are widely used to study the dynamics, ionization, and chemical abundances of the gas in and around AGNs \citep[e.g.][]{hamman1993,VM97,ferland1996,Ha99,villar-martin1999b,villar1999,villar2000,dietrich2000,Bre2000,dietrich2001,hamman2002,baldwin2003,VM1,nagao2006a,nagao2006b,jiang2007,Hu4,juarez2009,humphrey2009,sandy2017,marckelson2018,xu2018}.

The gas phase metallicity in AGNs has been studied extensively, particularly in the case of the BLR of quasars. Using fluxes of several UV emission lines within the framework of photoionization models, previous works have reported that the BLR metallicity is typically higher than the solar value (Z/Z$_{\odot}$ $\simeq$ 4 to 8; e.g. \citealt{dietrich2000,dietrich2001,hamman2002,dietrich2003,nagao2006b,jiang2007,juarez2009,xu2018}), reaching as much as Z/Z$_{\odot}$ $\simeq$ 15 in extreme cases \citep[e.g.][]{baldwin2003,bentz2004}. Although the gas that makes up the BLR represents only a tiny fraction of the total gas in the host galaxy, it has been concluded that intense star formation at an early epoch is required to explain such high levels of chemical enrichment in many cases \citep{dietrich2001,hamman2002,baldwin2003,dietrich2003,simon2010,xu2018}.

Another possibility presented by the literature is to explore the gas phase metallicity in NLRs.
In contrast to BLRs, the size of these regions are comparable to the size of their host galaxies which makes it a good tracer of chemical properties on galactic scales.
However, as only a few optically selected high-z type II quasars (QSO2s) have been discovered, the literature has as main target the study of the Extended Emission Line Regions (EELR) of HzRGs, name commonly used for the most extended NLRs which have gas density lower than the less extended NLRs (n$_{H}$ $\lesssim$ 100 cm$^{-3}$) and sizes that extend to several hundred kpc  \citep{fosbury1982,serego1988,Mc90,Vo,Pe98,Fr2001,Re,Hu1,Cantalupo2014,swinbank,borisova2016,Cai2017}.
There have been a number of attempts at estimating the chemical abundances in the EELR of HzRGs, where the rest-frame UV lines have been widely used. For example, using the flux ratios  \ion{N}{V}/\ion{He}{II} and \ion{N}{V}/\ion{C}{IV} of two HzRGs at z $\sim$ 2.5, \citet{villar1999} find large flux ratios values suggesting that N is overabundant in the ionized gas of these objects. By focusing on the same flux ratios and using sequences in metallicity, \citet{Bre2000} and \citet{Ve2001} reported nitrogen abundances in the range 0.4 $\lesssim$ Z/Z$_{\odot}$ $\lesssim$ 4.0 or possibly even much higher. Given the correlation defined by the sample, they suggest a metallicity evolution in nitrogen abundance. Based on rest-frame optical spectra obtained by sensitive near-infrared spectroscopy, \citet{iwamuro2003} report that HzRGs tend to show sub-solar metallicities (Z/Z$_{\odot}$ $\sim$ 0.2), significantly lower than values previously reported. Such a disagreement with previous results must surely be due to the use of different modelling methodologies or different sets of lines which can give significantly different metallicity estimates. An additional tool for the metallicity determination is the \ion{N}{IV]} diagram (e.g. diagrams using the \ion{N}{IV]} emission line relative to \ion{He}{II} and \ion{C}{IV}). This diagram poses a problem in that photoionization models imply it should be less sensitive to ionization parameter (U) than the \ion{N}{V} diagram (e.g. diagrams using the \ion{N}{V} emission line relative to \ion{He}{II} and \ion{C}{IV}). However, the models systematically overpredict \ion{N}{IV]}/\ion{He}{II} and \ion{N}{IV]}/\ion{C}{IV} by a factor of $\sim$ 2 \citep[e.g.][]{Ve2001,Hu4}. Taken at face value, the observed \ion{N}{IV]} flux ratios would imply lower N abundance than the \ion{N}{V} flux ratios by a factor of $\sim$ 2. By using a large set of lines, \citet{marckelson2018} report a high gas metallicity ( Z/Z$_{\odot}$ = 2.1) for the EELR of the HzRG MRC 0943--242. In addition, they find a substantial range in ionization level across the object, validated by the analysis of the 2D ionization properties, with a clear spatial correlation between the radio hotspots and UV emission lines indicative of relatively low ionization.

In this paper, we present a study of the UV emission lines of a sample of 145 QSO2s from Sloan Digital Sky Survey III Baryon Oscillation Spectroscopic Survey (SDSS BOSS; \citealt{Al}) in the redshift range 2.0 $<$ z $<$ 4.3, and compare against a grid of AGN photoionization models in order to obtain new information about the ionization conditions and chemical enrichment history of QSO2s at high-z. In addition, we use rest-frame optical emission line fluxes from the literature \citep{greene2014}, which partially overlaps with the SDSS BOSS sample of QSO2s. To complement our study making comparisons of two different populations of objects, we also use emission-line fluxes of HzRGs from the literature \citep{cimatti1998,Ve2001,Hu4}.

This paper is organized as follows. In \S\ref{sample}, we describe the sample and data used in this study. In \S\ref{analysis}, we describe the methods for measuring the emission line parameters we use in our analysis. In \S\ref{models}, we present the photoionization model grid used to build our diagnostic diagrams, and in \S\ref{discussion} we describe and discuss the results of our comparison between the data and models. Finally, in \S\ref{conclusions} we summarize our conclusions.

\section{Sample and Emission Line Data}
\label{sample}

\subsection{SDSS BOSS QSO2s}
\label{sdss}

The objects studied in this paper are a subsample of quasars from the spectroscopic quasar sample of the BOSS of the SDSS Data Release 9 (see \citealt{gunn2006,Al} and references therein). \citet{Al} selected 452 candidate QSO2s in the redshift range 2.0 $<$ z $<$ 4.3, using selection criteria that included 5$\sigma$ detections in both Ly$\alpha$ and \ion{C}{IV}, with both lines also requiring to have requiring to have full width at half-maximum (FWHM) $<$ 2000 km s$^{-1}$. These authors further refined their QSO2s candidate sample by removing the two main contaminants in their original sample, narrow line Seyfert 1 (NLS1) galaxies and broad absorption line (BAL) quasars, and then splitting the remaining objects into two categories: i) those with the characteristics of an obscured quasar, i.e., narrow emission lines, no associated absorption and weak continuum (Class A; 145 objects); and ii) those with one or more characteristics of an unobscured quasar, i.e., a broad (BLR) component to an emission line, a BAL, or a strong blue continuum (Class B; 307 objects). In this study, we consider only the 145 Type II Class A quasar candidates of \citet{Al}, in order to avoid unnecessary contamination from unobscured objects.

In addition, we make use of rest-frame optical line fluxes of QSO2 candidates published by \citet{greene2014}, who used Triplespec on the Apache Point Observatory 3.5m Telescope (APO), Folded-port Infrared Echellete (FIRE) on the 6.5m Baade-Magellan Telescope, and the Gemini Near-Infrared Spectrograph (GNIRS) on the Gemini North 8.1m Telescope. The sample of \citet{greene2014} consists of 19 candidate QSO2s at 2.0 $<$ z $<$ 3.4 \citep[see][Table 2]{greene2014}; here we use the line fluxes of 16 candidate QSO2s which are also present in the Class A subsample discussed above. In the interest of simplicity, we henceforth refer to the candidate QSO2s studied here simply as 'QSO2s'.

\subsection{Radio Galaxies at 2.27 $\leq$ z $\leq$ 3.56}
\label{keck}

As a comparison sample, we used emission-line fluxes of HzRGs at 2.27 $\leq$ z $\leq$ 3.56 from the literature (see Table \ref{hzrgs}). The objects were observed using the Low Resolution Imaging Spectrometer (LRIS) in spectropolarimetry mode at the Keck II 10-m telescope, the Infrared Spectrometer and Array Camera (ISAAC) in long-slit mode and Focal Reducer and Spectrograph (FORS1) at the Very Large Telescope (VLT), OH Airglow Suppressor (OHS) spectrograph mounted on the Subaru telescope and ESO Multi-Mode Instrument (EMMI) at the New Technology Telescope (NTT). The line flux measurements and more detailed information on the observation are given by \citet{cimatti1998}, \citet{Ve2001} and \citet{Hu4}. \citeauthor{Hu4} also presented near-IR spectroscopy for a partially overlapping sample of HzRGs in a similar redshift range, which we make use of 11 of the 16 HzRGs in this sample.
A number of previous publications have presented further analyses of these objects \citep[e.g.][]{Mc1992,Rot1995,knopp1997,cimatti1998,villar1999,Pe1999,overzier2001,Pe2001,VM4,iwamuro2003,egami2003,VM1,Hu2,broderick2007,Hu3,Gu,sandy2017,marck2018}.

In the special case of the radio galaxy MRC 0943--242 (z = 2.92), we have used the UV and optical lines from the X-SHOOTER study of \citet{marckelson2018}. However, given that some UV emission lines were not detected by the X-SHOOTER instrument we also make use of the UV emission lines from \cite{Ve2001} in order to complement the X-SHOOTER data.

Two of the HzRGs in this sample appear to be intermediate objects. MRC 2025--218 (z = 2.63) appears to be a broad line radio galaxy (BLRG), on account of its broad \ion{C}{IV} \citep[e.g.][]{villar1999} and H$\alpha$ \cite[e.g.][]{larkin2000,Hu4} lines. In addition, MRC 1558-003 (z = 2.57) shows broad (FWHM = 12000 km s$^{-1}$) H$\alpha$ \citep{Hu4}, while its permitted UV lines remain narrow (e.g., FWHM$\sim$1000 km s$^{-1}$), but as this paper focuses on the UV lines we classify it as a type 2 radio galaxy.

\subsection{Core-Extremely Red Quasars}

We also include four 'core extremely red quasars' (core-ERQs; see \citealt{ross2015}, \citealt{zaka2016}, \citealt{hamann2017}, \citealt{villar-martin2019}) in the redshift range 2.3 $<$ $z$ $<$ 2.6. It has been proposed that core-ERQs are in a strongly dust-obscured evolutionary stage, where the central engine is in the process of removing gas and dust from its vicinity via powerful outflows, after which it will evolve into a classical unobscured quasar \citep{zaka2016}. Recently, however, \citet{villar-martin2019} have challenged this scenario, proposing instead that the observed properties are instead the consequence of orientation and the effects of the high bolometric luminosities of the central engine. Whatever their true nature, it is clear that core-ERQs represent an important phase in the co-evolution of a quasar and its host galaxy. We make use of the UV line ratios published by \citet[see Table 1]{villar-martin2019}.

\section{Data Analysis}
\label{analysis}

For the purpose of fitting the emission line parameters (e.g. flux and wavelength) we created a {\scshape python} routine that minimizes the sum of the squares of the difference between the model and data using the {\scshape lmfit} algorithm \citep{lmfit}. Gaussian profiles were used to fit the emission lines, with additional Gaussian profiles being added when necessary to obtain a good fit. We also use Voigt profiles to model any detected absorption features within the velocity profiles of Ly$\alpha$ or \ion{C}{IV}. The continuum emission, although not used in our later analysis, was modeled on a line-by-line basis using a single power law.

For doublets, such as \ion{C}{IV} $\lambda \lambda$1548,1551 (hereafter \ion{C}{IV}) or \ion{N}{V} $\lambda \lambda$1239,1243 (hereafter \ion{N}{V}), we use a single Gaussian for each doublet component, with both components constrained to have equal FWHM, and a fixed wavelength separation corresponding to the theoretical value. In the case of resonant lines, we adopt a fixed flux ratio of blue:red = 2:1 corresponding to the optically thin case. We have repeated our fits using optically-thick (blue:red $\sim$1:1) and find no significant difference to the summed flux of the two doublet components.

In the case of density sensitive doublets such as \ion{Si}{III]} $\lambda \lambda$1883,1892 (hereafter \ion{Si}{III]}) and \ion{C}{III]} $\lambda \lambda$1907,1909 (hereafter \ion{C}{III]}), we adopt the low density flux ratio (blue:red$\ga$1:1). Repeating our fits using a doublet ratio corresponding to high (BLR) density made no significant differences to our flux measurements of sum of the doublet flux.

In the special cases where multiple lines are blended together, such as \ion{O}{VI} $\lambda \lambda$1032,1038 + \ion{C}{II} $\lambda$1037 (hereafter \ion{O}{VI}$+$\ion{C}{II}) or \ion{Si}{IV} $\lambda \lambda$1393,1402 + \ion{O}{IV]}$\lambda\lambda$1401,1407 (hereafter \ion{Si}{IV}$+$\ion{O}{IV]}), each line was fitted with a single Gaussian, with all lines constrained to have equal FWHM, and a fixed wavelength separation corresponding to the theoretical value. However, the flux ratio between different species was allowed to vary freely. In such cases, we provide the flux of the summed blend.

The forbidden emission line \ion{[Ne}{IV]} $\lambda$1602 (hereafter \ion{[Ne}{IV]}1602) was detected only in 3 objects (see Fig. \ref{neiv1602}), but considering its importance as a diagnostic of electron density when used together with \ion{[Ne}{IV]} $\lambda$2422, we have estimated its 3$\sigma$ upper limit in cases where the 2422 \AA~line was detected.

As shown by \citet{humphrey2019}, the Ly$\alpha$ emission from the narrow line region can be significantly contaminated by the \ion{O}{V]} $\lambda \lambda$1213.8,1218.3 doublet, particularly when other high-ionzation UV lines such as \ion{N}{V} or \ion{C}{IV} are strong. This potential contamination, together with the fact that Ly$\alpha$ emission can also be strongly affected by line transfer effects \citep[e.g.][]{marck2018}, reduces the quantitative usefulness of the line. As such, Ly$\alpha$ will not be considered in our later analysis. However, for the sake of completeness we include in our table of results the best fit flux of the overall Ly$\alpha$ + \ion{O}{V]} + \ion{He}{II} $\lambda$1215.1 blend, without attempting to correct for the effects described above.

Throughout this paper we shall abbreviate the other emission lines as follows: Ly$\alpha$ for Ly$\alpha$ $\lambda$1216, \ion{O}{I} for \ion{O}{I} $\lambda$1304, \ion{N}{IV]} for \ion{N}{IV]} $\lambda$1487, \ion{He}{II} for \ion{He}{II} $\lambda$1640, \ion{O}{III]} for \ion{O}{III]} $\lambda$1661, \ion{[Ne}{IV]} for \ion{[Ne}{IV]} $\lambda$2422 (see Figures \ref{neiv2422} and \ref{neiv2422_02} showing the detection), \ion{Mg}{II} for \ion{Mg}{II} $\lambda \lambda$2796, 2803 (see Table \ref{table01}), H$\beta$ for H$\beta$ $\lambda$4863, \ion{[O}{III]} for \ion{[O}{III]} $\lambda$5007, \ion{[N}{II]} for \ion{[N}{II]} $\lambda$6585 and H$\alpha$ for H$\alpha$ $\lambda$6565.

\begin{table*}
	\caption{Summary of the results obtained with the diagrams. PC $=$ Primary behaviour of Carbon, SC $=$ Secondary behaviour of Carbon, PCSi10 $=$ Primary behaviour of Carbon with Si/O $=$ 10$\times$Solar, SCSi10 $=$ Secondary behaviour of Carbon with Si/O $=$ 10$\times$Solar.}
	\label{summary}
	\resizebox{\textwidth}{!}{%
		\begin{tabular}{|c|c|c|l|}
			\hline
			Fig. number & Diagrams & Adundance & \multicolumn{1}{c|}{Summary of results} \\ \hline
			\ref{NV01} & \ion{N}{V}/\ion{C}{IV} $vs.$ \ion{N}{V}/\ion{He}{II} & \textbf{PC} & \begin{tabular}[c]{@{}l@{}}A combination of n$_H$ $\gtrsim$ 10$^{7}$ cm$^{-3}$ and O/H $\gtrsim$ 4$\times$Solar is required (QSO2s and core-ERQs).\\ The HzRGs require gas metallicity $\gtrsim$ 0.5$\times$Solar, with some objects requiring Solar or super-Solar metallicity.\end{tabular} \\ \hline
			\ref{NV01s} & \ion{N}{V}/\ion{C}{IV} $vs.$ \ion{N}{V}/\ion{He}{II} & \textbf{SC} & \begin{tabular}[c]{@{}l@{}}The models fail to reproduce the positions of the core-ERQs and also of $\sim$ 30\% of QSO2s.\end{tabular} \\ \hline
			\ref{NV02} & \ion{N}{V}/\ion{C}{IV} $vs.$ \ion{C}{IV}/\ion{C}{III]} & \textbf{PC} & \begin{tabular}[c]{@{}l@{}}The observed \ion{C}{IV}/\ion{C}{III]} ratio suggests significant variation in U between objects.\\ There is no evidence for a systematic difference in the N/C abundance ratio between the QSO2s and the HzRGs.\end{tabular} \\ \hline
			\ref{NV02s} & \ion{N}{V}/\ion{C}{IV} $vs.$ \ion{C}{IV}/\ion{C}{III]} & \textbf{SC} & \begin{tabular}[c]{@{}l@{}}The N/C abundance ratio ceases to be a useful metallicity indicator.\end{tabular} \\ \hline
			\ref{NIV} & \ion{N}{IV]}/\ion{C}{IV} $vs.$ \ion{C}{IV}/\ion{C}{III]} & \textbf{PC} & \begin{tabular}[c]{@{}l@{}}It strongly suggests a significant variation in the N/C abundance ratio between objects.\end{tabular} \\ \hline
			\ref{NIVs} & \ion{N}{IV]}/\ion{C}{IV} $vs.$ \ion{C}{IV}/\ion{C}{III]} & \textbf{SC} & \begin{tabular}[c]{@{}l@{}}The N/C abundance ratio ceases to be a useful metallicity indicator.\end{tabular} \\ \hline
			\ref{CIV} & \ion{C}{IV}/\ion{C}{III]} $vs.$ \ion{C}{IV}/\ion{He}{II} & \textbf{PC} & \begin{tabular}[c]{@{}l@{}}A combination of n$_H$ $\gtrsim$ 10$^{6}$ cm$^{-3}$ and  Z/Z$_{\odot}$ $=$ 2.0 appears to provide the closest fit to the high flux ratio of QSO2s and core-ERQs.\\ A combination of n$_H$ $=$ 10$^{2}$ cm$^{-3}$ and Z/Z$_{\odot}$ $>$ 2.0 appears to provide the best fit to the observed flux ratios of most of HzRGs.\end{tabular} \\ \hline
			\ref{CIVs} & \ion{C}{IV}/\ion{C}{III]} $vs.$ \ion{C}{IV}/\ion{He}{II} & \textbf{SC} & \begin{tabular}[c]{@{}l@{}}A combination of intermediate gas density (n$_H$ $<$ 10$^{6}$ cm$^{-3}$) and Z/Z$_{\odot}$ $\gtrsim$ 4.0 is required for the QSO2s.\\ The core-ERQs strongly suggest high gas densities and metallicities well above solar.\\ Most of the diagrams predict very high \ion{C}{IV}/\ion{He}{II} ratio compared with the observed \ion{C}{IV}/\ion{He}{II} ratio of the HzRGs.\end{tabular} \\ \hline
			\ref{SiIII} & \ion{Si}{III]}/\ion{C}{III]} $vs.$ \ion{C}{IV}/\ion{C}{III]} & \textbf{PC} & \begin{tabular}[c]{@{}l@{}}Models using the Solar Si/O abundance ratio fail to explain the observed \ion{Si}{III]}/\ion{C}{III]} flux ratios of the QSO2s and HzRGs.\end{tabular} \\ \hline
			\ref{SiIIIs} & \ion{Si}{III]}/\ion{C}{III]} $vs.$ \ion{C}{IV}/\ion{C}{III]} & \textbf{SC} & \begin{tabular}[c]{@{}l@{}}Models also fail to explain the observed \ion{Si}{III]}/\ion{C}{III]} flux ratios of the QSO2s and HzRGs.\end{tabular} \\ \hline
			\ref{SiIIIx10} & \ion{Si}{III]}/\ion{C}{III]} $vs.$ \ion{C}{IV}/\ion{C}{III]} & \textbf{PCSi10} & \begin{tabular}[c]{@{}l@{}}Models with a Si/O abundance ratio of 10 $\times$ Solar suggest that Si/C is an order of magnitude (or more) above their Solar values.\end{tabular} \\ \hline
			\ref{SiIIIx10s} & \ion{Si}{III]}/\ion{C}{III]} $vs.$ \ion{C}{IV}/\ion{C}{III]} & \textbf{SCSi10} & \begin{tabular}[c]{@{}l@{}}Models with a Si/O abundance ratio of 10 $\times$ Solar suggest that Si/C is an order of magnitude (or more) above their Solar values.\end{tabular} \\ \hline
			\ref{OIII} & \ion{Si}{III]}/\ion{O}{III]} $vs.$ \ion{C}{IV}/\ion{He}{II} & \textbf{PC} & \begin{tabular}[c]{@{}l@{}}Consistent with the results obtained with the diagrams using \ion{Si}{III]}/\ion{C}{III]} flux ratios.\end{tabular} \\ \hline
			\ref{OIIIs} & \ion{Si}{III]}/\ion{O}{III]} $vs.$ \ion{C}{IV}/\ion{He}{II} & \textbf{SC} & \begin{tabular}[c]{@{}l@{}}Consistent with the results obtained with the diagrams using \ion{Si}{III]}/\ion{C}{III]} flux ratios.\end{tabular} \\ \hline
			\ref{OIIIx10} & \ion{Si}{III]}/\ion{O}{III]} $vs.$ \ion{C}{IV}/\ion{He}{II} & \textbf{PCSi10} & \begin{tabular}[c]{@{}l@{}}Models with a Si/O abundance ratio of 10 $\times$ Solar suggest that Si/O is an order of magnitude (or more) above their Solar values.\end{tabular} \\ \hline
			\ref{OIIIx10s} & \ion{Si}{III]}/\ion{O}{III]} $vs.$ \ion{C}{IV}/\ion{He}{II} & \textbf{SCSi10} & \begin{tabular}[c]{@{}l@{}}Models with a Si/O abundance ratio of 10 $\times$ Solar suggest that Si/O is an order of magnitude (or more) above their Solar values.\end{tabular} \\ \hline
			\ref{neiv01} & \ion{[Ne}{IV]}1602/\ion{[Ne}{IV]}2422 $vs.$ \ion{C}{IV}/\ion{He}{II} & \textbf{PC} & \begin{tabular}[c]{@{}l@{}}The \ion{[Ne}{IV]}1602/\ion{[Ne}{IV]}2422 ratio which imply n$_e$ $\la$ 10$^6$ cm$^{-3}$ for most of the QSO2s.\\ There are two QSO2s where the \ion{[Ne}{IV]}1602/\ion{[Ne}{IV]}2422 ratio implies n$_e$ $\ga$ 10$^6$ cm$^{-3}$.\end{tabular} \\ \hline
			\ref{neiv01s} & \ion{[Ne}{IV]}1602/\ion{[Ne}{IV]}2422 $vs.$ \ion{C}{IV}/\ion{He}{II} & \textbf{SC} & \begin{tabular}[c]{@{}l@{}}Consistent with the results using PC.\end{tabular} \\ \hline
			\ref{neiv02} & \ion{[Ne}{IV]}2422/\ion{C}{IV} $vs.$ \ion{C}{IV}/\ion{He}{II} & \textbf{PC} & \begin{tabular}[c]{@{}l@{}}The models fail to reach the positions of most of the plotted QSO2s, which lie above and/or to the right of the models.\end{tabular} \\ \hline
			\ref{neiv02s} & \ion{[Ne}{IV]}2422/\ion{C}{IV} $vs.$ \ion{C}{IV}/\ion{He}{II} & \textbf{SC} & \begin{tabular}[c]{@{}l@{}}Models produce much higher CIV/HeII ratios even at low gas density, allowing the positions of the QSO2s to be generally well reproduced in this diagram.\end{tabular} \\ \hline
			\ref{neiv03} & \ion{[Ne}{IV]}2422/\ion{He}{II} $vs.$ \ion{C}{IV}/\ion{He}{II} & \textbf{PC} & \begin{tabular}[c]{@{}l@{}}The models fail to reach the positions of most of the plotted QSO2s, which lie above and/or to the right of the models.\end{tabular} \\ \hline
			\ref{neiv03s} & \ion{[Ne}{IV]}2422/\ion{He}{II} $vs.$ \ion{C}{IV}/\ion{He}{II} & \textbf{SC} & \begin{tabular}[c]{@{}l@{}}Models produce much higher CIV/HeII ratios even at low gas density, allowing the positions of the QSO2s to be generally well reproduced in this diagram.\end{tabular} \\ \hline
			\ref{neiv02} & \ion{[N}{II]}/\ion{H}{$\alpha$} $vs.$ \ion{C}{IV}/\ion{He}{II} & \textbf{PC} & \begin{tabular}[c]{@{}l@{}}Some QSO2s appear consistent with intermediate gas density and super-Solar gas metallicity.\\ The HzRGs appear consistent with low gas density and super-Solar gas metallicity.\end{tabular} \\ \hline
			\ref{neiv02s} & \ion{[N}{II]}/\ion{H}{$\alpha$} $vs.$ \ion{C}{IV}/\ion{He}{II} & \textbf{SC} & \begin{tabular}[c]{@{}l@{}}Models with super-Solar gas metallicity (Z/Z$_{\odot}$ $\gtrsim$ 2.0) appears to provide the closest fit to the observed flux ratios of most of the QSO2s.\end{tabular} \\ \hline
			\ref{neiv03} & \ion{[N}{II]}/\ion{H}{$\alpha$} $vs.$ \ion{N}{V}/\ion{He}{II} & \textbf{PC} & \begin{tabular}[c]{@{}l@{}}Models with super-Solar gas metallicity (Z/Z$_{\odot}$ $\gtrsim$ 2.0) appears to provide the closest fit to the observed flux ratios of most of the QSO2s.\\ There is no evidence for a systematic difference in the \ion{[N}{II]}/\ion{H}{$\alpha$} ratio between the QSO2s and the HzRGs.\end{tabular} \\ \hline
			\ref{neiv03s} & \ion{[N}{II]}/\ion{H}{$\alpha$} $vs.$ \ion{N}{V}/\ion{He}{II} & \textbf{SC} & \begin{tabular}[c]{@{}l@{}}Models fail to reach the positions of most of the QSO2s.\end{tabular} \\ \hline
		\end{tabular}%
	}
\end{table*}

\section{Photoionization Models}
\label{models}

In order to characterize the chemical abundances, density and ionization properties of the observed emission line gas, a grid of photoionization models was computed using the multipurpose code MAPPINGS Ie \citep{Bi85,Fe97}. In the interest of simplicity, we have adopted a single-slab, radiation bounded and isochoric geometry. The main input parameters were:
\begin{itemize}

	\item Spectral energy distribution (SED)\footnote{The ionizing continuum is characterized as a power law with $S_v \propto v^{\alpha}$ where $v$ is frequency (Hz) and $S_v$ is the flux density (erg s$^{-1}$ Hz$^{-1}$).} of the photoionizing continuum radiation, set to $\alpha$ = - 1.0 with a high-energy cut-off of at 1.0 $\times$ 10$^{3}$ eV \citep[e.g.][]{francis1993,VM97,villar1999,Bre2000,Ve2001,marckelson2018}. We have also briefly explored using a softer SED ($\alpha$ = - 1.5), since it is conceivable that some AGNs at z$\gtrsim$ 2 may have softer ionizing SEDs \citep[e.g.][]{binette2003,VM2}, but these models provided a considerably worse fit to the observed line ratios of the QSO2s, and thus will not be considered further.

	\item Hydrogen density n$_{H}$, with the values n$_{H}$ = 10$^{2}$, 10$^{4}$, 10$^{6}$, 10$^{7}$ or 10$^{8}$ cm$^{-3}$.

	\item Ionization parameter $U$, defined as the ratio of the ionizing photon density to the hydrogen density at the irradiated surface of a cloud\footnote{We adopt the definition $Q$/($4\pi r^{2} n_{H} c$), where $Q$ is the isotropic ionizing photon number luminosity of the source, r is the distance of the cloud from the ionizing source, $n_{H}$ is the hydrogen density and c is the speed of light.}. It was set to vary from  $U$ = 2$\times$10$^{-3}$ to 1.0.

	\item Gas metallicity $Z/Z_{\odot}$, which we define as the O/H abundance ratio divided by its Solar value as determined by \cite{asplund2006}, and for which we use the values Z/Z$_{\odot}$ = 0.5, 1.0, 2.0, 3.0, 4.0, or 5.0. For our default solar abundances set, we adopt the solar abundances of \cite{asplund2006}. With the exceptions of nitrogen and sometimes carbon and silicon, we maintain all heavy elements at their solar abundance ratio with oxygen. In the case of N, we have adopted secondary N behaviour (see \S \ref{nitrogen}), such that N/H $\propto$ O/H at $Z/Z_{\odot}$$<$0.3 and N/O $\propto$ O/H at $Z/Z_{\odot}$$\ge$0.3, to take into account its expected behaviour at moderate to high metallicity (\citealt{villar1999,henry2000}). To explore the potential impact of secondary behaviour of C, every model in grid has been run twice: once with C/O fixed at its solar ratio (primary behaviour; see  Table \ref{PC}) and once with C/O $\propto$ N/O (secondary behaviour; Table \ref{SC}; see \citep{nicholls2017}). To explore the impact of supersolar Si abundances, some of our models were also rerun with Si/O = 10 $\times$ its solar value (see Tables \ref{PCSi10} and \ref{SCSi10}). In total, our grid contains 130 individual models.
\end{itemize}

\section{Results and Discussion}
\label{discussion}

\subsection{The diagnostic diagrams}
\begin{figure*}

	\includegraphics[width=\columnwidth,height=1.50in,keepaspectratio]{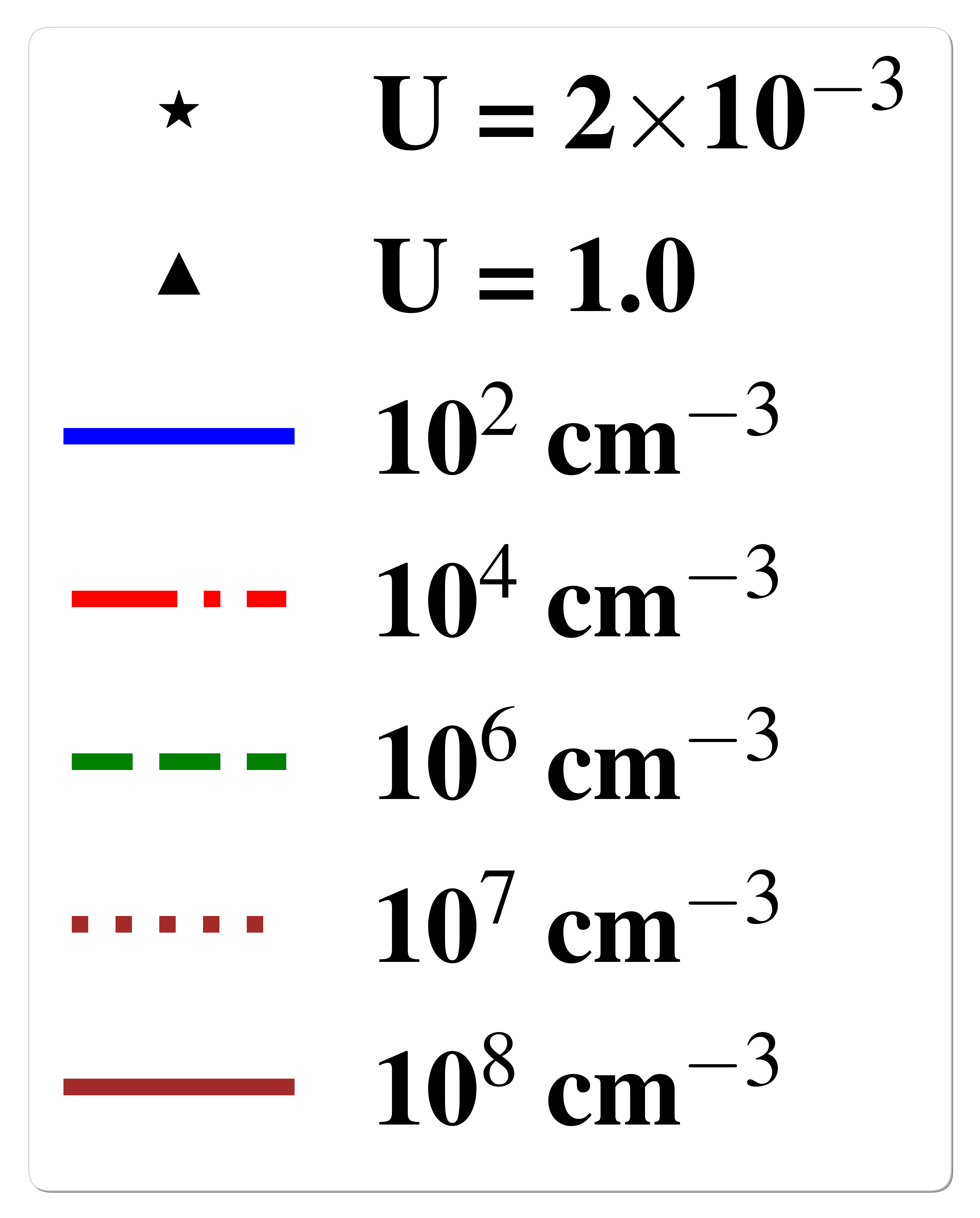}
	\includegraphics[width=\columnwidth,height=1.5in,keepaspectratio]{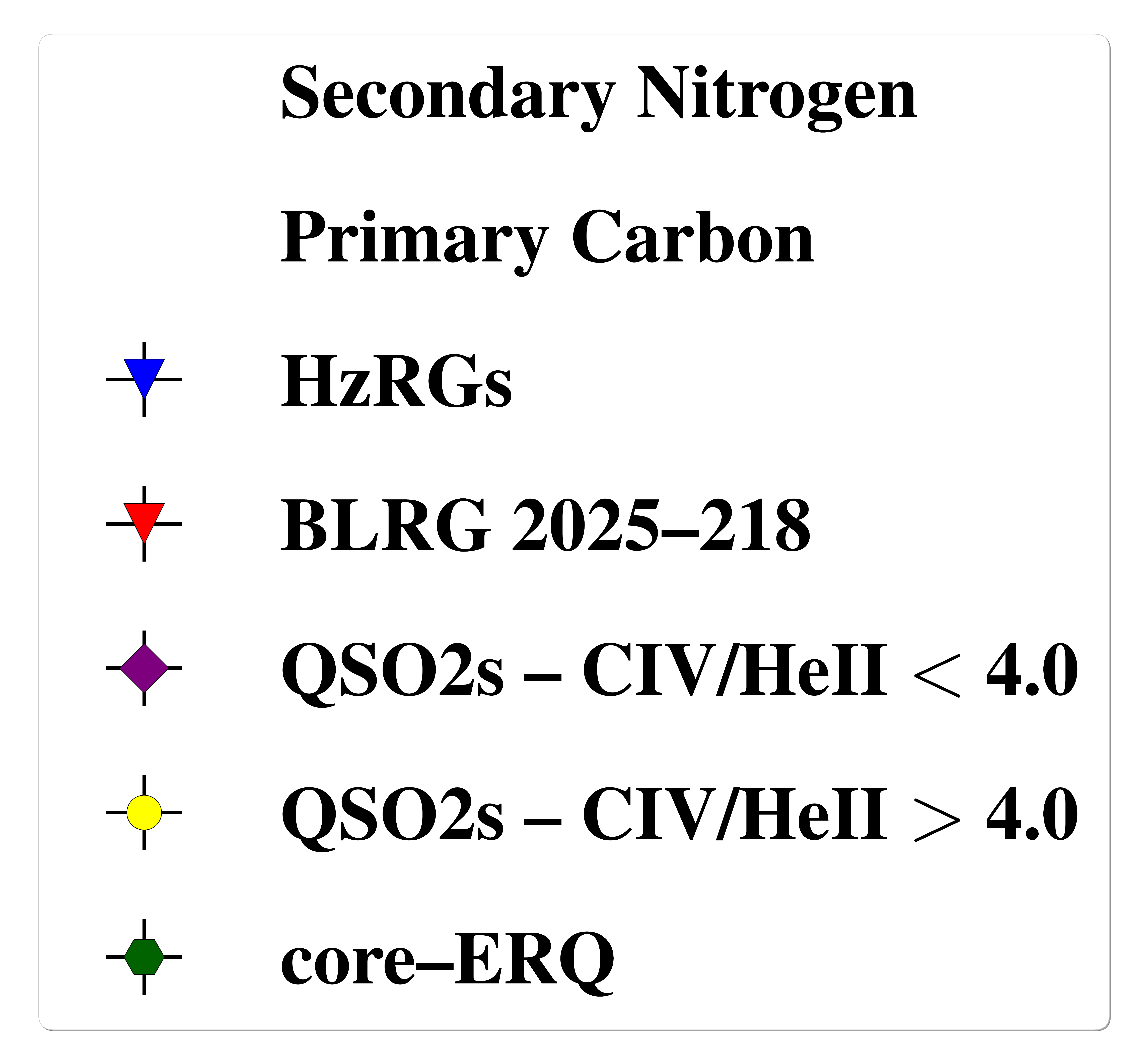}
	\quad
	\includegraphics[width=7.0in,height=7.0in]{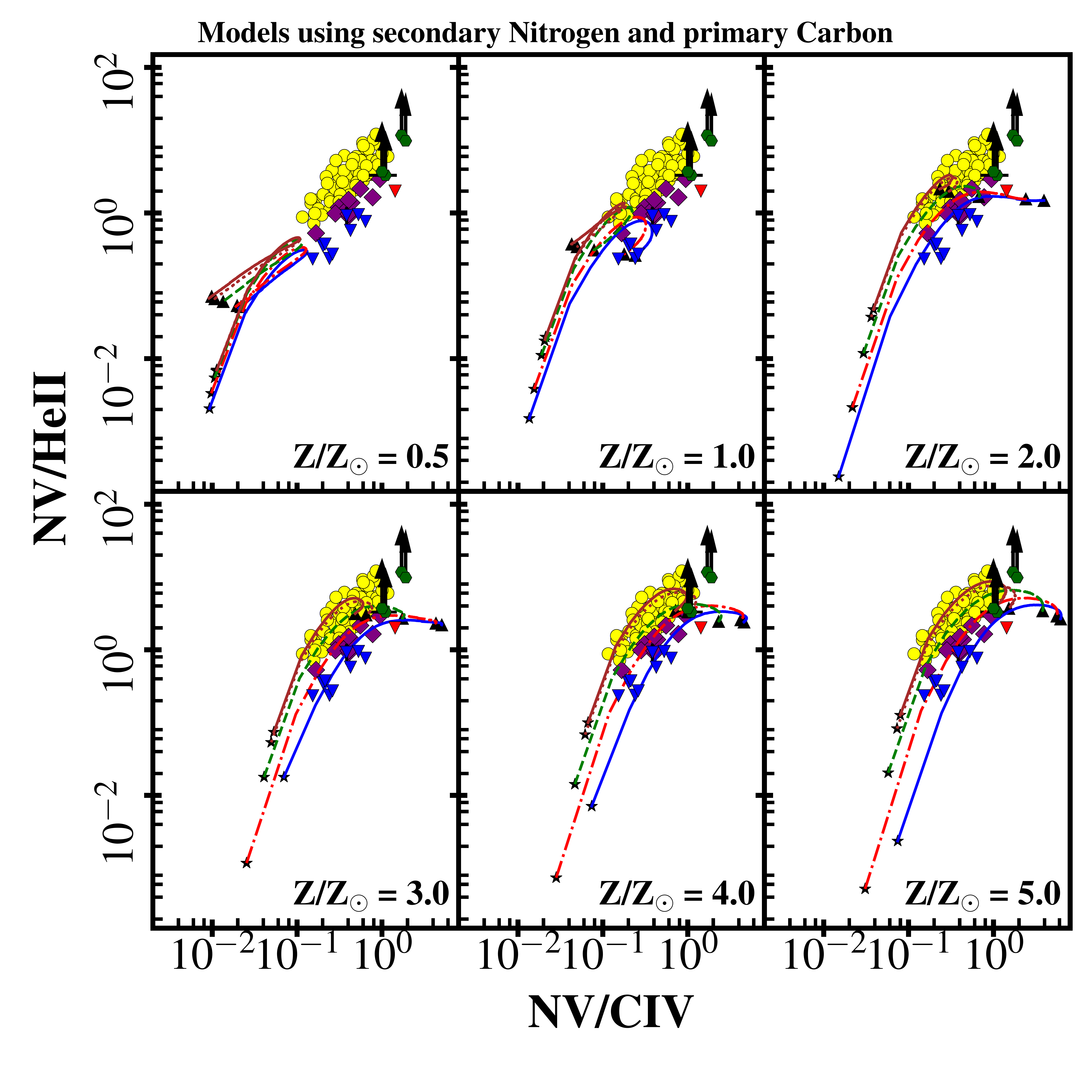}

	\caption{Comparison of the observed emission line ratios from the SDSS type II quasars divided in objects with \ion{C}{IV}/\ion{He}{II} $<$ 4 and \ion{C}{IV}/\ion{He}{II} $>$ 4 (purple filled diamond and yellow filled circles, respectively), Keck II HzRGs from \citet{Ve2001} (blue filled triangles), BLRG 2025--218 (red filled triangle) from \citet{Hu6} and core--ERQ (green filled hexagon) from \citet{villar-martin2019} with photoionization models using ionizing continuum power law index $\alpha$ = -1.0. Each diagram presents a different gas metallicity, i.e., Z/Z$_{\odot}$ = 0.5, 1.0, 2.0, 3.0, 4.0, 5.0. Curves with different colors represent the hydrogen gas density (n$_{H}$). At the end of each sequence, a solid black triangle corresponds to the lowest ionization parameter (U = 2$\times$10$^{-3}$) while the solid black star corresponds to the maximum value of the ionization parameter (U = 1.0).}
	\label{NV01}
\end{figure*}

\begin{figure*}

	\includegraphics[width=\columnwidth,height=1.50in,keepaspectratio]{Fig/legend.pdf}
	\includegraphics[width=\columnwidth,height=1.5in,keepaspectratio]{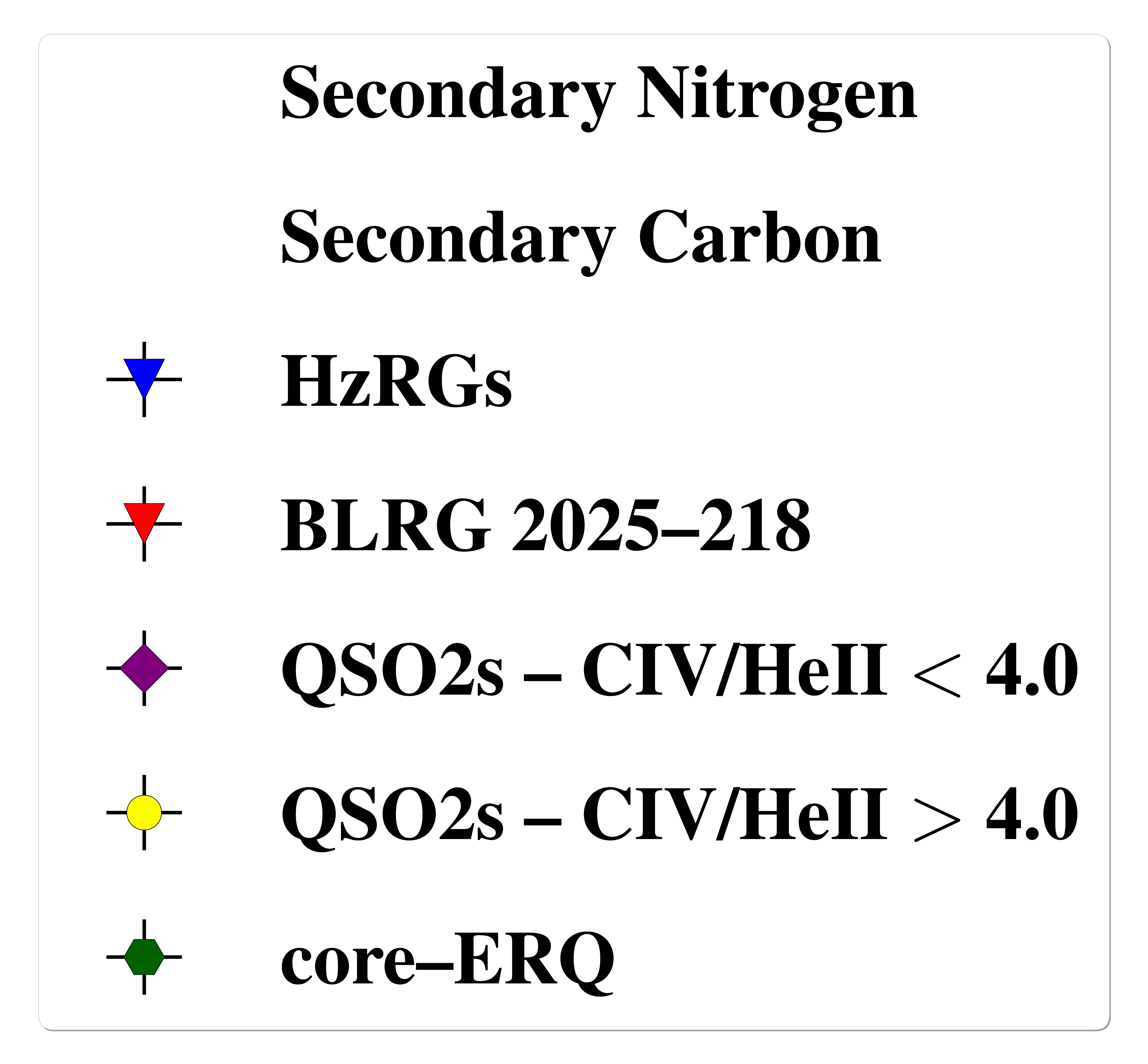}
	\quad
	\includegraphics[width=7.0in,height=7.0in]{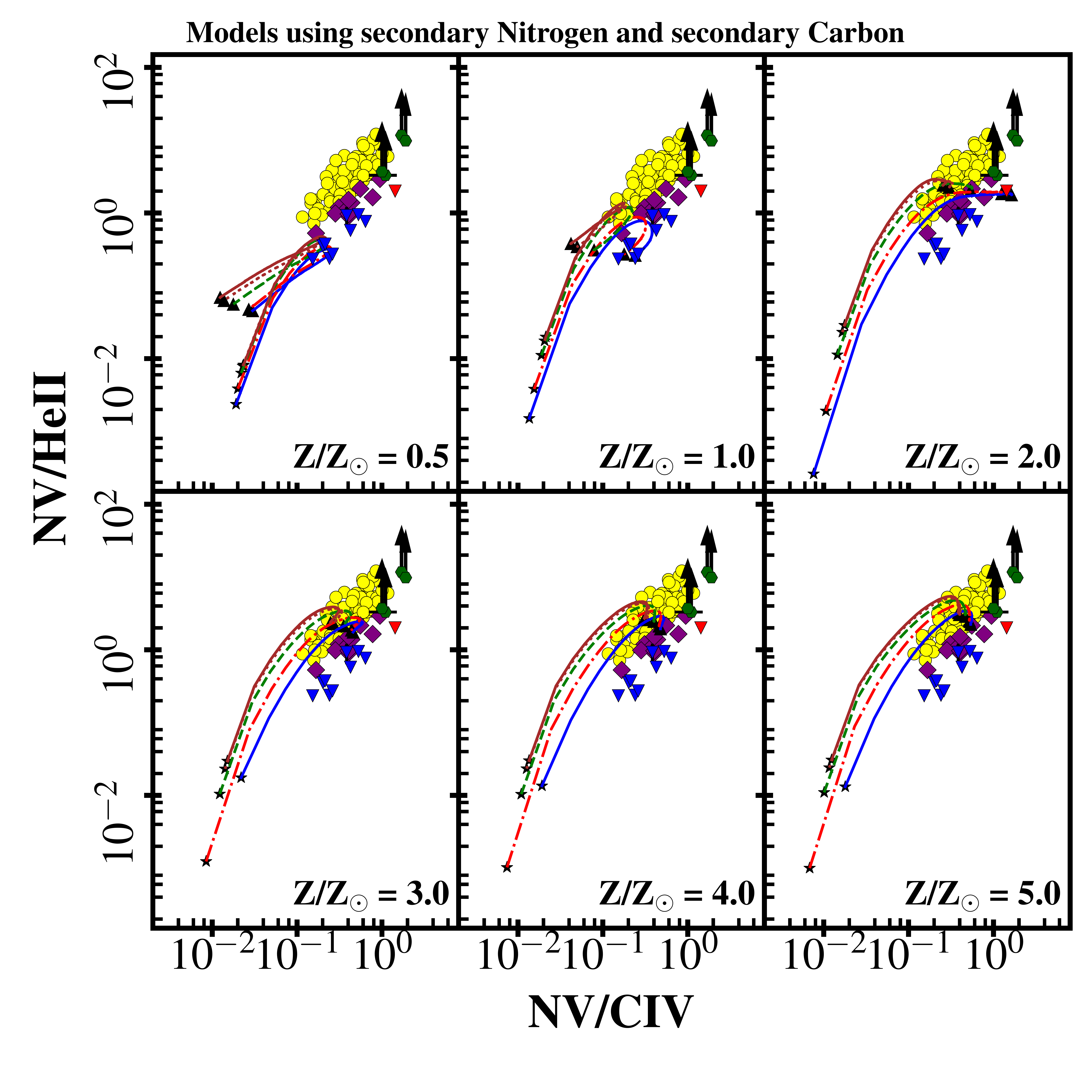}

	\caption{Comparison of the observed emission line ratios from the SDSS type II quasars divided in objects with \ion{C}{IV}/\ion{He}{II} $<$ 4 and \ion{C}{IV}/\ion{He}{II} $>$ 4 (purple filled diamond and yellow filled circles, respectively), Keck II HzRGs from \citet{Ve2001} (blue filled triangles), BLRG 2025--218 (red filled triangle) from \citet{Hu6} and core--ERQ (green filled hexagon) from \citet{villar-martin2019} with photoionization models using ionizing continuum power law index $\alpha$ = -1.0. Nitrogen and Carbon are a secondary elements in which their abundances are proportional to the square of the metallicity. Each diagram presents a different gas metallicity, i.e., Z/Z$_{\odot}$ = 0.5, 1.0, 2.0, 3.0, 4.0, 5.0. Curves with different colors represent the hydrogen gas density (n$_{H}$). At the end of each sequence, a solid black triangle corresponds to the lowest ionization parameter (U = 2$\times$10$^{-3}$) while the solid black star corresponds to the maximum value of the ionization parameter (U = 1.0).}
	\label{NV01s}
\end{figure*}

\begin{figure*}

	\includegraphics[width=\columnwidth,height=1.60in,keepaspectratio]{Fig/legend.pdf}
	\includegraphics[width=\columnwidth,height=1.6in,keepaspectratio]{Fig/legend_01a.pdf}
	\quad
	\includegraphics[width=7.0in,height=7.0in]{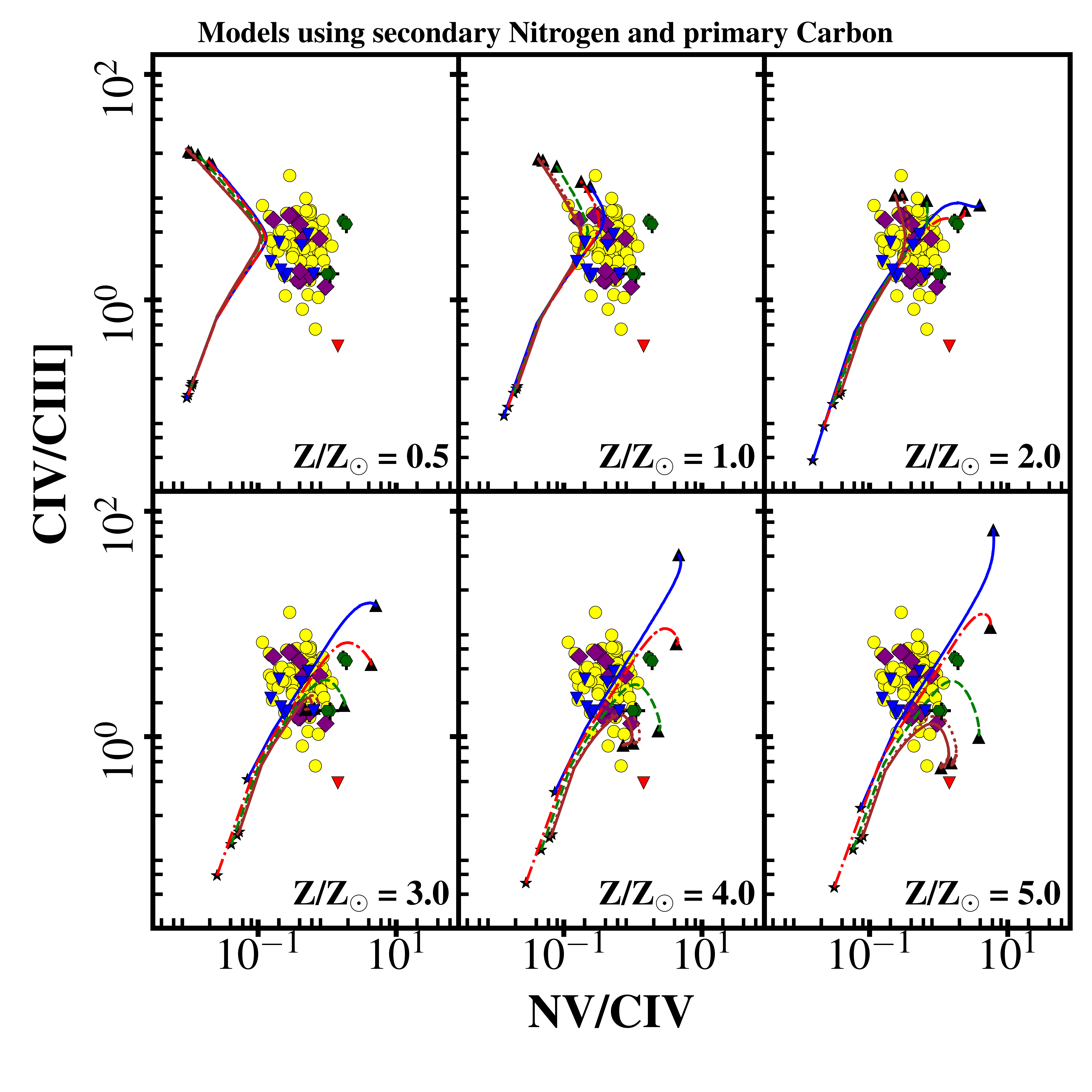}

	\caption{Comparison of the observed emission line ratios from the SDSS type II quasars divided in objects with \ion{C}{IV}/\ion{He}{II} $<$ 4 and \ion{C}{IV}/\ion{He}{II} $>$ 4 (purple filled diamond and yellow filled circles, respectively), Keck II HzRGs from \citet{Ve2001} (blue filled triangles), BLRG 2025--218 (red filled triangle) from \citet{Hu6} and core--ERQ (green filled hexagon) from \citet{villar-martin2019} with photoionization models using ionizing continuum power law index $\alpha$ = -1.0. Each diagram presents a different gas metallicity, i.e., Z/Z$_{\odot}$ = 0.5, 1.0, 2.0, 3.0, 4.0, 5.0. Curves with different colors represent the hydrogen gas density (n$_{H}$). At the end of each sequence, a solid black triangle corresponds to the lowest ionization parameter (U = 2$\times$10$^{-3}$) while the solid black star corresponds to the maximum value of the ionization parameter (U = 1.0).}
	\label{NV02}
\end{figure*}

\begin{figure*}

	\includegraphics[width=\columnwidth,height=1.50in,keepaspectratio]{Fig/legend.pdf}
	\includegraphics[width=\columnwidth,height=1.5in,keepaspectratio]{Fig/legend_02a.pdf}
	\quad
	\includegraphics[width=7.0in,height=7.0in]{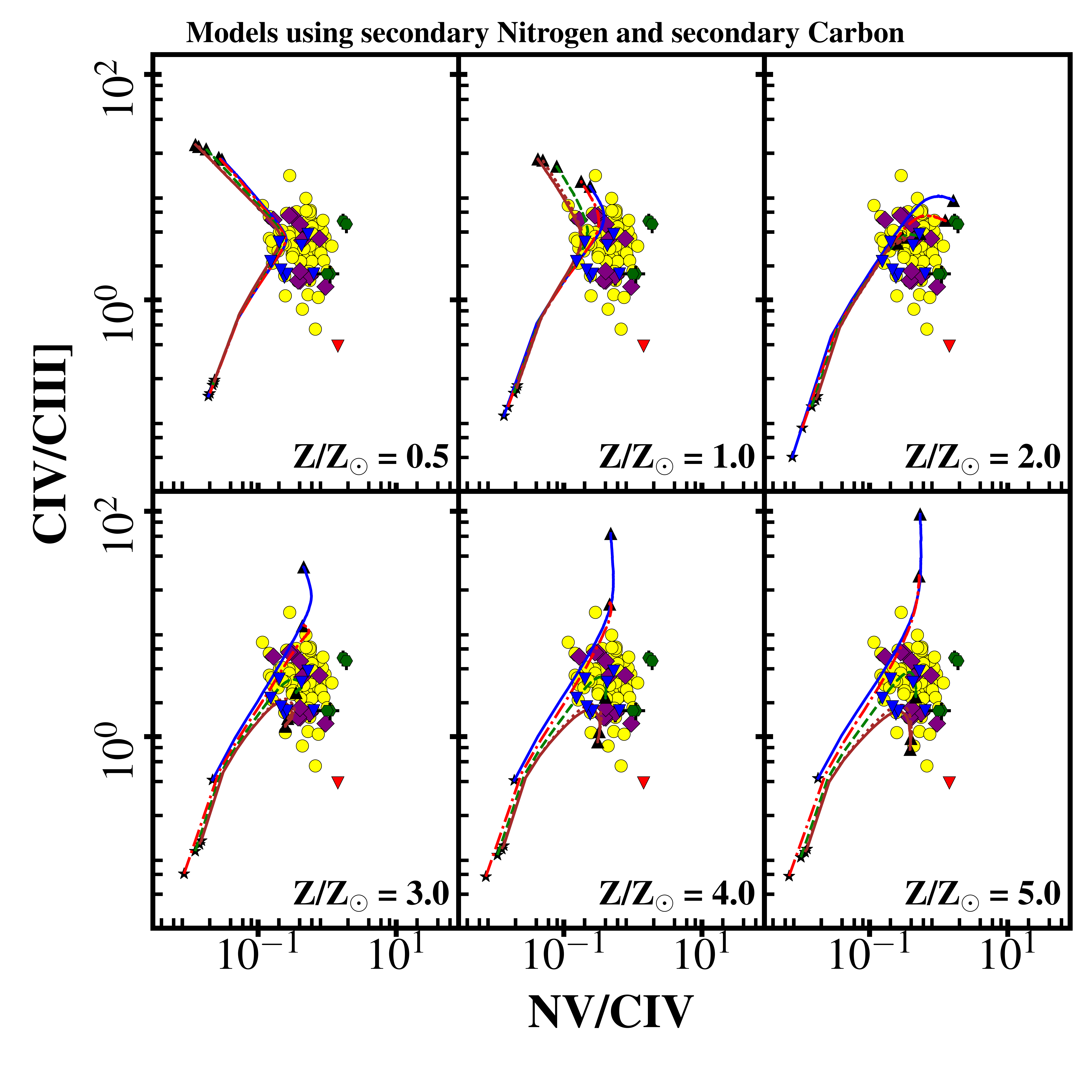}

	\caption{Comparison of the observed emission line ratios from the SDSS type II quasars divided in objects with \ion{C}{IV}/\ion{He}{II} $<$ 4 and \ion{C}{IV}/\ion{He}{II} $>$ 4 (purple filled diamond and yellow filled circles, respectively), Keck II HzRGs from \citet{Ve2001} (blue filled triangles), BLRG 2025--218 (red filled triangle) from \citet{Hu6} and core--ERQ (green filled hexagon) from \citet{villar-martin2019} with photoionization models using ionizing continuum power law index $\alpha$ = -1.0. Carbon is a secondary element in which its abundance is proportional to the square of the metallicity. Each diagram presents a different gas metallicity, i.e., Z/Z$_{\odot}$ = 0.5, 1.0, 2.0, 3.0, 4.0, 5.0. Curves with different colors represent the hydrogen gas density (n$_{H}$). At the end of each sequence, a solid black triangle corresponds to the lowest ionization parameter (U = 2$\times$10$^{-3}$) while the solid black star corresponds to the maximum value of the ionization parameter (U = 1.0).}
	\label{NV02s}
\end{figure*}

\begin{figure*}

	\includegraphics[width=\columnwidth,height=1.50in,keepaspectratio]{Fig/legend.pdf}
	\includegraphics[width=\columnwidth,height=1.5in,keepaspectratio]{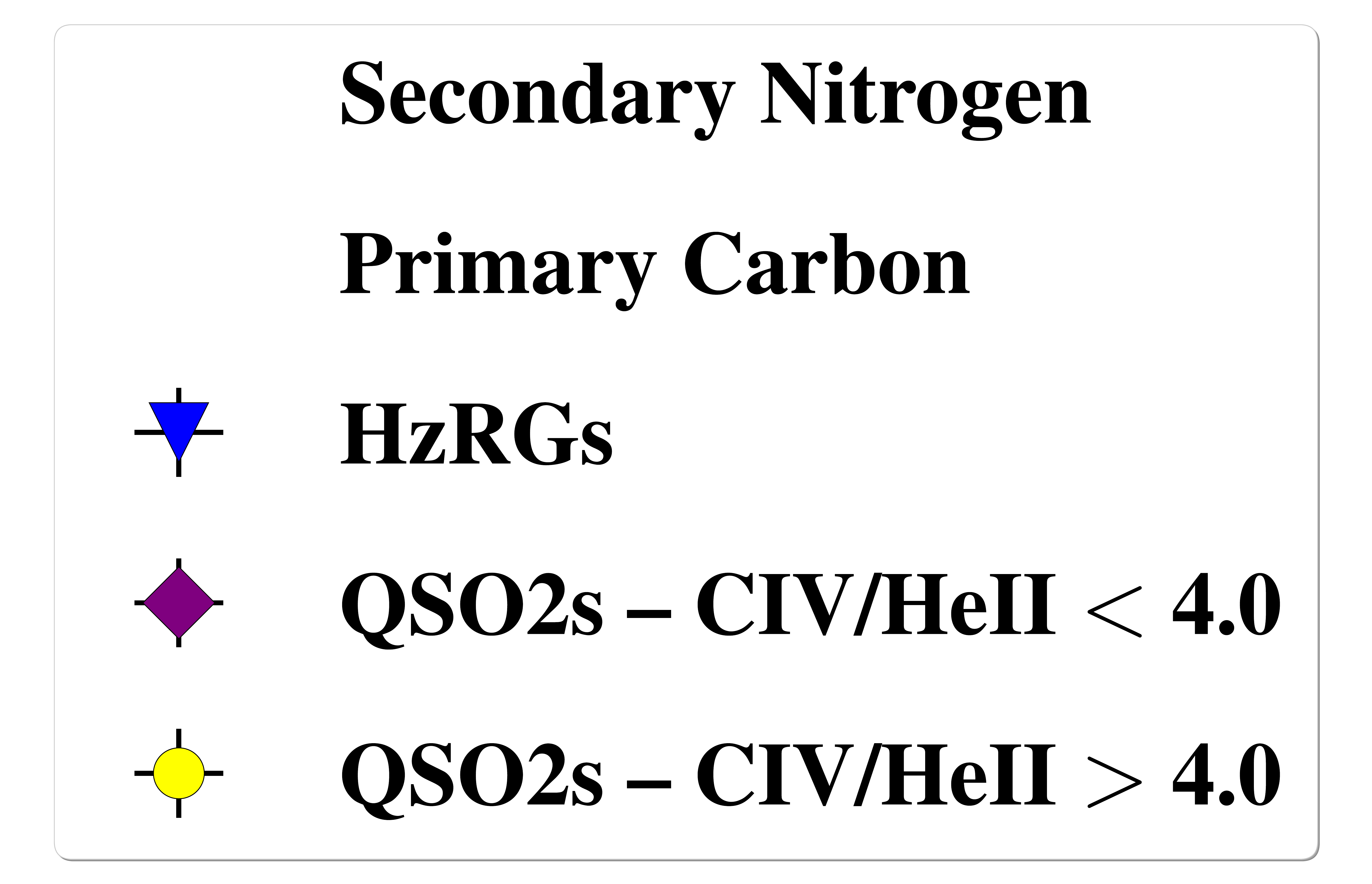}

	\quad
	\includegraphics[width=7.0in,height=7.0in]{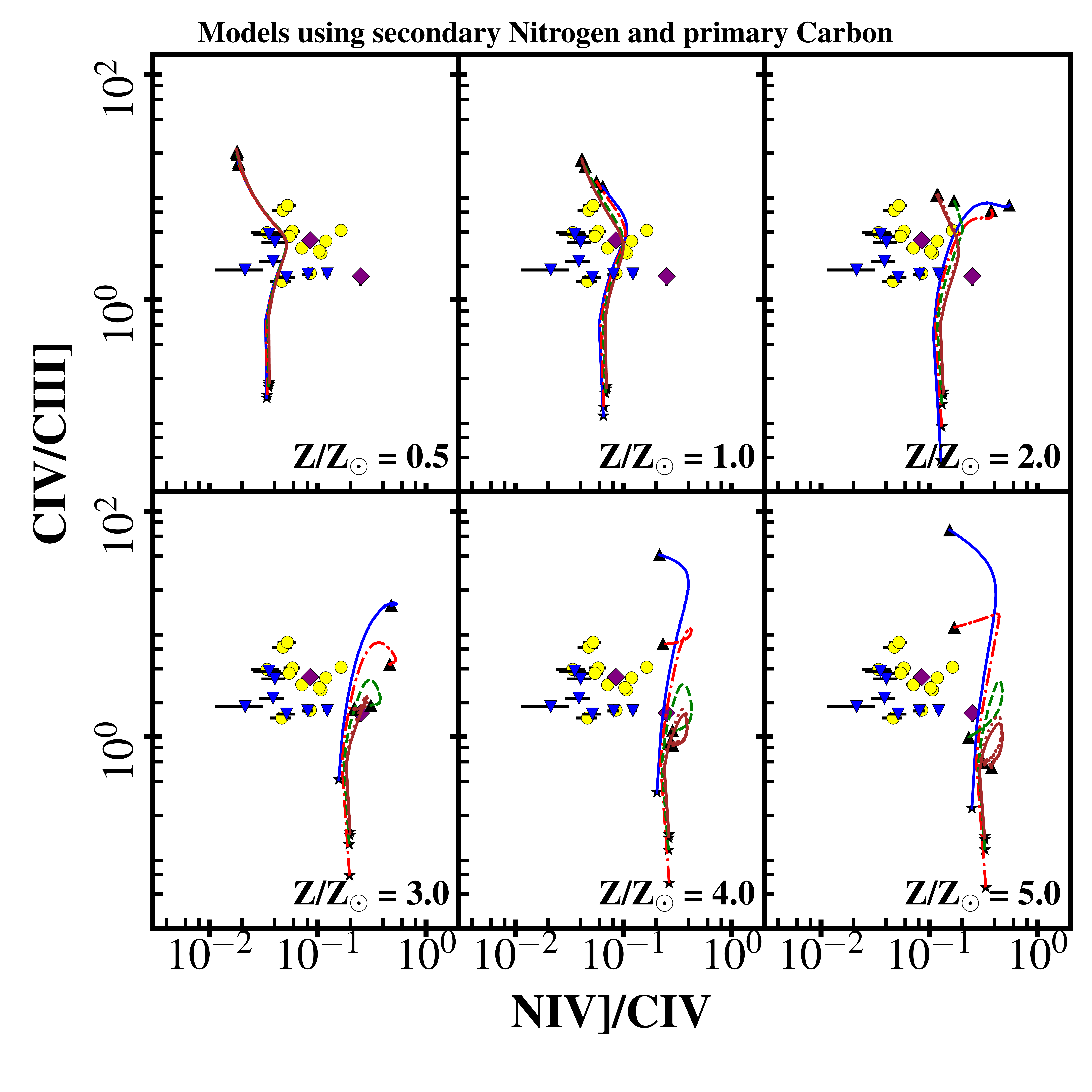}

	\caption{Comparison of the observed emission line ratios from the SDSS type II quasars divided in objects with \ion{C}{IV}/\ion{He}{II} $<$ 4 and \ion{C}{IV}/\ion{He}{II} $>$ 4 (purple filled diamond and yellow filled circles, respectively) and Keck II HzRGs from \citet{Ve2001} (blue filled triangles) with photoionization models using ionizing continuum power law index $\alpha$ = -1.0. Each diagram presents a different gas metallicity, i.e., Z/Z$_{\odot}$ = 0.5, 1.0, 2.0, 3.0, 4.0, 5.0. Curves with different colors represent the hydrogen gas density (n$_{H}$). At the end of each sequence, a solid black triangle corresponds to the lowest ionization parameter (U = 2$\times$10$^{-3}$) while the solid black star corresponds to the maximum value of the ionization parameter (U = 1.0).}
	\label{NIV}
\end{figure*}

\begin{figure*}

	\includegraphics[width=\columnwidth,height=1.50in,keepaspectratio]{Fig/legend.pdf}
	\includegraphics[width=\columnwidth,height=1.5in,keepaspectratio]{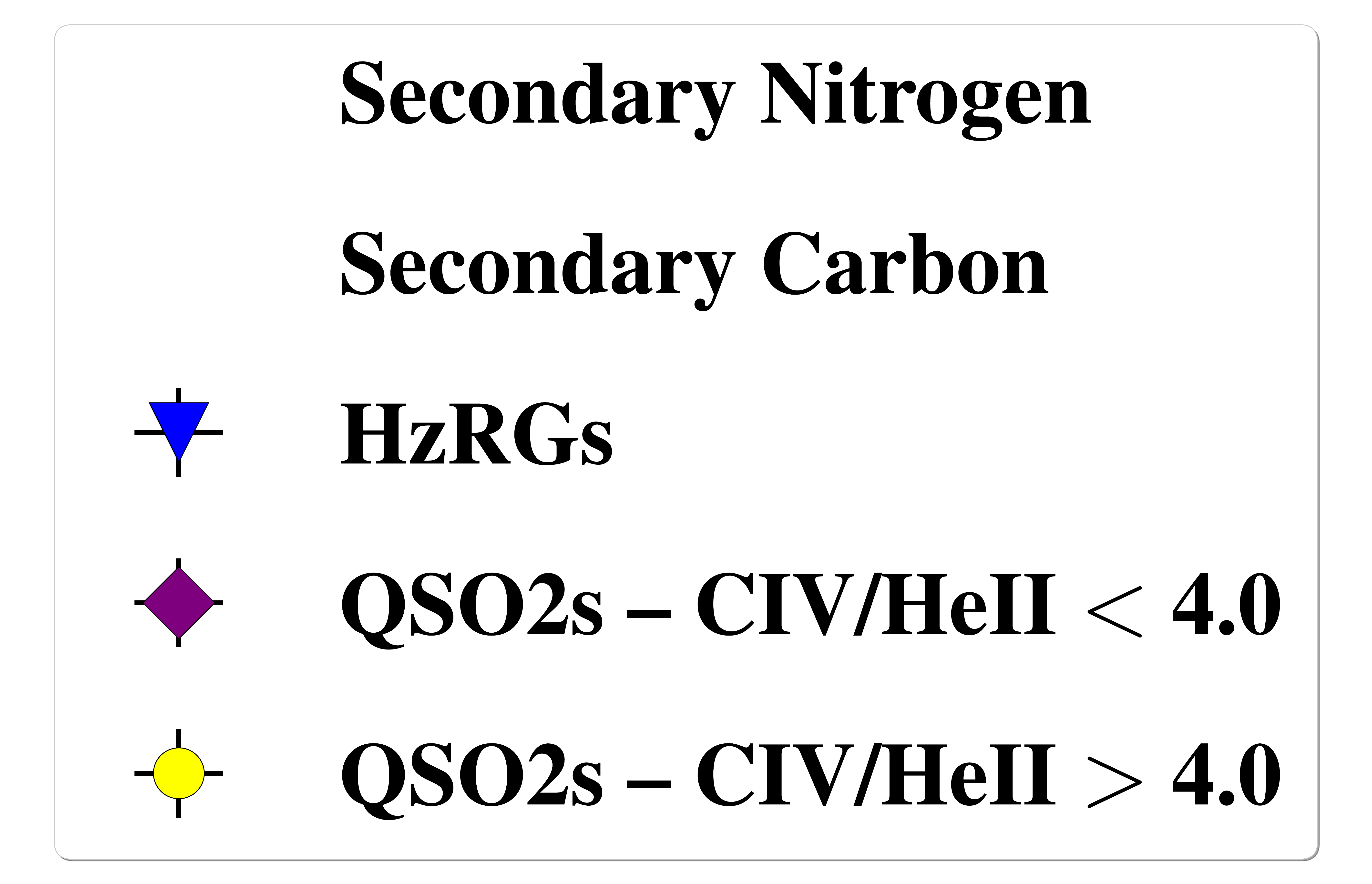}
	\quad
	\includegraphics[width=7.0in,height=7.0in]{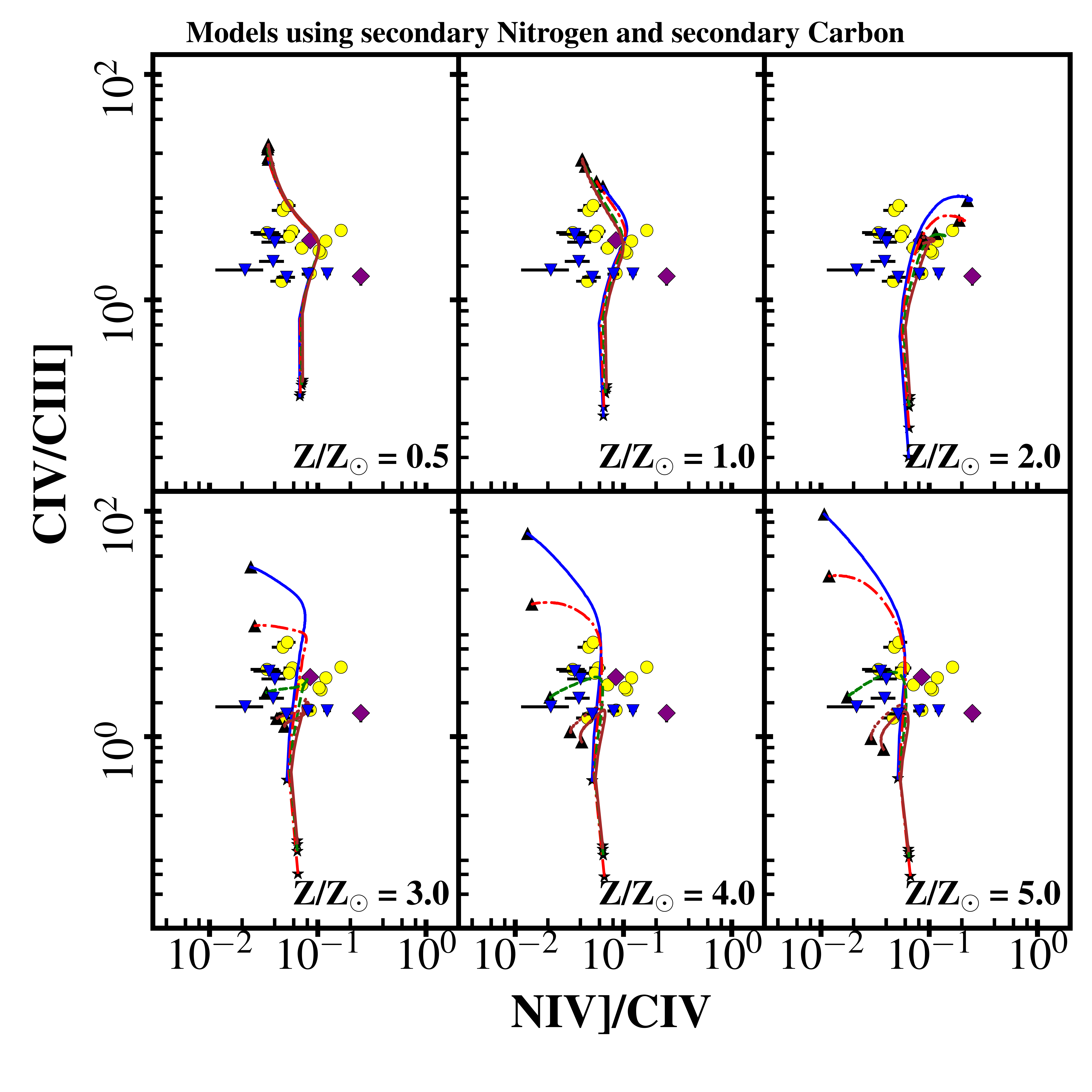}

	\caption{Comparison of the observed emission line ratios from the SDSS type II quasars divided in objects with \ion{C}{IV}/\ion{He}{II} $<$ 4 and \ion{C}{IV}/\ion{He}{II} $>$ 4 (purple filled diamond and yellow filled circles, respectively) and Keck II HzRGs from \citet{Ve2001} (blue filled triangles) with photoionization models using ionizing continuum power law index $\alpha$ = -1.0. Carbon is a secondary element in which its abundance is proportional to the square of the metallicity. Each diagram presents a different gas metallicity, i.e., Z/Z$_{\odot}$ = 0.5, 1.0, 2.0, 3.0, 4.0, 5.0. Curves with different colors represent the hydrogen gas density (n$_{H}$). At the end of each sequence, a solid black triangle corresponds to the lowest ionization parameter (U = 2$\times$10$^{-3}$) while the solid black star corresponds to the maximum value of the ionization parameter (U = 1.0).}
	\label{NIVs}
\end{figure*}

In Figures \ref{NV01}--\ref{CIVs} and \ref{SiIII}--\ref{nii03s}, we show diagnostic diagrams with our grid of photoionization models and observed emission line ratios of the QSO2s, HzRGs and core-ERQs. In these Figures, each line represents a sequence in ionization parameter U at fixed values of n$_{H}$ and $Z$. The six panels in each figure show the models at a different gas metallicity $Z$. The measured line ratios of the QSO2s are shown by purple filled diamonds (\ion{C}{IV}/\ion{He}{II} $<$ 4) or yellow filled circles (\ion{C}{IV}/\ion{He}{II} $>$ 4). The HzRGs are shown by blue filled triangles; however, for the special case of the BLRG 2025--218, we use instead a red filled triangle. The core-ERQs are shown by green filled hexagon. Arrows indicate 3$\sigma$ upper or lower limits (see for example Figs. \ref{neiv01} and \ref{neiv01s}). For a summary of the results obtained with each diagram, see Table \ref{summary}.

\subsection{Gas chemical abundances}
\label{chemical}

\subsubsection{Nitrogen abundance}
\label{nitrogen}

Nitrogen is mainly produced by the CN cycle of the CNO reactions which catalyze hydrogen burning in stars. It is mostly a secondary element being synthesized from the C and O already present in the stars, and the increase in the abundance of N should be proportional to the initial C and O content, which will be proportional to the square of the metal content in a galaxy \citep{hamman1992,matteucci1993,matteucci1996,cristina2003,nicholls2017}. As such, line ratios involving N are valuable tracers of chemical enrichment, since this element is selectively enhanced by secondary processing in stellar populations.

The study of the gas metallicity using nitrogen lines (more specifically \ion{N}{V}) over \ion{C}{IV} or \ion{He}{II} has been applied to the BLRs and NLRs of high-z quasars \citep[e.g.][]{hamman1993,ferland1996,Ha99,dietrich2000,dietrich2001,hamman2002,baldwin2003,nagao2006a,nagao2006b,xu2018} and also HzRGs \citep[e.g.][]{vO1994,VM97,villar1999,Bre2000,Ve2001,overzier2001,Hu4,sandy2017,marckelson2018}.
Such studies have generally concluded that the BLR or NLR of many high-z quasars and HzRGs have solar or supersolar gas metallicities.

Looking first at the \ion{N}{V}-diagram (Fig. \ref{NV01} and \ref{NV01s}), the observational data shows an order of magnitude range in \ion{N}{V}/\ion{C}{IV} and \ion{N}{V}/\ion{He}{II}, with the data running along three parallel correlations: (i) the HzRGs; (ii) the QSO2s with \ion{C}{IV}/\ion{He}{II} $<$ 4 offset vertically to higher \ion{N}{V}/\ion{He}{II} values; (iii) the QSO2s with \ion{C}{IV}/\ion{He}{II} $>$ 4 and the core-ERQs vertically offset to still higher values of \ion{N}{V}/\ion{He}{II}.
In this diagram, there is a degeneracy between nitrogen abundance and U, such that a sequence in U can mimic a sequence in nitrogen abundance, and vice versa \citep[see also][]{Hu4}. The trend defined by the QSO2s sample runs parallel to the loci of varying U and varying metallicity, suggesting that U and/or metallicity varies significantly between the quasars. Based on their position relative to the models, QSO2s at the low \ion{N}{V}/\ion{He}{II} and \ion{N}{V}/\ion{C}{IV} end of the trend require N/H of at least 0.25$\times$Solar, corresponding to O/H = 0.5$\times$solar if our adopted relation between N/H and O/H is correct (Fig. \ref{NV01} and \ref{NV01s}, top left panel)\footnote{Strictly speaking, using \ion{N}{V}/\ion{He}{II} in this way we obtain the N/He ratio, which we use as a proxy for N/H assuming He/H $=$ 0.085.}.

Conversely, to explain the QSO2s at the high end of the \ion{N}{V}/\ion{He}{II} and \ion{N}{V}/\ion{C}{IV} trend (including the core-ERQs from \citealt{villar-martin2019}), we find that a combination of high gas density (n$_H$ $\gtrsim$ 10$^{7}$ cm$^{-3}$) and N/H $\gtrsim$ 16$\times$solar is required, corresponding to O/H $\gtrsim$ 4$\times$solar. A potential alternative to n$_H$ $\gtrsim$ 10$^{7}$ cm$^{-3}$ could be that this subset of QSO2s and the four core-ERQs have a nitrogen abundance that is a factor of $\ga$2 higher than assumed in our highest metallicity models. While there is potentially a large variation in N/H throughout the QSO2 sample, from N/H $\sim$ 0.25 to N/H $\sim$ 16$\times$solar, the degeneracy between U and N/H inherent to the \ion{N}{V}-diagram means that the QSO2 data are also compatible with a range in U and a smaller (or no) variation in N/H between objects. For instance, our sequence in U with 4$\times$solar metallicity is able to reproduce the full range of \ion{N}{V}/\ion{He}{II} and \ion{N}{V}/\ion{C}{IV} shown by the QSO2s.

When using secondary behaviour for carbon, the \ion{N}{V}-diagram changes such that even at high-U and high metallicity, the models fail to reproduce the positions of the core-ERQs and also of $\sim$ 30\% of QSO2s with the highest values of \ion{N}{V}/\ion{C}{IV} (Fig. \ref{NV01s}). This is because our adopted secondary behaviour of C and N results in a constant solar N/C abundance ratio at all values of gas metallicity, preventing the supersolar N/C ratios that give rise to the highest \ion{N}{V}/\ion{C}{IV} line ratios in Fig. \ref{NV01}. Nevertheless, Fig. \ref{NV01} and Fig. \ref{NV01s} give consistent results in terms of \ion{N}{V}/\ion{He}{II} and the N/H abundance ratio, because this ratio is unaffected by the choice between using the primary or secondary behaviour of carbon.

In Figs. \ref{NV02} and \ref{NV02s} we show the \ion{N}{V}/\ion{C}{IV} $vs.$ \ion{C}{IV}/\ion{C}{III]} diagram, which one might expect to provide information about the possible degeneracy between N abundance and U.
However, no obvious correlation is seen between the \ion{N}{V}/\ion{C}{IV} ratio and the U-sensitive \ion{C}{IV}/\ion{C}{III]} ratio. The observed CIV/CIII] ratio varies by more than an order of magnitude in the QSO2 sample, suggesting significant variation in U between objects. However, the degeneracy between U, metallicity and density in this diagram makes it challenging to determine the extent to which U might vary in the sample, or whether U genuinely varies at all.
Given the complex behaviour of the U-sequences, which are highly dependent on the density and metallicity of the gas we suggest that  at least two out of U, Z and n$_H$ vary significantly in the QSO2 sample, in order to produce the observed distribution of points in this diagram.

The observed flux ratio \ion{N}{IV]}/\ion{C}{IV} (see Figs. \ref{NIV} and \ref{NIVs}) shows significant scatter perpendicular to the model loci, which is difficult to explain by variation in U or n$_H$, and which strongly suggests a significant variation in the N/C abundance ratio between objects. However, it is unclear whether the relative variation is due to differences in the ISM abundance ratio, differential depletion of N and C into dust grains, or interstellar absorption of \ion{C}{IV}. Previous modeling which used primary carbon behaviour systematically overpredicted \ion{N}{IV]}/\ion{He}{II} and \ion{N}{IV]}/\ion{C}{IV} by a factor of $\sim$ 2 \citep[e.g.][]{Ve2001,Hu4} but, interestingly, using secondary behaviour for both N and C appears to alleviate this problem involving this so-called \ion{N}{IV]}-problem.

Another interesting result from our models is that when secondary behaviour is adopted for carbon, the N/C abundance ratio (and consequently the \ion{N}{V}/\ion{C}{IV} and \ion{N}{IV]}/\ion{C}{IV} flux ratios) ceases to be a useful metallicity indicator. This is because, at least in the scheme we have adopted for the secondary behaviour of N and C, the N/C abundance ratio remains constant across the full range of gas metallicity.

Although the gas phase metallicity of the HzRGs has been examined in several previous studies \citep{villar1999,Ve2001,Hu4}, we revisit this topic here in light of our new analysis and models. Consistent with previous authors \citep{villar1999,Ve2001,Hu4}, we find that the HzRGs require N/H $\gtrsim$ 0.25 $\times$solar, implying gas metallicity $\gtrsim$ 0.5$\times$solar, with some objects requiring solar or supersolar N/H, and by implication in our N abundance variation scheme, solar or supersolar metallicity. Interestingly, the BLRG 2025-218 shows higher \ion{N}{V}/\ion{He}{II} and \ion{N}{V}/\ion{C}{IV} ratios than any of the HzRGs, with its \ion{N}{V}/\ion{He}{II} ratio being consistent with the group of QSO2s that have \ion{C}{IV}/\ion{He}{II} $<$ 4.
As argued by \citet{Hu4}, the positions of the HzRGs in the \ion{N}{V}-diagram are also consistent with little or no variation in N/H, and a range in ionization parameter U.
We argue that the vertical offset between the QSO2s and HzRGs implies that the QSO2s typically have significantly higher gas density (at least 2 orders of magnitude), and/or that the QSO2s have significantly higher N/H (a factor of $\sim$ 4) than the HzRGs. On the other hand, we note that there is no evidence for a systematic difference in the N/C abundance ratio between the QSO2s and the HzRGs.

\begin{figure*}

	\includegraphics[width=\columnwidth,height=1.50in,keepaspectratio]{Fig/legend.pdf}
	\includegraphics[width=\columnwidth,height=1.5in,keepaspectratio]{Fig/legend_01a.pdf}
	\quad
	\includegraphics[width=7.0in,height=7.0in]{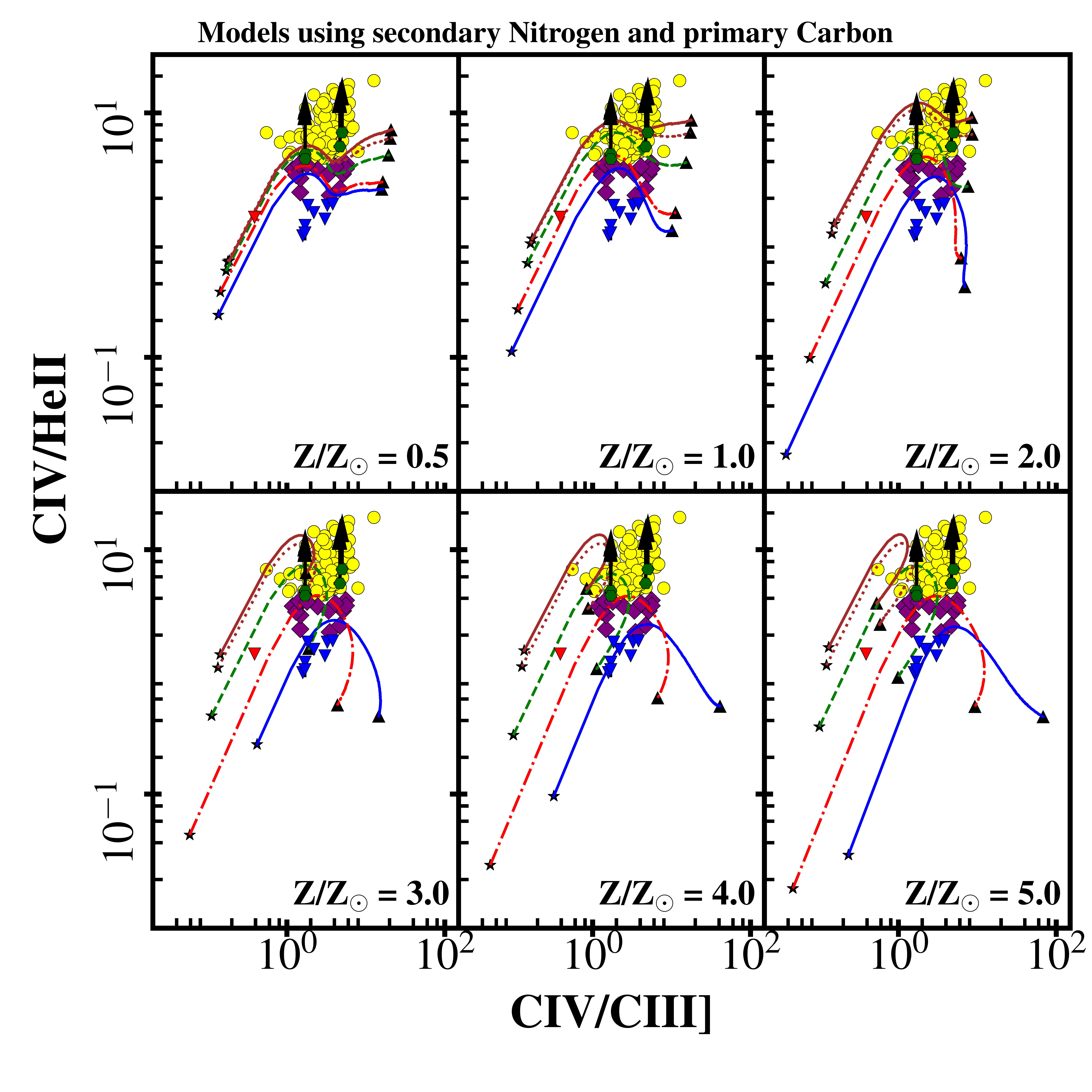}

	\caption{Comparison of the observed emission line ratios from the SDSS type II quasars divided in objects with \ion{C}{IV}/\ion{He}{II} $<$ 4 and \ion{C}{IV}/\ion{He}{II} $>$ 4 (purple filled diamond and yellow filled circles, respectively), Keck II HzRGs from \citet{Ve2001} (blue filled triangles), BLRG 2025--218 (red filled triangle) from \citet{Hu6} and core--ERQ (green filled hexagon) from \citet{villar-martin2019} with photoionization models using ionizing continuum power law index $\alpha$ = -1.0. Each diagram presents a different gas metallicity, i.e., Z/Z$_{\odot}$ = 0.5, 1.0, 2.0, 3.0, 4.0, 5.0. Curves with different colors represent the hydrogen gas density (n$_{H}$). At the end of each sequence, a solid black triangle corresponds to the lowest ionization parameter (U = 2$\times$10$^{-3}$) while the solid black star corresponds to the maximum value of the ionization parameter (U = 1.0).}
	\label{CIV}
\end{figure*}

\begin{figure*}

	\includegraphics[width=\columnwidth,height=1.50in,keepaspectratio]{Fig/legend.pdf}
	\includegraphics[width=\columnwidth,height=1.5in,keepaspectratio]{Fig/legend_02a.pdf}
	\quad
	\includegraphics[width=7.0in,height=7.0in]{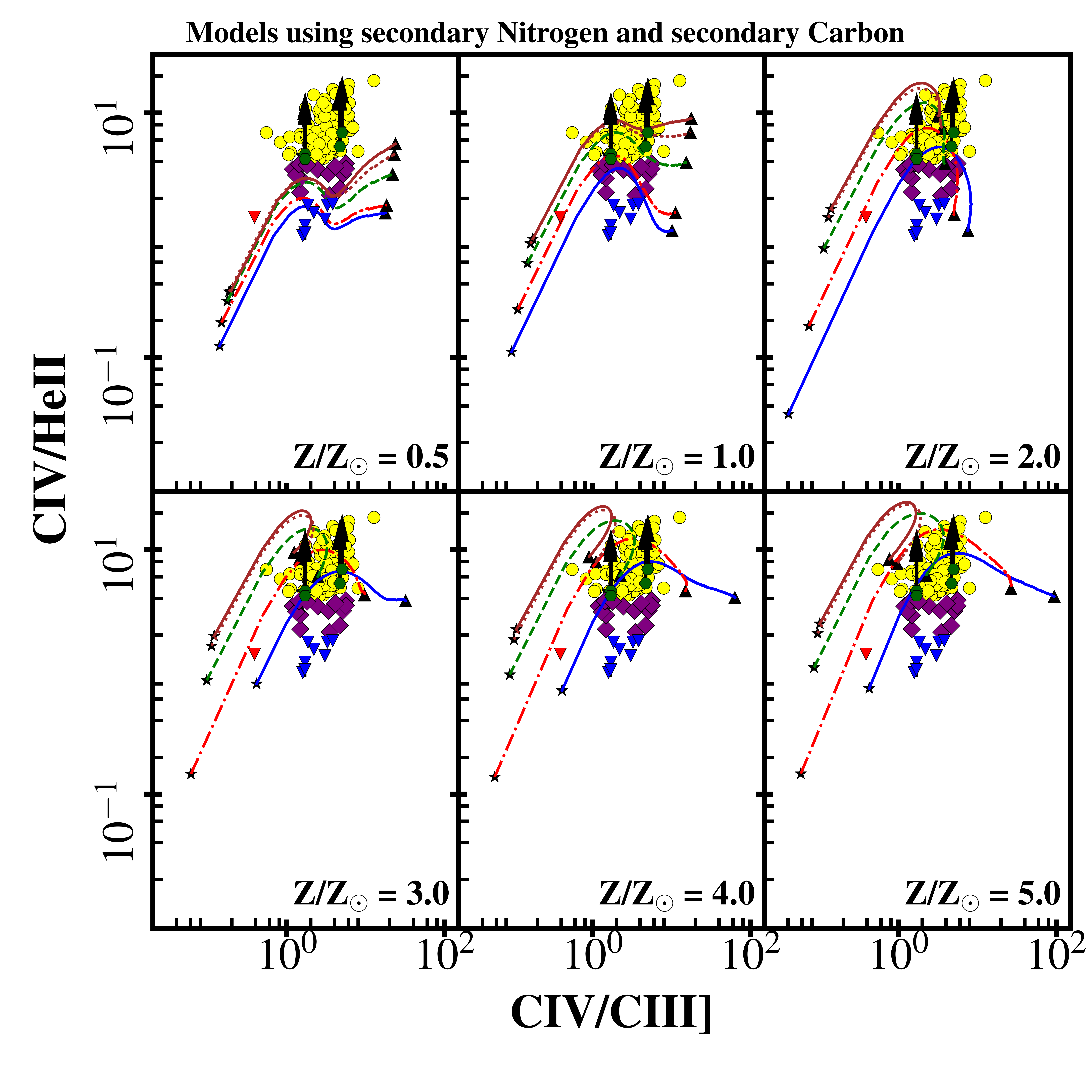}

	\caption{Comparison of the observed emission line ratios from the SDSS type II quasars divided in objects with \ion{C}{IV}/\ion{He}{II} $<$ 4 and \ion{C}{IV}/\ion{He}{II} $>$ 4 (purple filled diamond and yellow filled circles, respectively), Keck II HzRGs from \citet{Ve2001} (blue filled triangles), BLRG 2025--218 (red filled triangle) from \citet{Hu6} and core--ERQ (green filled hexagon) from \citet{villar-martin2019} with photoionization models using ionizing continuum power law index $\alpha$ = -1.0. Carbon is a secondary element in which its abundance is proportional to the square of the metallicity. Each diagram presents a different gas metallicity, i.e., Z/Z$_{\odot}$ = 0.5, 1.0, 2.0, 3.0, 4.0, 5.0. Curves with different colors represent the hydrogen gas density (n$_{H}$). At the end of each sequence, a solid black triangle corresponds to the lowest ionization parameter (U = 2$\times$10$^{-3}$) while the solid black star corresponds to the maximum value of the ionization parameter (U = 1.0).}
	\label{CIVs}
\end{figure*}

\subsubsection{Carbon abundance}
\label{carbon}

Carbon is also a product of the CNO cycle which can be produced by stars of all masses. Although it is substantially produced in low and intermediate mass stars where the bulk of nitrogen also originates, carbon is also expected to derive from supernovae as is also the case for the so-called $\alpha$-elements (e.g. O, Ne, Mg; \citealt{matteucci1993,matteucci1996}).

In Figures \ref{CIV} and \ref{CIVs}, we show the \ion{C}{IV}-diagram. The observational data shows an order of magnitude range in \ion{C}{IV}/\ion{He}{II} and \ion{C}{IV}/\ion{C}{III]}, with some QSO2s showing very high values of \ion{C}{IV}/\ion{He}{II} ($>$ 10). The core-ERQs show very high \ion{C}{IV}/\ion{He}{II}, with \ion{C}{IV}/\ion{C}{III]} consistent with the range of values spanned by the QSO2s.
Although some of this variation could be due to the presence of a range in U in the sample, the large dispersion in \ion{C}{IV}/\ion{He}{II} at a given value of \ion{C}{IV}/\ion{C}{III]} appears inconsistent with being due to variation in U alone. We argue that at least two out of U, Z ,C/H, or n$_H$ would need to have a significant variation in order to simultaneously explain the distribution of the observational data in this diagram. In the case of the QSO2s with \ion{C}{IV}/\ion{He}{II} $>$ 4 and the core-ERQs, we find that a combination of high gas density (n$_H$ $\gtrsim$ 10$^{6}$ cm$^{-3}$) and supersolar gas metallicity (Z/Z$_{\odot}$ $=$ 2.0) provides the best overall match to the line ratios (see Fig. \ref{CIV}).

Using secondary behaviour of C we find that a combination of intermediate gas density (n$_H$ $<$ 10$^{6}$ cm$^{-3}$) and Z/Z$_{\odot}$ $\gtrsim$ 4.0 is required for the QSO2s. In addition, \ion{C}{IV}/\ion{He}{II} does appear to show a simple relationship with Z, such that higher Z results in higher \ion{C}{IV}/\ion{He}{II} ratios. The large \ion{C}{IV}/\ion{He}{II} values shown by the core-ERQs strongly suggest high gas densities, and metallicities well above solar.

The HzRGs (including the BLRG 2025-218) have systematically weaker \ion{C}{IV}/\ion{He}{II} (versus UV and rest-frame optical lines) than most of the QSO2s. On the other hand, we note that there is no evidence for a systematic difference in the \ion{C}{IV}/\ion{C}{III]} ratio between the QSO2s and the HzRGs, with the exception of the BLRG 2025-218, which shows the lowest \ion{C}{IV}/\ion{C}{III]} ratio. Using primary behaviour of C, a combination of low gas density (n$_H$ $=$ 10$^{2}$ cm$^{-3}$) and Z/Z$_{\odot}$ $>$ 2.0 appears to provide the best fit to the observed flux ratios of most of HzRGs, which is different to the BLRG 2025-218 that appears consistent with gas density 10$^{4}$ cm$^{-3}$ $<$ n$_H$ $<$ 10$^{6}$ cm$^{-3}$ and Z/Z$_{\odot}$ $\geq$ 1.0. In the case of secondary C behaviour we note that, with exception of the diagram with Z/Z$_{\odot}$ = 0.5, the U-sequences with n$_H$ $=$ 10$^{2}$ cm$^{-3}$ predict very high \ion{C}{IV}/\ion{He}{II} ratio compared with the observed \ion{C}{IV}/\ion{He}{II} ratio of the HzRGs.

\subsubsection{Silicon abundance}
\label{silicon}

The abundance of Si relative to other metals is particularly interesting because its behaviour deviates somewhat from that of O and other $\alpha$-elements. In addition to being produced by short-lived massive stars, Si is also produced on very long timescales ($\sim$ 1 Gyr) by Type Ia supernovae, which means the Si abundance can in principle continue to increase long after star formation and O-production has ceased \citep[e.g.][]{matteucci1993}.

In Figs. \ref{SiIII} and \ref{SiIIIs}, we compare the observed flux ratios of \ion{Si}{III]}/\ion{C}{III]} and \ion{C}{IV}/\ion{C}{III]} with the prediction of the photoionization models. Our models using the solar Si/O abundance ratio fail to explain the observed \ion{Si}{III]}/\ion{C}{III]} flux ratios of the QSO2s and HzRGs that are detected in \ion{Si}{III]}, with the observed ratios being a factor of $\sim$10 higher than produced in these models. On the other hand, our models with a Si/O abundance ratio of 10 $\times$ solar dramatically reduce this discrepancy, suggesting that Si/O and Si/C are an order of magnitude (or more) above their solar values (see Figs. \ref{SiIIIx10} and \ref{SiIIIx10s}). The \ion{C}{IV}/\ion{He}{II} vs. \ion{Si}{III]}/\ion{O}{III]} diagram is consistent with this. We find no systematic difference in the \ion{Si}{III]}/\ion{C}{III]} or \ion{Si}{III]}/\ion{O}{III]} flux ratios between the QSO2 and HzRG samples.

These results are independent of whether the primary or secondary behaviour of carbon is used. It is also important to be aware that these results only apply to quasars or radio galaxies for which \ion{Si}{III]} emission has been detected and, as such, there may be a bias towards the presence of objects with very high Si abundances in these diagrams.

\subsection{Gas Density}
\label{density}

As mentioned already in Section \ref{chemical}, a subset of the QSO2s show very high values of \ion{C}{IV}/\ion{He}{II} and \ion{N}{V}/\ion{He}{II}, with one possible explanation being that their NLR has a relatively high gas density (e.g., $\ga$10$^6$ cm$^{-3}$), although very high N and C abundances may also explain the ratios. Here, we consider new line ratio diagnostics to break the degeneracy between density and abundance effects, and discuss which of the two effects is most likely the cause of the very high \ion{C}{IV}/\ion{He}{II} and \ion{N}{V}/\ion{He}{II} ratios.

For this, we propose three additional density-sensitive ratios: \ion{[Ne}{IV]}2422/\ion{C}{IV}, \ion{[Ne}{IV]}2422/\ion{He}{II}, and \ion{[Ne}{IV]}1602/\ion{[Ne}{IV]}2422. New diagnostic diagrams using these ratios are shown in Figs. \ref{neiv01} -- \ref{neiv03s}. The line \ion{[Ne}{IV]}2422 is often detected in photoionized regions associated with AGNs at high-z \citep[e.g.][]{Ve2001}, and as a forbidden line (critical density = 1.0$\times$10$^6$ cm$^{-3}$ for \textit{T}$_{e}$ = 10$^{4}$ K) it becomes collisionally de-excited at n$_e$ $\ga$ 10$^6$ cm$^{-3}$. This means its flux ratio to permitted lines (e.g., \ion{[Ne}{IV]}2422/\ion{C}{IV} or \ion{[Ne}{IV]}2422/\ion{He}{II}) decreases as density increases.

Similarly, the forbidden line \ion{[Ne}{IV]}1602, with a critical density of 1.0$\times$10$^8$ cm$^{-3}$ (for \textit{T}$_{e}$ = 10$^{4}$ K), also suffers collisional de-excitation at densities of $\ga$ 10$^8$ cm$^{-3}$, and its ratio to permitted lines also decreases as density increases. However, this line is usually very weak and thus detections are rare \citep[e.g.][]{Ve2001}, although the advent of the next generation of extremely large telescopes should make detection of this line in type 2 active galaxies more commonplace.

In addition, the ratio between \ion{[Ne}{IV]}1602 and \ion{[Ne}{IV]}2422 is sensitive to density because the two lines have different critical densities, with collisional de-excitation of \ion{[Ne}{IV]}2422 starting at lower densities than \ion{[Ne}{IV]}1602. As a result, the \ion{[Ne}{IV]}1602/\ion{[Ne}{IV]}2422 flux ratio increases as density increases.

In Figs. \ref{neiv03} and \ref{neiv03s} we show \ion{C}{IV}/\ion{He}{II} $vs.$ \ion{[Ne}{IV]}2422/\ion{He}{II}. When using the primary behaviour of carbon (Fig. \ref{neiv03}), we find that the models fail to reach the positions of most of the plotted QSO2s, which lie above and/or to the right of the models. Although high densities do produce sufficiently high values of \ion{C}{IV}/\ion{He}{II}, this comes at the cost of a substantial decrease in \ion{[Ne}{IV]}2422/\ion{He}{II} due to collisional de-excitation of the \ion{[Ne}{IV]} line. In other words, \ion{C}{IV}/\ion{He}{II} and \ion{[Ne}{IV]}2422/\ion{He}{II} are simultaneously too high to be explained by these models, regardless of which density or metallicity is used.

However, we find that when using secondary behaviour for carbon (Fig. \ref{neiv03s}), our high metallicity models (Z/Z$_\odot$ $\ga$ 2) produce much higher CIV/HeII ratios even at low gas density ($\la$ 10$^4$ cm$^{-3}$), allowing the positions of the QSO2s to be generally well reproduced in this diagram. We obtain essentially the same result from the \ion{C}{IV}/\ion{He}{II} $vs.$ \ion{[Ne}{IV]}2422/\ion{C}{IV} diagram (see Figs. \ref{neiv02} and \ref{neiv02s}).

In the \ion{C}{IV}/\ion{He}{II} $vs.$ \ion{[Ne}{IV]}1602/\ion{[Ne}{IV]}2422 diagram (Figs. \ref{neiv01} and \ref{neiv01s}), most of the QSO2s have upper limits on the \ion{[Ne}{IV]}1602/\ion{[Ne}{IV]}2422 ratio which imply n$_e$ $\la$ 10$^6$ cm$^{-3}$. However, there are two QSO2s where \ion{[Ne}{IV]}1602 was detected and \ion{[Ne}{IV]}2422 was not, and their lower limits on \ion{[Ne}{IV]}1602/\ion{[Ne}{IV]}2422 requires n$_e$ $\ga$ 10$^6$ cm$^{-3}$, implying these two are intermediate or high density objects.

Overall, the positions of the HzRGs in the \ion{C}{IV}/\ion{He}{II} $vs.$ \ion{[Ne}{IV]}2422/\ion{He}{II} and \ion{C}{IV}/\ion{He}{II} $vs.$ \ion{[Ne}{IV]}2422/\ion{C}{IV} diagrams are well explained using models with n$_e$ $\la$ 10$^4$ cm$^{-3}$, consistent with previous conclusions about the gas density in the extended gas of HzRGs \citep[e.g.][]{Bre2000,Ve2001,Hu4}. Furthermore, we find that the single HzRG which has a measurement of both \ion{[Ne}{IV]}1602 and \ion{[Ne}{IV]}2422 (TXS 0828+193; \citealt{Ve2001}) lies very close to the n$_e$ = 100 cm$^{-3}$ density models, consistent with a low gas density. Based on the emission line ratios of the BLRG 2025-218, we find that intermediate gas density (10$^{4}$ cm$^{-3}$ $<$ n$_H$ $<$ 10$^{6}$ cm$^{-3}$) appears to be required, consistent with previous work \citep[e.g.][]{villar1999}.

Interestingly, in the \ion{C}{IV}/\ion{He}{II} $vs.$ \ion{[Ne}{IV]}2422/\ion{He}{II} diagram (Figs. \ref{neiv03} and \ref{neiv03s}) the HzRGs and QSO2s show no systematic difference in \ion{[Ne}{IV]}2422/\ion{He}{II}, suggesting the NLR of the plotted QSO2s and the HzRGs have similar gas densities. The systematic vertical offset of the QSO2s above the HzRGs then suggests that the QSO2s have substantially higher carbon abundance.

Of course, the conclusions derived from the \ion{[Ne}{IV]} lines apply only to the objects for which we have detected one such line, and it is plausible that many other QSO2s in this sample are undetected in \ion{[Ne}{IV]} as a result of collisional de-excitation in high-density gas. Nonetheless, this analysis shows that at least some of the QSO2s in this sample, including some objects with very high \ion{C}{IV}/\ion{He}{II}, have low density NLRs with high carbon abundances. On the other hand, at least two of the QSO2s do appear to have intermediate or high density (n$_e$ $\ga$ 10$^6$ cm$^{-3}$), and thus it is likely that this sample of QSO2s encompasses objects with a wide range in gas density.

\begin{figure}
	\includegraphics[width=\columnwidth,height=3.5in,keepaspectratio]{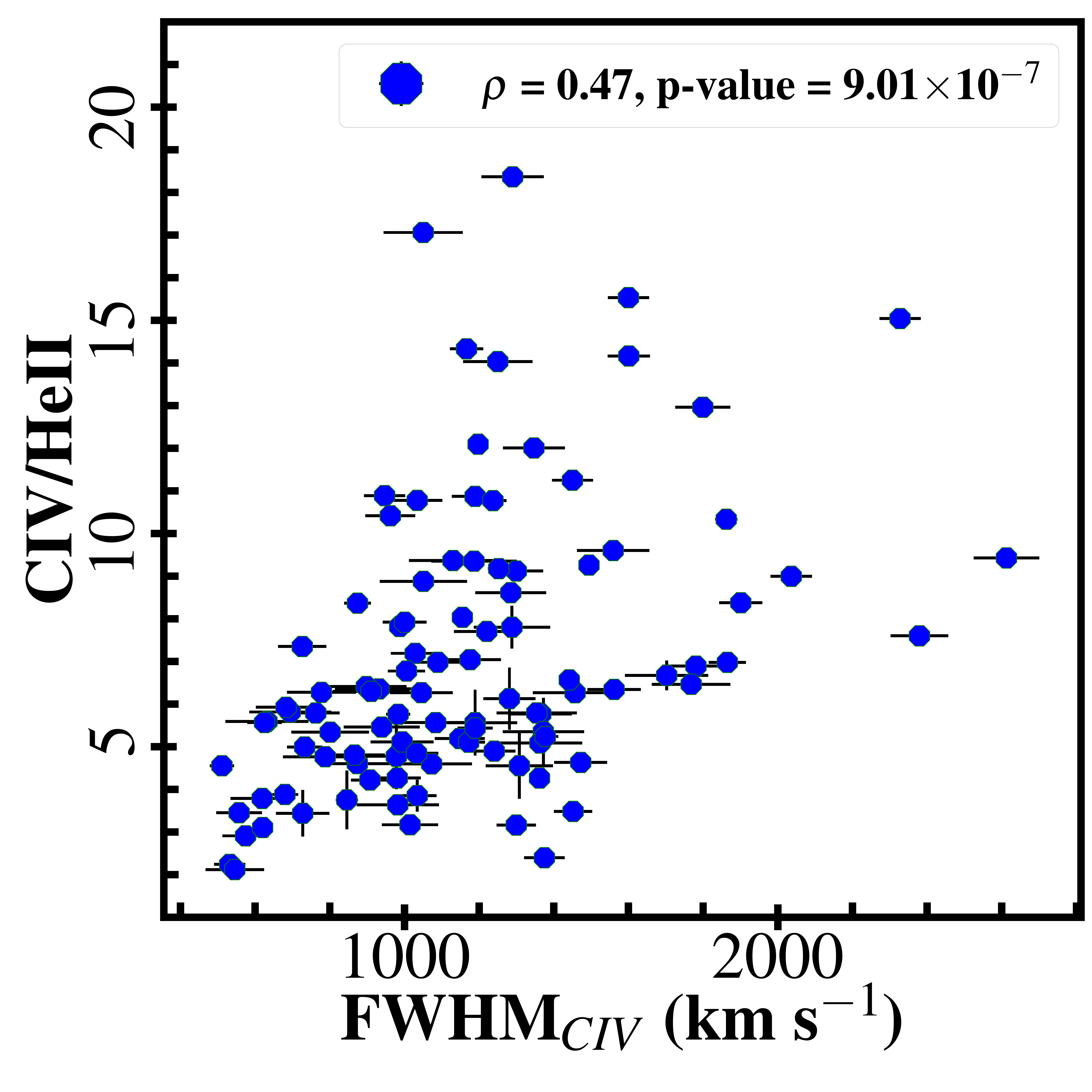}
	\caption{Variation of \ion{C}{IV}/\ion{He}{II} flux ratio as a function of the FWHM of \ion{C}{IV} with $\rho$ and p-value representing the Spearman's rank correlation coefficient and t-distribution, respectively.}
	\label{fwhm}
\end{figure}

Investigating the possible correlation between the \ion{C}{IV}/\ion{He}{II} flux ratio and the FWHM of the \ion{C}{IV} emission line, we find that the Spearman correlation ($\rho$) and the t-distribution (p-value) indicate a reasonable positive relationship between the \ion{C}{IV}/\ion{He}{II} flux ratio and FWHM$_{CIV}$ with $\rho$ = 0.47 and p-value = 9.01$\times$10$^{-7}$ (see Fig. \ref{fwhm}). The diagram shows that there is a correlation between higher FWHM$_{CIV}$ and higher \ion{C}{IV}/\ion{He}{II} flux ratio, which is consistent with some of the QSO2s having a contribution from a region with density and kinematic properties intermediate between the classical NLR and BLR. A possible alternative explanation is that outflows are present in some objects, broadening \ion{C}{IV} and enhancing the \ion{C}{IV}/\ion{He}{II} due to shock ionization.

The extremely high values of \ion{N}{V}/\ion{He}{II} and \ion{C}{IV}/\ion{He}{II} shown by the core-ERQs strongly suggest high gas densities (e.g., $\ga$10$^6$ cm$^{-3}$). Given the overlap between the positions of the core-ERQs and the QSO2s with \ion{C}{IV}/\ion{He}{II} $>$ 4 in our diagnostic diagrams \citep[see also][]{villar-martin2019}, we argue that both groups of quasars have a broadly similar nature: obscured AGN whose geometry and/or orientation allows part of the intermediate/high density outer BLR to be viewed, in addition to the low density NLR. As noted by \citet{villar-martin2019}, the core-ERQs lie at the high \ion{N}{V}/\ion{C}{IV} edge of the QSO2 cloud (see Fig. \ref{NV01}), implying they have have some of the highest values of N abundance and/or ionization parameter of any obscured quasars.

\section{Revised selection criteria for QSO2s with low density NLR}
\label{new-crit}

In the previous sections we found that a significant fraction (71\%) of the 145 Class A candidate QSO2s selected by \citet{Al} show UV line ratios consistent with intermediate/high gas densities. Here we propose refined selection criteria for QSO2s, using density-sensitive UV line flux ratios.

\begin{figure}
	\centering
	\subfloat[\ion{[Ne}{IV]}2422/\ion{He}{II} vs. \ion{C}{IV}/\ion{He}{II}]{
		\includegraphics[width=\columnwidth,height=2.8in,keepaspectratio]{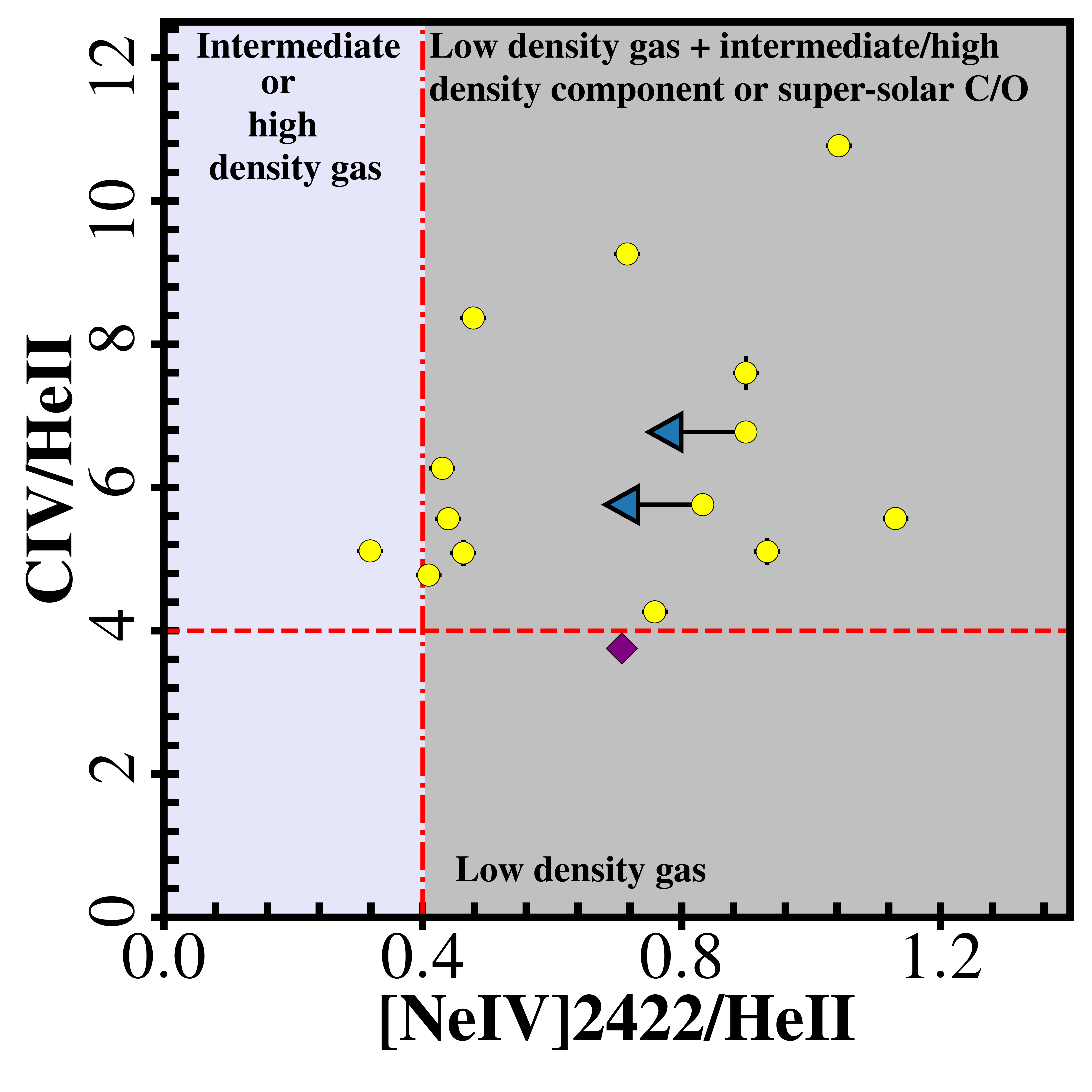}\label{crit01}}
	\quad
	\subfloat[\ion{[Ne}{IV]}2422/\ion{C}{IV} vs. \ion{C}{IV}/\ion{He}{II}]{
		\includegraphics[width=\columnwidth,height=2.8in,keepaspectratio]{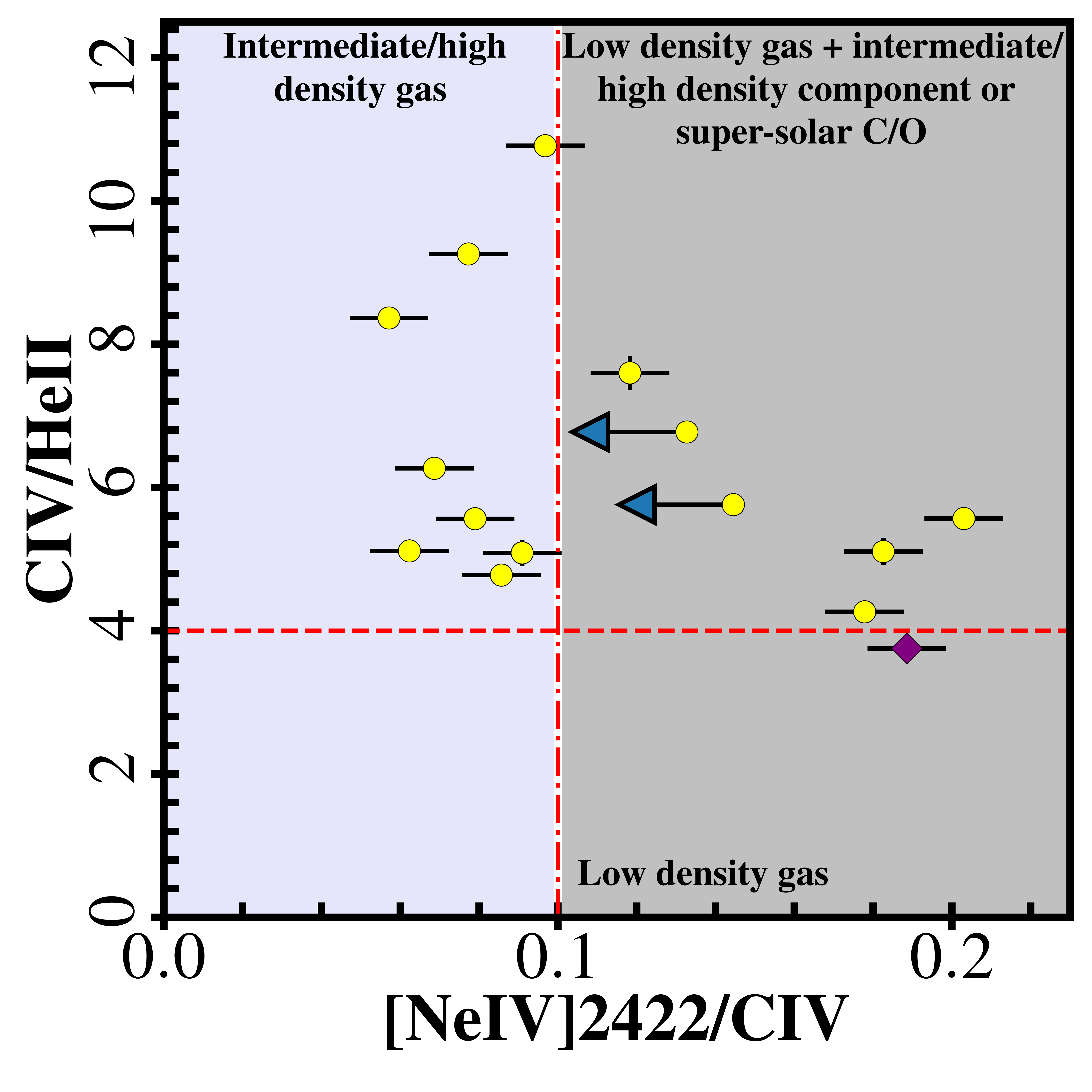}\label{crit02}}
	\quad
	\subfloat[\ion{[Ne}{IV]}1602/\ion{[Ne}{IV]}2422 vs. \ion{C}{IV}/\ion{He}{II}]{
		\includegraphics[width=\columnwidth,height=2.8in,keepaspectratio]{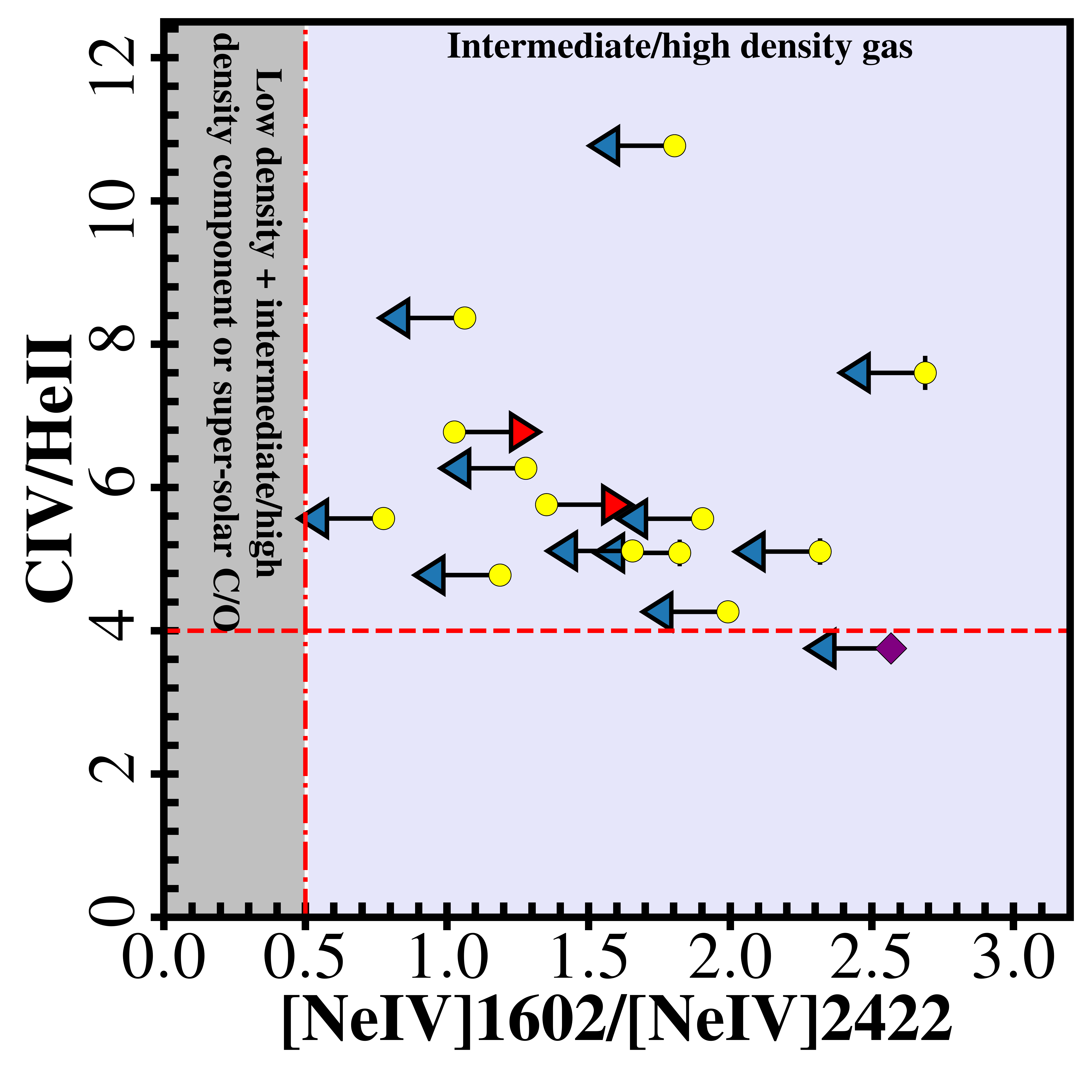}\label{crit03}}

	\caption{Proposed new criteria to the QSO2s sample to produce our revised list of QSO2s.}
	\label{criteria}

\end{figure}

\begin{itemize}
	\item Class 1: Highly likely QSO2 candidates with \ion{C}{IV}/\ion{He}{II} $<$ 4 and \ion{[Ne}{IV]}2422/\ion{He}{II} $>$ 0.4 and \ion{[Ne}{IV]}2422/\ion{C}{IV} $>$ 0.1 consistent with low electron density (see Figs. \ref{crit01} and \ref{crit02}). One object (0.7\%) from the QSO2 sample meets these requirements (SDSSJ213557.35-032130.0; see Table \ref{class_tab}).
	\item Class 2: Good QSO2 candidates with \ion{C}{IV}/\ion{He}{II} $<$ 4, but have no useful information from \ion{[Ne}{IV]} (\ion{[Ne}{IV]} not detected and the \ion{[Ne}{IV]}2422/\ion{He}{II} and \ion{[Ne}{IV]}2422/\ion{C}{IV} upper limits are consistent with \ion{[Ne}{IV]}2422/\ion{He}{II} $>$ 0.4 and \ion{[Ne}{IV]}2422/\ion{C}{IV} $>$ 0.1; see Figs. \ref{crit01} and \ref{crit02}).
	Fourteen objects (9.7\%) from the QSO2 sample meet these requirements (see Table \ref{class_tab}).
	\item Class 3: QSO2s with \ion{[Ne}{IV]}2422/\ion{He}{II} and \ion{[Ne}{IV]}2422/\ion{C}{IV} ratios consistent with low densities, but with \ion{C}{IV}/\ion{He}{II} $>$ 4 consistent with high gas density and/or high carbon abundance. As discussed in section \ref{density}, such objects could be objects with an intermediate orientation such that the low density NLR and more central intermediate/high density are both observed. On the other hand, they may be QSO2s with a low density NLR and a high carbon abundance.
	Six objects (4.1\%) from the QSO2 sample meet these requirements (see Table \ref{class_tab}).
	\item Class 4: Ambiguous, potentially intermediate/high density objects. QSO2s with \ion{C}{IV}/\ion{He}{II} $>$ 4 but \ion{[Ne}{IV]}2422/\ion{He}{II} $<$ 0.4 and \ion{[Ne}{IV]}2422/\ion{C}{IV} $<$ 0.1 (or no information about \ion{[Ne}{IV]}2422; see Figs. \ref{crit01} and \ref{crit02}). A few objects appear consistent with this criterion, including those two using \ion{[Ne}{IV]}2422/\ion{He}{II} and \ion{[Ne}{IV]}2422/\ion{C}{IV} upper limits.
	However, by rejecting those QSO2s that have CIV/HeII > 4 but no information about \ion{[Ne}{IV]}2422/\ion{He}{II} or \ion{[Ne}{IV]}2422/\ion{C}{IV}, it is possible that some QSO2s with genuine low density NLRs and high carbon abundance are being rejected alongside high density objects.
	Seventy-seven objects (53\%) from the QSO2 sample meet these requirements (see Table \ref{class_tab} and \ref{class_tab2}).
	\item Class 5: Likely intermediate or high density objects. QSO2s with \ion{[Ne}{IV]}1602/\ion{[Ne}{IV]}2422 $>$ 0.5 (see Fig. \ref{crit03}). Two objects (1.4\%) from the QSO2 sample meet these requirements (see Table \ref{class_tab2}).

\end{itemize}

\section{Summary}
\label{conclusions}

We have studied the ultraviolet emission line ratios of a sample of 145 QSO2s from SDSS BOSS, and have compared them against powerful radio galaxies at z $\sim$ 2.5 and against a grid of AGN photoionization models. The main results are as follows:

\begin{itemize}
	\item Most of the QSO2s show unusually high \ion{C}{IV}/\ion{He}{II} ratios ($>$4, e.g. 86 out of 145 QSO2s), with some showing extreme values of this ratio ($>$ 10, e.g. 17 out of 145 QSO2s), consistent with BLR-like gas densities ($>$10$^{7}$ cm$^{-3}$) or super-Solar C/O abundance ratios (i.e., secondary carbon).

	\item  The degeneracy between high density and high C abundance can be broken using the \ion{C}{IV}/\ion{He}{II} vs \ion{[Ne}{IV]}/\ion{He}{II} diagram. Both ratios are density sensitive, but the cloud of QSO2s is positioned where none of the models using the primary C behaviour can reach. However, models using the secondary C behaviour can reach the cloud of QSO2s, provided the gas has  super-Solar metallicity.

	\item The UV emission line ratios of the core-ERQs place them within the cloud defined by the QSO2s that have extreme values for the \ion{C}{IV}/\ion{He}{II} ratio (\ion{C}{IV}/\ion{He}{II} $>$ 4), suggesting that the observed emission line gas in both classes of object has broadly similar physical conditions, i.e., strong contamination from intermediate/high density gas in the outer BLR, likely due to orientation and/or geometrical effects \citep[see also][]{villar-martin2019}.

	\item Looking at the HzRGs and QSO2s, we suggest that the differences between them is due to (i) an orientation effect where the QSO2s are viewed at an angle that reveals high/intermediate density gas in the outskirts of the BLR, and/or (ii) the extension of high/intermediate density gas above the plane of the obscuring torus \cite{villar-martin2019}.

	\item We have identified \ion{[Ne}{IV]} $\lambda$2422 as a useful indicator for constraining the gas density. Where this line has been detected (15 QSO2s), its strength relative to other lines implies gas densities $<$ 10$^{6}$ cm$^{-3}$. However, two further objects have a \ion{[Ne}{IV]} $\lambda$1602 / \ion{[Ne}{IV]} $\lambda$2422 ratio that implies densities $>$ 10$^{6}$ cm$^{-3}$, indicating that the density of the narrow-line emitting gas varies between the QSO2s of this sample. In addition, we have proposed new UV-line based criteria to select genuine QSO2s with low-density narrow line regions, and to identify moderate to high-density impostors.

	\item Solar or super-Solar nitrogen abundances and gas metallicities are required in most of the QSO2s (e.g. the \ion{N}{V}/\ion{He}{II} ratio).

	\item We have compared the UV emission line ratios of the QSO2s to high-redshift radio galaxies at z $>$ 2, and find that the QSO2s are systematically offset to higher \ion{C}{IV}/\ion{He}{II}, \ion{C}{III]}/\ion{He}{II} and \ion{N}{V}/\ion{He}{II} ratios, suggesting systematically higher gas density and/or systematically higher C and N abundances. However, we find no evidence for a systematic difference in the N/C abundance ratio between the two types of object (e.g., from the \ion{N}{V}/\ion{C}{IV} ratio).

	\item Scatter in the \ion{N}{IV]}/\ion{C}{IV} ratio implies that there is a significant scatter in the N/C abundance ratio within the combined QSO2 and HzRG sample, although it is unclear whether the scatter is due to differences in the ISM abundance ratio or differential depletion of N and C into dust grains.

	\item Interestingly, we find that adopting secondary behaviour for both N and C alleviates the "\ion{N}{IV]} problem" reported by a number of previous studies. However, we also note that the secondary behaviour of carbon (in addition to Nitrogen) is likely to prevent \ion{N}{V}/\ion{C}{IV} and \ion{N}{IV]}/\ion{C}{IV} from being useful metallicity indicators.

	\item A subset of the QSO2s and HzRGs appear to be "silicon-loud", with \ion{Si}{III]} relative fluxes suggesting Si/C and Si/O are an order of magnitude above their Solar values (see Figs. from \ref{SiIII} to \ref{OIIIx10s}). We suggest that this is due to a higher than expected contribution from type Ia supernovae to the chemical enrichment of the ISM of the galaxes, resulting either from the continuation of type Ia supernovae for $\sim$1 Gyr after the end of the last major star formation event, or from a stellar initial mass function with a substantially higher binary fraction than in the Milky Way.

\end{itemize}

\section*{Acknowledgements}

MS acknowledges support from the National Council of Research and Development (CNPq) under the process of number 248617/2013-3. MS thanks Montse Villar-Mart\'in and Luc Binette for helpful and valuable discussions. AH acknowledges FCT support through fellowship SFRH/BPD/107919/2015 and contract DL 57/2016/CP1364/CT0002. PL acknowledges FCT support through the contract DL 57/2016/CP1364/CT0010. MS, AH and PL acknowledge the support of the European Community Programme (FP7/2007-2013) under grant agreement No. PIRSES-GA-2013-612701 (SELGIFS). This work was also supported by Funda\c{c}\~{a}o para a Ci\^{e}ncia e Tecnologia (FCT) through national funds (PTDC/FIS-AST/3214/2012 and UID/FIS/04434/2013), by FEDER through COMPETE (FCOMP-01-0124-FEDER-029170) and COMPETE2020 (POCI-01-0145-FEDER-007672), and by FCT/MCTES through national funds (PIDDAC) by grants UID/FIS/04434/2019 and PTDC/FIS-AST/29245/2017. SGM acknowledges PhD fellowship PD/BD/135228/2017, funded by the FCT via the PhD::SPACE programme (PD/00040/2012). We acknowledge the FCT-CAPES funded Transnational Cooperation project "Strategic Partnership in Astrophysics Portugal-Brazil".




\bibliographystyle{mnras}
\bibliography{mnras_template} 




\appendix

\section{Some extra material}

\begin{landscape}
	\begin{table}
		\centering
		\caption{List of flux ratios measurements of the rest-frame UV and optical of HzRGs from the literature. See \S \ref{keck} for more details.}
		\label{hzrgs}
		\resizebox{1.37\textwidth}{!}{%
			\begin{tabular}{cccccccccccccc}
				\hline
				& 4C+03.24$^{a}$    & 0828+193$^{a,b}$ & 4C+23.56a$^{a,b}$ & 0731+438$^{a}$   & 4C--00.54$^{a,b}$ & 4C+48.48$^{a}$   & 0211--122$^{a,b}$ & 4C+40.36$^{a,b}$ & 0406--244$^{b}$ & 1558--003$^{b}$  & 2025--218$^{b}$  & 2104--242$^{b}$  & 0943--242$^{a,c}$ \\ \hline
				Flux ratio                                   & $z$ = 3.56        & $z$ = 2.57       & $z$ = 2.48        & $z$ = 2.43       & $z$ =2.36         & $z$ = 2.34       & $z$ = 2.34        & $z$ = 2.27       & $z$ = 2.44      & $z$ = 2.53       & $z$ = 2.63       & $z$ = 2.49       & $z$ = 2.92        \\ \hline
				(OVI+CII)/HeII                               & --                & --               & --                & --               & --                & --               & --                & --               & --              & --               & --               & --               & 0.77 $\pm$ 0.12   \\
				Ly$\alpha$/HeII                              & 48.90 $\pm$ 12.30 & 8.40 $\pm$ 0.20  & 5.20 $\pm$ 0.40   & 14.10 $\pm$ 0.20 & 12.20 $\pm$ 0.30  & 13.20 $\pm$ 0.90 & 0.57 $\pm$ 0.02   & 14.68 $\pm$ 1.17 & 5.83 $\pm$ 0.59 & 15.44 $\pm$ 0.61 & 16.00 $\pm$ 3.77 & 13.00 $\pm$ 1.64 & 9.90 $\pm$ 0.43   \\
				NV/HeII                                      & 0.90 $\pm$ 0.30   & 0.4 $\pm$ 0.1    & 0.80 $\pm$ 0.10   & 0.23 $\pm$ 0.02  & 0.60 $\pm$ 0.10   & 0.27 $\pm$ 0.08  & 0.95 $\pm$ 0.09   & 0.40 $\pm$ 0.08  & $<$ 0.17        & 0.46 $\pm$ 0.10  & 2.00 $\pm$ 0.57  & 0.32 $\pm$ 0.04  & 0.30 $\pm$ 0.06   \\
				(SiIV+OIV{]})/HeII                       & 1.20 $\pm$ 0.60   & 0.30             & 0.10              & 0.32 $\pm$ 0.01  & 0.50              & 0.30             & 0.20              & 0.30             & 0.25 $\pm$ 0.09 & 0.41 $\pm$ 0.10  & --               & 0.39 $\pm$ 0.05  & 4.04              \\
				NIV{]}/HeII                              & --                & 0.07 $\pm$ 0.01  & 0.15 $\pm$ 0.04   & 0.06 $\pm$ 0.01  & --                & 0.08 $\pm$ 0.02  & 0.06 $\pm$ 0.01   & 0.04 $\pm$ 0.01  & $<$ 0.08        & 0.14 $\pm$ 0.04  & --               & --               & 1.09 $\pm$ 0.14   \\
				CIV/HeII                                     & 2.40 $\pm$ 0.70   & 1.80 $\pm$ 0.10  & 1.20 $\pm$ 0.10   & 1.53 $\pm$ 0.04  & 1.40 $\pm$ 0.10   & 1.04 $\pm$ 0.07  & 1.80 $\pm$ 0.07   & 1.80 $\pm$ 0.14  & 0.75 $\pm$ 0.10 & 2.67 $\pm$ 0.14  & 1.40 $\pm$ 0.49  & 0.80 $\pm$ 0.13  & 1.47 $\pm$ 0.11   \\
				{[}NeIV{]}1602/HeII                  & --                & 0.02             & --                & --               & --                & --               & --                & --               & --              & --               & --               & --               & --                \\
				OIII{]}/HeII                             & --                & 0.18             & 0.30              & 0.20             & 0.17              & 0.11             & 0.08              & 0.16 $\pm$ 0.01  & 0.17 $\pm$ 0.08 & 0.43 $\pm$ 0.08  & --               & 0.1 $\pm$ 0.03   & 3.45              \\
				SiIII{]}/HeII                            & --                & 0.18             & --                & 0.09             & --                & --               & --                & --               & --              & 0.13 $\pm$ 0.03  & 1.80 $\pm$ 1.25  & --               & 0.05              \\
				CIII{]}/HeII                             & --                & 0.5 $\pm$ 0.1    & 0.7 $\pm$ 0.1     & 0.70 $\pm$ 0.02  & 0.45 $\pm$ 0.03   & 0.62 $\pm$ 0.04  & 0.50 $\pm$ 0.03   & 0.95 $\pm$ 0.17  & 0.75 $\pm$ 0.10 & 0.75 $\pm$ 0.10  & 3.60 $\pm$ 1.40  & --               & 0.54 $\pm$ 0.04   \\
				{[}NeIV{]}2422/HeII                  & --                & 0.50             & 0.90              & 0.60             & 1.00              & 0.60             & 0.30              & 0.50             & 0.83 $\pm$ 0.26 & 0.56 $\pm$ 0.11  & --               & --               & --                \\
				{[}OIII{]}5008/H$\alpha$             & --                & --               & 3.40 $\pm$ 0.31   & --               & 3.43 $\pm$ 0.29   & --               & 3.27 $\pm$ 0.56   & 3.69 $\pm$ 0.06  & --              & 4.80 $\pm$ 1.47  & 1.60 $\pm$ 0.57  & 4.00 $\pm$ 0.64  & --                \\
				{[}OIII{]}5008/H$\beta$              & --                & 9.53 $\pm$ 0.58  & 9.71 $\pm$ 4.20   & --               & $>$  17.15        & --               & $>$ 6.54          & 10.85 $\pm$ 1.61 & 12.6 $\pm$ 0.89 & 8.00 $\pm$ 2.71  & $>$ 5.33         & 13.33 $\pm$ 4.75 & --                \\
				{[}NII{]}/H$\alpha$                  & --                & --               & 0.21 $\pm$ 0.07   & --               & 0.29 $\pm$ 0.05   & --               & $<$ 0.2           & 1.6              & --              & 0.70 $\pm$ 0.29  & 0.60 $\pm$ 0.27  & 0.30 $\pm$ 0.06  & --                \\
				{[}OII{]}3728/{[}OIII{]}5008 & --                & 0.18 $\pm$ 0.01  & 0.27 $\pm$ 0.06   & --               & 0.15 $\pm$ 0.03   & --               & 0.12 $\pm$ 0.03   & 0.39 $\pm$ 0.02  & 0.21 $\pm$ 0.02 & 0.31 $\pm$ 0.05  & $<$ 1.50         & 0.15 $\pm$ 0.03  & 0.32 $\pm$ 0.01                \\ \hline
				\multicolumn{14}{c}{References: $^{a}$\citet{Ve2001}, $^{b}$\citet{Hu4}, $^{c}$\citet{marckelson2018}.} \\
				\hline
				\hline
			\end{tabular}%
		}
	\end{table}
\end{landscape}

\begin{landscape}
	\begin{center}
		\begin{table}
			\caption{List of the first forty type II quasars followed by the detected emission lines. The line fluxes obtained with the fitting routine are measured in $\times$10$^{-17}$ erg cm$^{-2}$ s$^{-1}$ . See \S\ref{analysis} for more details.}
			\label{table01}
			\resizebox{1.38\textwidth}{!}{%
				\begin{tabular}{ccccccccccccccccc}
					\hline
					\hline
					SDSS name           & z    & Class & OVI+CII          & Ly$\alpha$        & NV              & OI             & SiIV+OIV] & NIV]     & CIV             & [NeIV]1602 & HeII           & OIII]    & SiIII]   & CIII]     & [NeIV]2422 & MgII             \\ \hline
					J011506.65-015307.0 & 2.33 &   4   &                  & 587.7 $\pm$ 11.8  & 44.7 $\pm$ 3.9  & 26.4 $\pm$ 2.9 & 9.3 $\pm$ 1.9   &                & 149.8 $\pm$ 4.8 &                        & 17.9 $\pm$ 2.2 &                & 10.6 $\pm$ 0.7 & 37.3 $\pm$ 0.7  &                        &                  \\
					J022051.68-012403.3 & 2.60 &  4    &                  & 1171.7 $\pm$ 18.3 & 130.3 $\pm$ 4.8 & 22.6 $\pm$ 3.9 & 36.0 $\pm$ 4.8  &                & 203.7 $\pm$ 7.2 &                        & 21.6 $\pm$ 3.3 &                &                & 47.1 $\pm$ 2.7  &                        &                  \\
					J075119.09+130202.4 & 2.64 &   4   &                  & 558.4 $\pm$ 3.8   & 68.4 $\pm$ 4.5  & 28.5 $\pm$ 3.8 & 38.1 $\pm$ 3.8  &                & 129.6 $\pm$ 5.2 &                        & 27.2 $\pm$ 3.9 &                &                & 116.6 $\pm$ 5.0 &                        &                  \\
					J075656.49+362507.4 & 2.87 &   5   & 104.9 $\pm$ 7.1  & 912.9 $\pm$ 16.6  & 122.5 $\pm$ 4.9 & 26.6 $\pm$ 3.2 & 20.7 $\pm$ 1.9  &                & 195.9 $\pm$ 0.3 & $<$17.2                & 31.3 $\pm$ 4.2 &                & 82.8 $\pm$ 2.3 & 45.0 $\pm$ 0.7  & 13.5 $\pm$ 1.6         &                  \\
					J081950.96+111507.9 & 2.81 &   4   & 94.5 $\pm$ 4.4   & 787.0 $\pm$ 13.7  & 56.7 $\pm$ 3.5  & 5.9 $\pm$ 1.9  & 38.3 $\pm$ 2.0  &                & 161.8 $\pm$ 4.4 &                        & 20.7 $\pm$ 3.3 &                &                & 39.5 $\pm$ 3.0  &                        &                  \\
					J084005.00+344832.1 & 2.55 &   4   &                  & 648.7 $\pm$ 5.4   & 27.1 $\pm$ 2.3  & 21.3 $\pm$ 2.5 & 8.5 $\pm$ 1.5   & 11.8 $\pm$ 1.8 & 164.6 $\pm$ 9.8 &                        & 18.1 $\pm$ 2.1 & 21.8 $\pm$ 1.9 & 13.7 $\pm$ 0.7 & 57.1 $\pm$ 0.5  &                        &                  \\
					J095118.93+450432.4 & 2.45 &   5   &                  & 1396.5 $\pm$ 19.6 & 45.3 $\pm$ 3.6  & 46.5 $\pm$ 3.9 & 51.5 $\pm$ 4.8  &                & 288.8 $\pm$ 5.8 & $<$17.5                & 34.5 $\pm$ 5.0 &                &                & 55.2 $\pm$ 2.7  & 16.5 $\pm$ 1.1         & 77.4 $\pm$ 11.6  \\
					J100250.98+382532.3 & 2.73 &       & 129.6 $\pm$ 9.0  & 688.9 $\pm$ 13.8  & 96.8 $\pm$ 4.5  & 16.8 $\pm$ 1.6 &                 &                & 188.1 $\pm$ 3.7 & $<$38.4                & 20.3 $\pm$ 2.7 & 12.9 $\pm$ 3.4 &                & 112.4 $\pm$ 5.1 & 14.5 $\pm$ 1.0         &                  \\
					J100916.93+031128.9 & 2.69 &   4   & 169.5 $\pm$ 11.5 & 659.4 $\pm$ 10.6  & 136.9 $\pm$ 5.1 & 14.9 $\pm$ 3.2 & 24.7 $\pm$ 2.8  &                & 147.0 $\pm$ 3.0 &                        & 16.0 $\pm$ 2.4 &                &                & 41.0 $\pm$ 3.0  &                        &                  \\
					J112230.35+341538.0 & 2.40 &   5   &                  & 639.1 $\pm$ 9.6   & 52.9 $\pm$ 3.4  &                & 32.1 $\pm$ 4.5  &                & 148.4 $\pm$ 2.7 & $<$15.1                & 31.1 $\pm$ 3.5 &                &                & 37.6 $\pm$ 3.1  & 12.7 $\pm$ 1.8         &                  \\
					J114542.07+401318.4 & 3.30 &   4   & 129.6 $\pm$ 6.2  & 684.4 $\pm$ 3.7   & 134.0 $\pm$ 3.4 &                & 22.2 $\pm$ 3.6  & 11.2 $\pm$ 2.1 & 234.4 $\pm$ 2.2 &                        & 29.2 $\pm$ 2.0 & 9.9 $\pm$ 2.0  &                & 37.7 $\pm$ 3.2  &                        &                  \\
					J125148.53+060027.4 & 2.27 &   4   &                  & 766.0 $\pm$ 10.7  & 87.2 $\pm$ 5.7  & 39.2 $\pm$ 4.5 & 39.8 $\pm$ 4.4  &                & 169.4 $\pm$ 0.3 &                        & 29.4 $\pm$ 3.2 & 19.7 $\pm$ 3.2 &                & 38.2 $\pm$ 3.7  &                        & 43.9 $\pm$ 3.6   \\
					J135531.46+072950.6 & 3.15 &   5   & 112.9 $\pm$ 6.0  & 867.0 $\pm$ 17.2  & 58.1 $\pm$ 3.3  & 34.7 $\pm$ 3.4 & 43.9 $\pm$ 5.6  &                & 151.7 $\pm$ 2.6 & $<$22.8                & 27.3 $\pm$ 3.4 & 28.6 $\pm$ 4.0 &                & 65.5 $\pm$ 6.2  & 12.0 $\pm$ 0.7         &                  \\
					J150549.73+074309.0 & 3.32 &    4  & 319.6 $\pm$ 8.6  & 1314.3 $\pm$ 11.0 & 276.5 $\pm$ 4.1 & 28.9 $\pm$ 3.0 & 51.2 $\pm$ 4.5  &                & 392.7 $\pm$ 3.2 &                        & 67.8 $\pm$ 3.8 & 29.0 $\pm$ 3.7 &                & 83.4 $\pm$ 5.1  &                        &                  \\
					J153306.06+322649.6 & 2.71 &   4   & 160.2 $\pm$ 11.3 & 657.8 $\pm$ 9.1   & 137.8 $\pm$ 4.6 & 19.1 $\pm$ 1.1 & 47.3 $\pm$ 3.1  &                & 232.4 $\pm$ 9.6 &                        & 15.0 $\pm$ 2.8 & 17.5 $\pm$ 3.5 & 24.0 $\pm$ 0.7 & 61.1 $\pm$ 0.7  &                        &                  \\
					J160900.01+190534.8 & 2.54 &    4  & 204.2 $\pm$ 14.1 & 1791.3 $\pm$ 8.6  & 76.4 $\pm$ 3.6  & 65.6 $\pm$ 3.0 & 29.7 $\pm$ 1.8  & 12.8 $\pm$ 1.8 & 375.7 $\pm$ 6.9 &                        & 42.3 $\pm$ 3.1 & 45.4 $\pm$ 3.9 &                & 95.1 $\pm$ 3.0  &                        & 58.1 $\pm$ 8.1   \\
					J161059.96+260321.5 & 2.28 &   2   &                  & 703.5 $\pm$ 12.8  & 59.2 $\pm$ 6.4  & 44.2 $\pm$ 6.2 & 38.6 $\pm$ 6.2  &                & 190.8 $\pm$ 3.5 &                        & 49.5 $\pm$ 9.4 & 16.3 $\pm$ 2.8 &                & 34.1 $\pm$ 3.6  &                        & 44.1 $\pm$ 3.3   \\
					J162812.51+233734.8 & 2.39 &   4   &                  & 998.7 $\pm$ 11.2  & 106.4 $\pm$ 3.6 & 63.4 $\pm$ 4.1 & 75.2 $\pm$ 4.2  & 13.2 $\pm$ 1.3 & 283.2 $\pm$ 1.8 &                        & 42.5 $\pm$ 5.2 & 34.5 $\pm$ 4.3 &                & 193.5 $\pm$ 1.7 &                        & 209.7 $\pm$ 13.9 \\
					J163343.85+261026.2 & 3.07 &   4   & 50.8 $\pm$ 3.6   & 794.6 $\pm$ 9.4   & 54.0 $\pm$ 2.8  & 44.8 $\pm$ 3.4 &                 &                & 123.3 $\pm$ 0.3 & 24.1 $\pm$ 2.8         & 21.4 $\pm$ 3.4 &                & 17.2 $\pm$ 1.8 & 149.1 $\pm$ 1.8 & $<$17.8                &                  \\
					J222946.61+005540.5 & 2.37 &   4   &                  & 943.4 $\pm$ 17.1  & 80.5 $\pm$ 4.3  &                &                 &                & 153.6 $\pm$ 2.8 &                        & 17.1 $\pm$ 1.6 &                & 12.2 $\pm$ 0.8 & 57.8 $\pm$ 0.8  &                        &                  \\
					J001040.82+004550.5 & 2.72 &   3   & 63.7 $\pm$ 4.8   & 256.1 $\pm$ 3.9   & 34.3 $\pm$ 2.6  &                & 11.6 $\pm$ 1.7  &                & 69.9 $\pm$ 0.5  &                        & 16.4 $\pm$ 3.0 &                & 9.0 $\pm$ 0.6  & 23.8 $\pm$ 0.5  & 12.4 $\pm$ 1.6         &                  \\
					J001738.55-011838.7 & 3.22 &       &                  & 177.6 $\pm$ 4.9   & 16.7 $\pm$ 1.9  &                &                 &                & 31.4 $\pm$ 3.1  &                        &                &                &                &                 &                        &                  \\
					J001814.72+023258.8 & 2.90 &   2   & 16.7 $\pm$ 2.0   & 237.0 $\pm$ 3.6   &                 &                &                 &                & 25.3 $\pm$ 2.7  &                        & 7.3 $\pm$ 1.5  &                &                & 21.3 $\pm$ 2.2  &                        &                  \\
					J003605.26+001618.7 & 2.94 &       &                  & 82.8 $\pm$ 3.9    & 15.2 $\pm$ 2.5  &                &                 &                & 19.8 $\pm$ 2.2  &                        &                &                &                & 10.8 $\pm$ 1.8  &                        &                  \\
					J004423.20+035715.5 & 2.22 &   4   &                  & 560.6 $\pm$ 17.3  & 17.9 $\pm$ 3.9  &                &                 &                & 72.1 $\pm$ 2.4  &                        & 11.4 $\pm$ 2.0 & 12.5 $\pm$ 2.6 &                & 66.4 $\pm$ 2.4  &                        &                  \\
					J004600.48+000543.6 & 2.46 &   4   &                  & 267.3 $\pm$ 5.8   & 33.9 $\pm$ 2.9  &                & 14.6 $\pm$ 1.6  &                & 50.3 $\pm$ 2.5  &                        & 7.3 $\pm$ 2.0  &                &                & 91.4 $\pm$ 1.1  &                        & 25.8 $\pm$ 1.2   \\
					J004728.77+004020.3 & 3.06 &       & 101.6 $\pm$ 5.5  & 224.8 $\pm$ 7.1   & 36.0 $\pm$ 3.4  &                &                 &                & 94.4 $\pm$ 6.0  &                        &                &                &                &                 &                        &                  \\
					J012552.08-015218.3 & 3.18 &   4   & 83.1 $\pm$ 5.5   & 116.5 $\pm$ 2.5   & 85.0 $\pm$ 2.5  &                & 34.3 $\pm$ 3.6  &                & 107.2 $\pm$ 2.3 &                        & 8.3 $\pm$ 1.7  &                &                & 35.7 $\pm$ 3.6  &                        &                  \\
					J013747.84+000612.7 & 2.82 &       & 92.2 $\pm$ 7.2   & 567.8 $\pm$ 1.1   & 43.9 $\pm$ 3.3  & 30.1 $\pm$ 3.0 &                 &                & 121.6 $\pm$ 3.0 &                        &                &                &                &                 &                        &                  \\
					J014607.15+121119.9 & 3.14 &   4   & 67.6 $\pm$ 3.5   & 523.8 $\pm$ 7.3   & 60.9 $\pm$ 3.8  & 34.9 $\pm$ 3.2 &                 & 8.3 $\pm$ 1.5  & 97.2 $\pm$ 2.6  &                        & 8.9 $\pm$ 1.2  &                &                & 56.5 $\pm$ 2.1  &                        &                  \\
					J014841.83+055024.2 & 2.66 &       &                  & 233.2 $\pm$ 3.0   &                 &                &                 &                & 19.3 $\pm$ 1.9  &                        &                &                &                & 8.5 $\pm$ 1.6   &                        &                  \\
					J015458.32+015720.2 & 2.89 &       & 33.2 $\pm$ 4.0   & 187.1 $\pm$ 4.6   &                 &                &                 &                & 28.7 $\pm$ 2.7  &                        &                &                &                &                 &                        &                  \\
					J015700.14+073116.0 & 3.05 &   4   & 83.7 $\pm$ 5.2   & 590.1 $\pm$ 3.4   & 40.4 $\pm$ 3.0  &                &                 &                & 112.0 $\pm$ 3.5 &                        & 6.6 $\pm$ 1.2  &                &                & 18.5 $\pm$ 1.8  &                        &                  \\
					J020245.82+000848.4 & 2.22 &   4   &                  & 352.8 $\pm$ 8.2   &                 &                &                 &                & 75.0 $\pm$ 2.5  &                        & 10.4 $\pm$ 1.2 &                &                & 11.9 $\pm$ 2.5  &                        &                  \\
					J020643.64+010403.3 & 2.66 &   4   & 69.3 $\pm$ 9.9   & 465.6 $\pm$ 10.7  & 91.8 $\pm$ 7.4  & 65.9 $\pm$ 6.1 &                 &                & 123.2 $\pm$ 5.6 &                        & 27.1 $\pm$ 5.0 & 22.2 $\pm$ 4.9 & 29.1 $\pm$ 1.5 & 117.3 $\pm$ 1.4 &                        &                  \\
					J020728.19+033833.5 & 2.76 &   4   & 103.6 $\pm$ 6.1  & 404.2 $\pm$ 8.6   & 128.2 $\pm$ 1.9 & 18.4 $\pm$ 2.1 & 40.2 $\pm$ 2.7  &                & 150.7 $\pm$ 4.5 &                        & 23.8 $\pm$ 3.7 & 23.5 $\pm$ 3.7 & 29.5 $\pm$ 0.7 & 64.5 $\pm$ 0.7  &                        &                  \\
					J021834.53-033518.4 & 3.02 &       &                  & 157.4 $\pm$ 5.0   & 16.4 $\pm$ 2.1  &                &                 &                & 37.0 $\pm$ 3.7  &                        &                &                &                &                 &                        &                  \\
					J022447.84+032452.0 & 2.53 &       &                  & 249.8 $\pm$ 1.3   & 22.9 $\pm$ 2.8  &                &                 &                & 28.6 $\pm$ 1.3  &                        &                &                &                &                 &                        &                  \\
					J023210.88+002835.1 & 2.29 &       &                  & 51.8 $\pm$ 2.9    & 22.4 $\pm$ 2.8  &                &                 &                & 18.2 $\pm$ 0.9  &                        &                &                &                & 9.1 $\pm$ 1.6   &                        & 57.0 $\pm$ 9.0   \\
					J023337.89+002303.7 & 2.83 &       &                  & 145.4 $\pm$ 6.5   & 8.9 $\pm$ 1.6   & 3.2 $\pm$ 0.3  &                 &                & 19.1 $\pm$ 1.5  &                        &                &                &                &                 &                        &                  \\ \hline \hline
				\end{tabular}%
			}
		\end{table}
	\end{center}
\end{landscape}

\begin{table*}
	\caption{Set of chemical abundances used in MAPPINGS Ie. The abundances were obtained using primary behaviour of C (herefater PC).}
	\label{PC}
	\resizebox{0.7\textwidth}{!}{%
		\begin{tabular}{ccccccc}
			\hline
			Z/Z$_{\odot}$ & 0.5                   & 1.0                  & 2.0                   & 3.0                   & 4.0                    & 5.0                   \\
			\hline
			\hline
			H             & 1.0                   & 1.0                  & 1.0                   & 1.0                   & 1.0                    & 1.0                   \\
			He            & 8.5$\times$10$^{-2}$  & 8.5$\times$10$^{-2}$ & 8.5$\times$10$^{-2}$  & 8.5$\times$10$^{-2}$  & 8.5$\times$10$^{-2}$   & 8.5$\times$10$^{-2}$  \\
			C             & 1.35$\times$10$^{-4}$ & 2.7$\times$10$^{-4}$ & 5.4$\times$10$^{-4}$  & 8.1$\times$10$^{-4}$  & 1.08$\times$10$^{-3}$  & 1.35$\times$10$^{-3}$ \\
			N             & 1.7$\times$10$^{-5}$  & 6.8$\times$10$^{-5}$ & 2.72$\times$10$^{-4}$ & 6.12$\times$10$^{-4}$ & 1.088$\times$10$^{-3}$ & 1.7$\times$10$^{-3}$  \\
			O             & 2.45$\times$10$^{-4}$ & 4.9$\times$10$^{-4}$ & 9.8$\times$10$^{-4}$  & 1.47$\times$10$^{-3}$ & 1.96$\times$10$^{-3}$  & 2.45$\times$10$^{-3}$ \\
			Ne            & 4.25$\times$10$^{-5}$ & 8.5$\times$10$^{-5}$ & 1.7$\times$10$^{-4}$  & 2.55$\times$10$^{-4}$ & 3.4$\times$10$^{-4}$   & 4.25$\times$10$^{-4}$ \\
			Fe            & 1.6$\times$10$^{-5}$  & 3.2$\times$10$^{-5}$ & 6.4$\times$10$^{-5}$  & 9.6$\times$10$^{-5}$  & 1.28$\times$10$^{-4}$  & 1.6$\times$10$^{-4}$  \\
			Mg            & 2.0$\times$10$^{-5}$  & 4.0$\times$10$^{-5}$ & 8.0$\times$10$^{-5}$  & 1.2$\times$10$^{-4}$  & 1.6$\times$10$^{-4}$   & 2.0$\times$10$^{-4}$  \\
			Si            & 1.6$\times$10$^{-5}$  & 3.2$\times$10$^{-5}$ & 6.4$\times$10$^{-5}$  & 9.6$\times$10$^{-5}$  & 1.28$\times$10$^{-4}$  & 1.6$\times$10$^{-4}$  \\
			S             & 6.5$\times$10$^{-6}$  & 1.3$\times$10$^{-5}$ & 2.6$\times$10$^{-5}$  & 3.9$\times$10$^{-5}$  & 5.2$\times$10$^{-5}$   & 6.5$\times$10$^{-5}$  \\
			Ca            & 1.1$\times$10$^{-6}$  & 2.2$\times$10$^{-6}$ & 4.4$\times$10$^{-6}$  & 6.6$\times$10$^{-6}$  & 8.8$\times$10$^{-6}$   & 1.1$\times$10$^{-5}$  \\
			Ar            & 1.25$\times$10$^{-6}$ & 2.5$\times$10$^{-6}$ & 5.0$\times$10$^{-6}$  & 7.5$\times$10$^{-6}$  & 1.0$\times$10$^{-5}$   & 1.25$\times$10$^{-5}$ \\
			\hline
			\hline
		\end{tabular}%
	}
\end{table*}

\begin{table*}
	\caption{Set of chemical abundances used in MAPPINGS Ie. The abundances were obtained using secondary behaviour of C (herefater SC).}
	\label{SC}
	\resizebox{0.7\textwidth}{!}{%
		\begin{tabular}{ccccccc}
			\hline
			Z/Z$_{\odot}$ & 0.5                   & 1.0                  & 2.0                   & 3.0                   & 4.0                    & 5.0                   \\
			\hline
			\hline
			H             & 1.0                   & 1.0                  & 1.0                   & 1.0                   & 1.0                    & 1.0                   \\
			He            & 8.5$\times$10$^{-2}$  & 8.5$\times$10$^{-2}$ & 8.5$\times$10$^{-2}$  & 8.5$\times$10$^{-2}$  & 8.5$\times$10$^{-2}$   & 8.5$\times$10$^{-2}$  \\
			C             & 6.75$\times$10$^{-5}$ & 2.7$\times$10$^{-4}$ & 1.08$\times$10$^{-3}$ & 2.43$\times$10$^{-3}$ & 4.32$\times$10$^{-3}$  & 6.75$\times$10$^{-3}$ \\
			N             & 1.7$\times$10$^{-5}$  & 6.8$\times$10$^{-5}$ & 2.72$\times$10$^{-4}$ & 6.12$\times$10$^{-4}$ & 1.088$\times$10$^{-3}$ & 1.7$\times$10$^{-3}$  \\
			O             & 2.45$\times$10$^{-4}$ & 4.9$\times$10$^{-4}$ & 9.8$\times$10$^{-4}$  & 1.47$\times$10$^{-3}$ & 1.96$\times$10$^{-3}$  & 2.45$\times$10$^{-3}$ \\
			Ne            & 4.25$\times$10$^{-5}$ & 8.5$\times$10$^{-5}$ & 1.7$\times$10$^{-4}$  & 2.55$\times$10$^{-4}$ & 3.4$\times$10$^{-4}$   & 4.25$\times$10$^{-4}$ \\
			Fe            & 1.6$\times$10$^{-5}$  & 3.2$\times$10$^{-5}$ & 6.4$\times$10$^{-5}$  & 9.6$\times$10$^{-5}$  & 1.28$\times$10$^{-4}$  & 1.6$\times$10$^{-4}$  \\
			Mg            & 2.0$\times$10$^{-5}$  & 4.0$\times$10$^{-5}$ & 8.0$\times$10$^{-5}$  & 1.2$\times$10$^{-4}$  & 1.6$\times$10$^{-4}$   & 2.0$\times$10$^{-4}$  \\
			Si            & 1.6$\times$10$^{-5}$  & 3.2$\times$10$^{-5}$ & 6.4$\times$10$^{-5}$  & 9.6$\times$10$^{-5}$  & 1.28$\times$10$^{-4}$  & 1.6$\times$10$^{-4}$  \\
			S             & 6.5$\times$10$^{-6}$  & 1.3$\times$10$^{-5}$ & 2.6$\times$10$^{-5}$  & 3.9$\times$10$^{-5}$  & 5.2$\times$10$^{-5}$   & 6.5$\times$10$^{-5}$  \\
			Ca            & 1.1$\times$10$^{-6}$  & 2.2$\times$10$^{-6}$ & 4.4$\times$10$^{-6}$  & 6.6$\times$10$^{-6}$  & 8.8$\times$10$^{-6}$   & 1.1$\times$10$^{-5}$  \\
			Ar            & 1.25$\times$10$^{-6}$ & 2.5$\times$10$^{-6}$ & 5.0$\times$10$^{-6}$  & 7.5$\times$10$^{-6}$  & 1.0$\times$10$^{-5}$   & 1.25$\times$10$^{-5}$ \\
			\hline
			\hline
		\end{tabular}%
	}
\end{table*}

\begin{table*}
	\caption{Set of chemical abundances used in MAPPINGS Ie. The abundances were obtained using primary carbon with Si/O=10xSolar (or PCSi10).}
	\label{PCSi10}
	\resizebox{0.7\textwidth}{!}{%
		\begin{tabular}{ccccccc}
			\hline
			Z/Z$_{\odot}$ & 0.5                   & 1.0                  & 2.0                   & 3.0                   & 4.0                    & 5.0                   \\
			\hline
			\hline
			H             & 1.0                   & 1.0                  & 1.0                   & 1.0                   & 1.0                    & 1.0                   \\
			He            & 8.5$\times$10$^{-2}$  & 8.5$\times$10$^{-2}$ & 8.5$\times$10$^{-2}$  & 8.5$\times$10$^{-2}$  & 8.5$\times$10$^{-2}$   & 8.5$\times$10$^{-2}$  \\
			C             & 1.35$\times$10$^{-4}$ & 2.7$\times$10$^{-4}$ & 1.08$\times$10$^{-3}$ & 2.43$\times$10$^{-3}$ & 4.32$\times$10$^{-3}$  & 6.75$\times$10$^{-3}$ \\
			N             & 1.7$\times$10$^{-5}$  & 6.8$\times$10$^{-5}$ & 2.72$\times$10$^{-4}$ & 6.12$\times$10$^{-4}$ & 1.088$\times$10$^{-3}$ & 1.7$\times$10$^{-3}$  \\
			O             & 2.45$\times$10$^{-4}$ & 4.9$\times$10$^{-4}$ & 9.8$\times$10$^{-4}$  & 1.47$\times$10$^{-3}$ & 1.96$\times$10$^{-3}$  & 2.45$\times$10$^{-3}$ \\
			Ne            & 4.25$\times$10$^{-5}$ & 8.5$\times$10$^{-5}$ & 1.7$\times$10$^{-4}$  & 2.55$\times$10$^{-4}$ & 3.4$\times$10$^{-4}$   & 4.25$\times$10$^{-4}$ \\
			Fe            & 1.6$\times$10$^{-5}$  & 3.2$\times$10$^{-5}$ & 6.4$\times$10$^{-5}$  & 9.6$\times$10$^{-5}$  & 1.28$\times$10$^{-4}$  & 1.6$\times$10$^{-4}$  \\
			Mg            & 2.0$\times$10$^{-5}$  & 4.0$\times$10$^{-5}$ & 8.0$\times$10$^{-5}$  & 1.2$\times$10$^{-4}$  & 1.6$\times$10$^{-4}$   & 2.0$\times$10$^{-4}$  \\
			Si            & 1.6$\times$10$^{-4}$  & 3.2$\times$10$^{-4}$ & 6.4$\times$10$^{-4}$  & 9.6$\times$10$^{-4}$  & 1.28$\times$10$^{-3}$  & 1.6$\times$10$^{-3}$  \\
			S             & 6.5$\times$10$^{-6}$  & 1.3$\times$10$^{-5}$ & 2.6$\times$10$^{-5}$  & 3.9$\times$10$^{-5}$  & 5.2$\times$10$^{-5}$   & 6.5$\times$10$^{-5}$  \\
			Ca            & 1.1$\times$10$^{-6}$  & 2.2$\times$10$^{-6}$ & 4.4$\times$10$^{-6}$  & 6.6$\times$10$^{-6}$  & 8.8$\times$10$^{-6}$   & 1.1$\times$10$^{-5}$  \\
			Ar            & 1.25$\times$10$^{-6}$ & 2.5$\times$10$^{-6}$ & 5.0$\times$10$^{-6}$  & 7.5$\times$10$^{-6}$  & 1.0$\times$10$^{-5}$   & 1.25$\times$10$^{-5}$ \\
			\hline
			\hline
		\end{tabular}%
	}
\end{table*}

\begin{table*}
	\caption{Set of chemical abundances used in MAPPINGS Ie. The abundances were obtained using secondary carbon with Si/O=10xSolar (or SCSi10).}
	\label{SCSi10}
	\resizebox{0.7\textwidth}{!}{%
		\begin{tabular}{ccccccc}
			\hline
			Z/Z$_{\odot}$ & 0.5                   & 1.0                  & 2.0                   & 3.0                   & 4.0                    & 5.0                   \\
			\hline
			\hline
			H             & 1.0                   & 1.0                  & 1.0                   & 1.0                   & 1.0                    & 1.0                   \\
			He            & 8.5$\times$10$^{-2}$  & 8.5$\times$10$^{-2}$ & 8.5$\times$10$^{-2}$  & 8.5$\times$10$^{-2}$  & 8.5$\times$10$^{-2}$   & 8.5$\times$10$^{-2}$  \\
			C             & 6.75$\times$10$^{-5}$ & 2.7$\times$10$^{-4}$ & 1.08$\times$10$^{-3}$ & 2.43$\times$10$^{-3}$ & 4.32$\times$10$^{-3}$  & 6.75$\times$10$^{-3}$ \\
			N             & 1.7$\times$10$^{-5}$  & 6.8$\times$10$^{-5}$ & 2.72$\times$10$^{-4}$ & 6.12$\times$10$^{-4}$ & 1.088$\times$10$^{-3}$ & 1.7$\times$10$^{-3}$  \\
			O             & 2.45$\times$10$^{-4}$ & 4.9$\times$10$^{-4}$ & 9.8$\times$10$^{-4}$  & 1.47$\times$10$^{-3}$ & 1.96$\times$10$^{-3}$  & 2.45$\times$10$^{-3}$ \\
			Ne            & 4.25$\times$10$^{-5}$ & 8.5$\times$10$^{-5}$ & 1.7$\times$10$^{-4}$  & 2.55$\times$10$^{-4}$ & 3.4$\times$10$^{-4}$   & 4.25$\times$10$^{-4}$ \\
			Fe            & 1.6$\times$10$^{-5}$  & 3.2$\times$10$^{-5}$ & 6.4$\times$10$^{-5}$  & 9.6$\times$10$^{-5}$  & 1.28$\times$10$^{-4}$  & 1.6$\times$10$^{-4}$  \\
			Mg            & 2.0$\times$10$^{-5}$  & 4.0$\times$10$^{-5}$ & 8.0$\times$10$^{-5}$  & 1.2$\times$10$^{-4}$  & 1.6$\times$10$^{-4}$   & 2.0$\times$10$^{-4}$  \\
			Si            & 1.6$\times$10$^{-4}$  & 3.2$\times$10$^{-4}$ & 6.4$\times$10$^{-4}$  & 9.6$\times$10$^{-4}$  & 1.28$\times$10$^{-3}$  & 1.6$\times$10$^{-3}$  \\
			S             & 6.5$\times$10$^{-6}$  & 1.3$\times$10$^{-5}$ & 2.6$\times$10$^{-5}$  & 3.9$\times$10$^{-5}$  & 5.2$\times$10$^{-5}$   & 6.5$\times$10$^{-5}$  \\
			Ca            & 1.1$\times$10$^{-6}$  & 2.2$\times$10$^{-6}$ & 4.4$\times$10$^{-6}$  & 6.6$\times$10$^{-6}$  & 8.8$\times$10$^{-6}$   & 1.1$\times$10$^{-5}$  \\
			Ar            & 1.25$\times$10$^{-6}$ & 2.5$\times$10$^{-6}$ & 5.0$\times$10$^{-6}$  & 7.5$\times$10$^{-6}$  & 1.0$\times$10$^{-5}$   & 1.25$\times$10$^{-5}$ \\
			\hline
			\hline
		\end{tabular}%
	}
\end{table*}

\begin{figure*}

	\subfloat[]{
		\includegraphics[width=\columnwidth,height=2.0in,keepaspectratio]{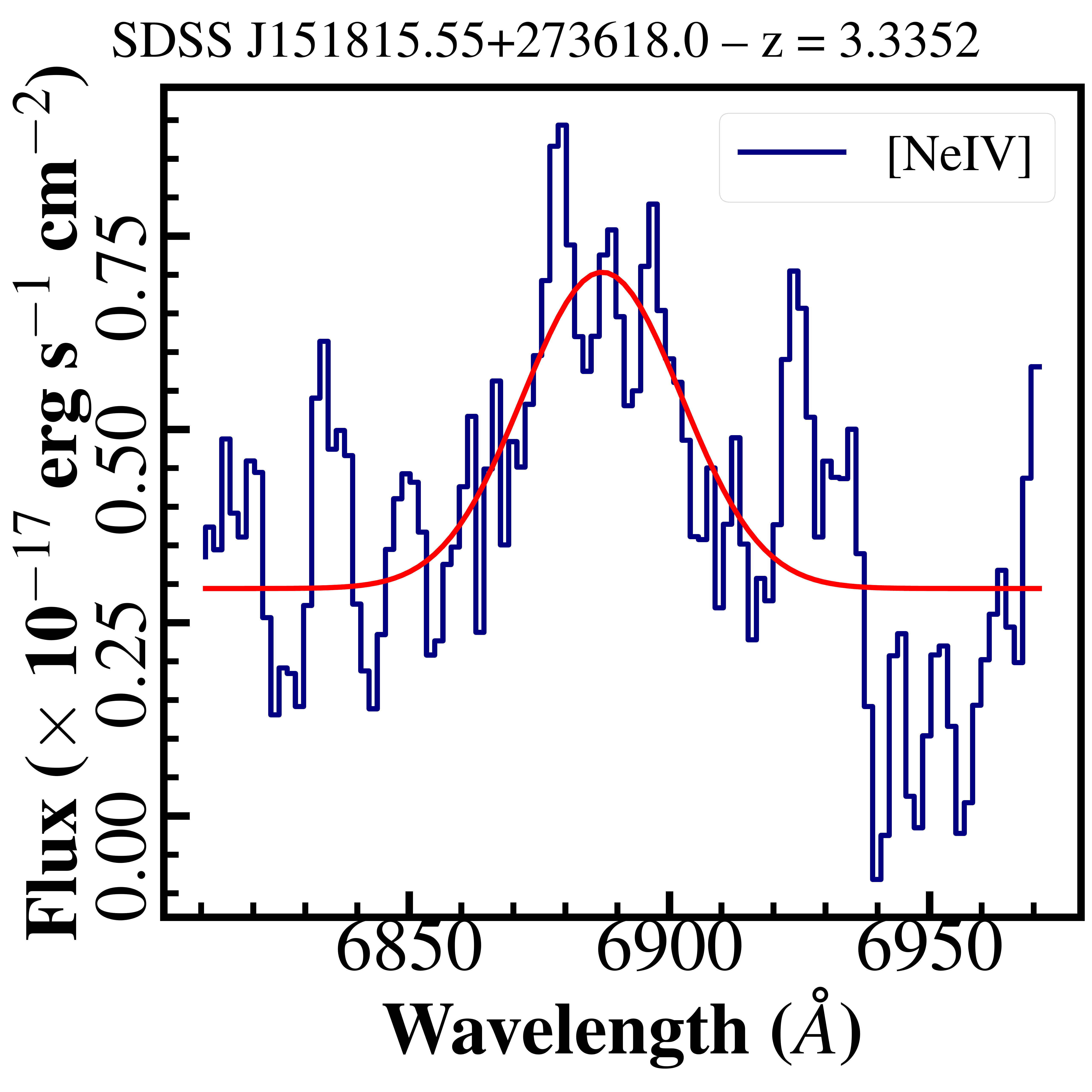}}
	\subfloat[]{
		\includegraphics[width=\columnwidth,height=2.0in,keepaspectratio]{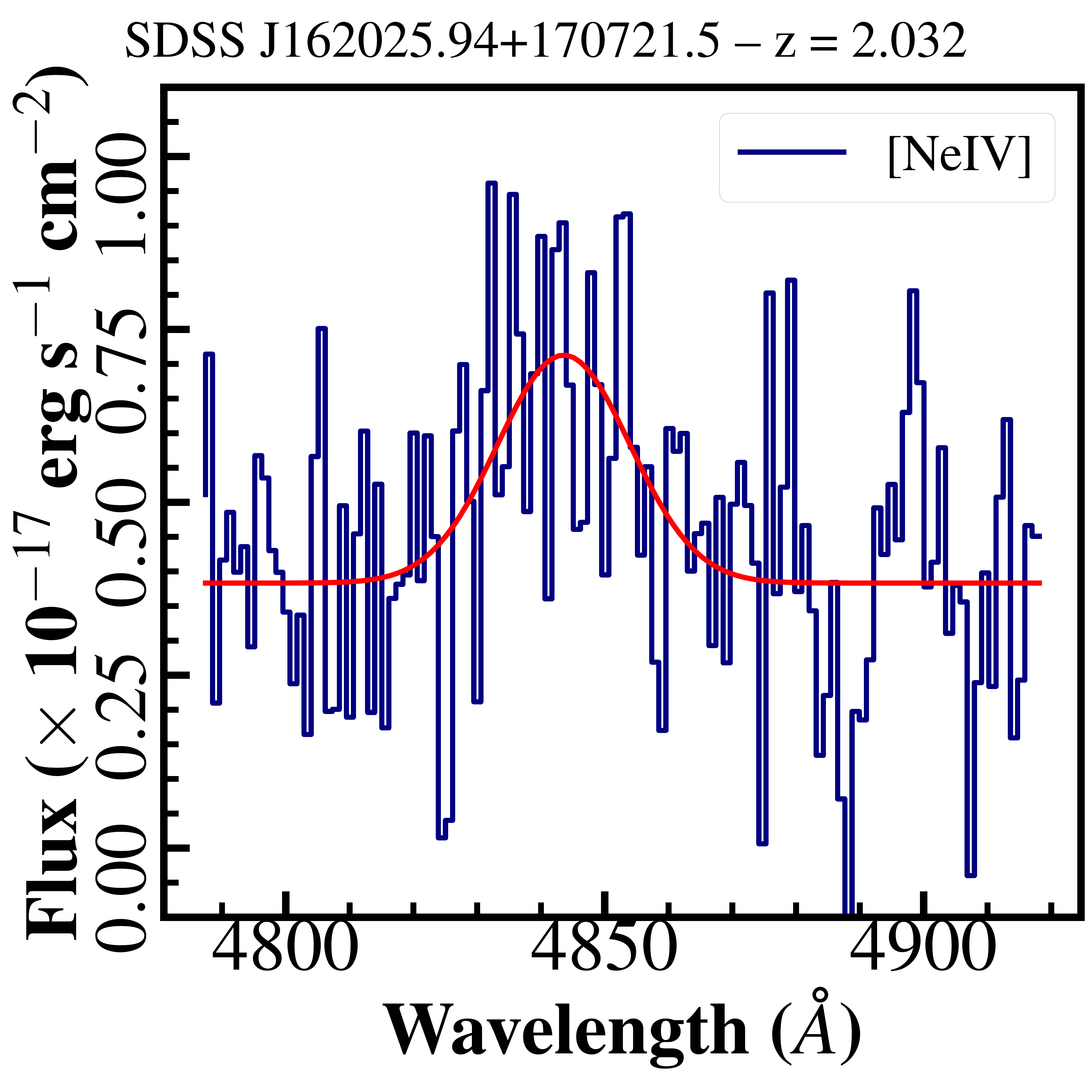}}
	\subfloat[]{
		\includegraphics[width=\columnwidth,height=2.0in,keepaspectratio]{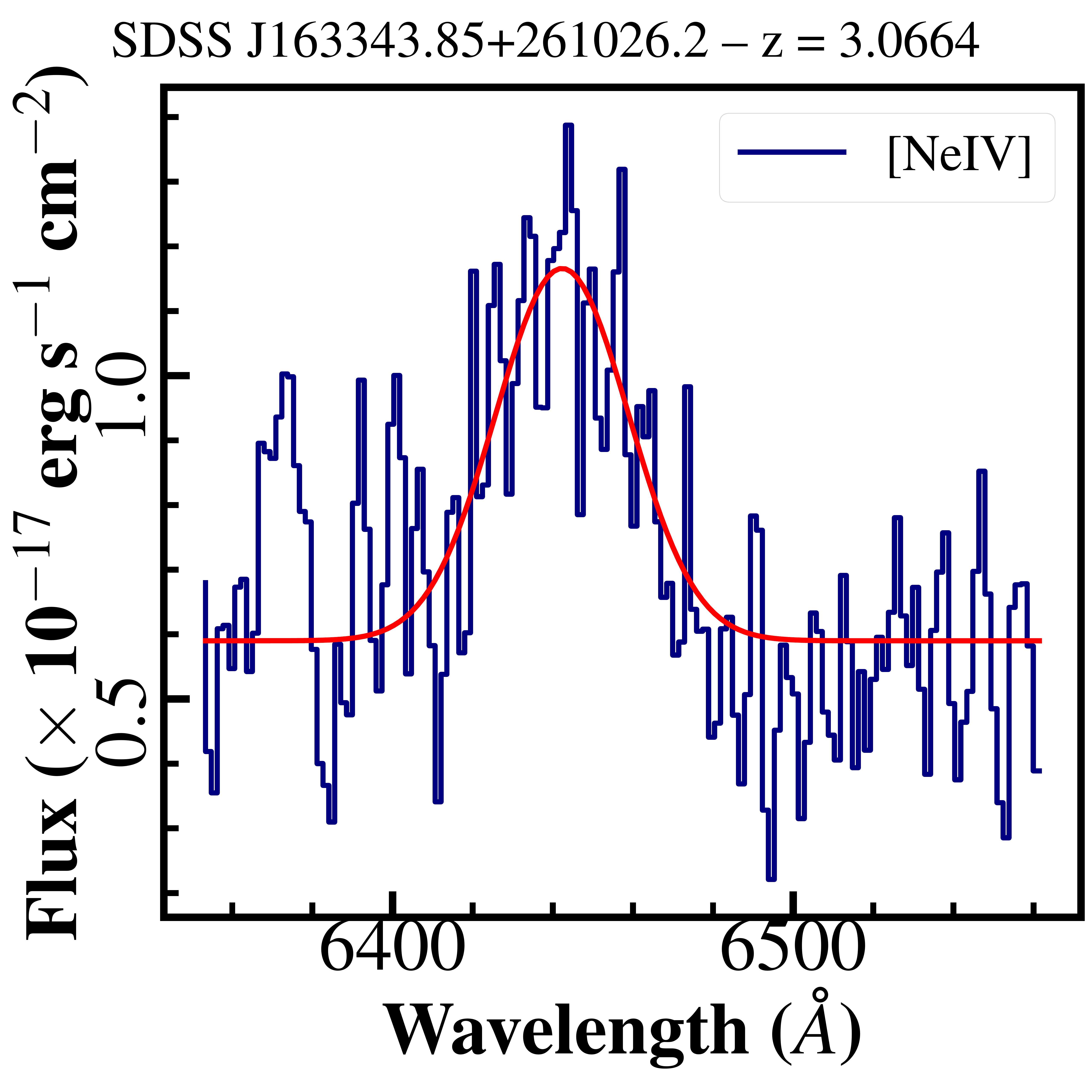}}

	\caption{Spectra of the emission line \ion{[Ne}{IV]} $\lambda$ 1602 detected in some type II quasars.}
	\label{neiv1602}
\end{figure*}

\begin{figure*}

	\subfloat[]{
		\includegraphics[width=\columnwidth,height=2.0in,keepaspectratio]{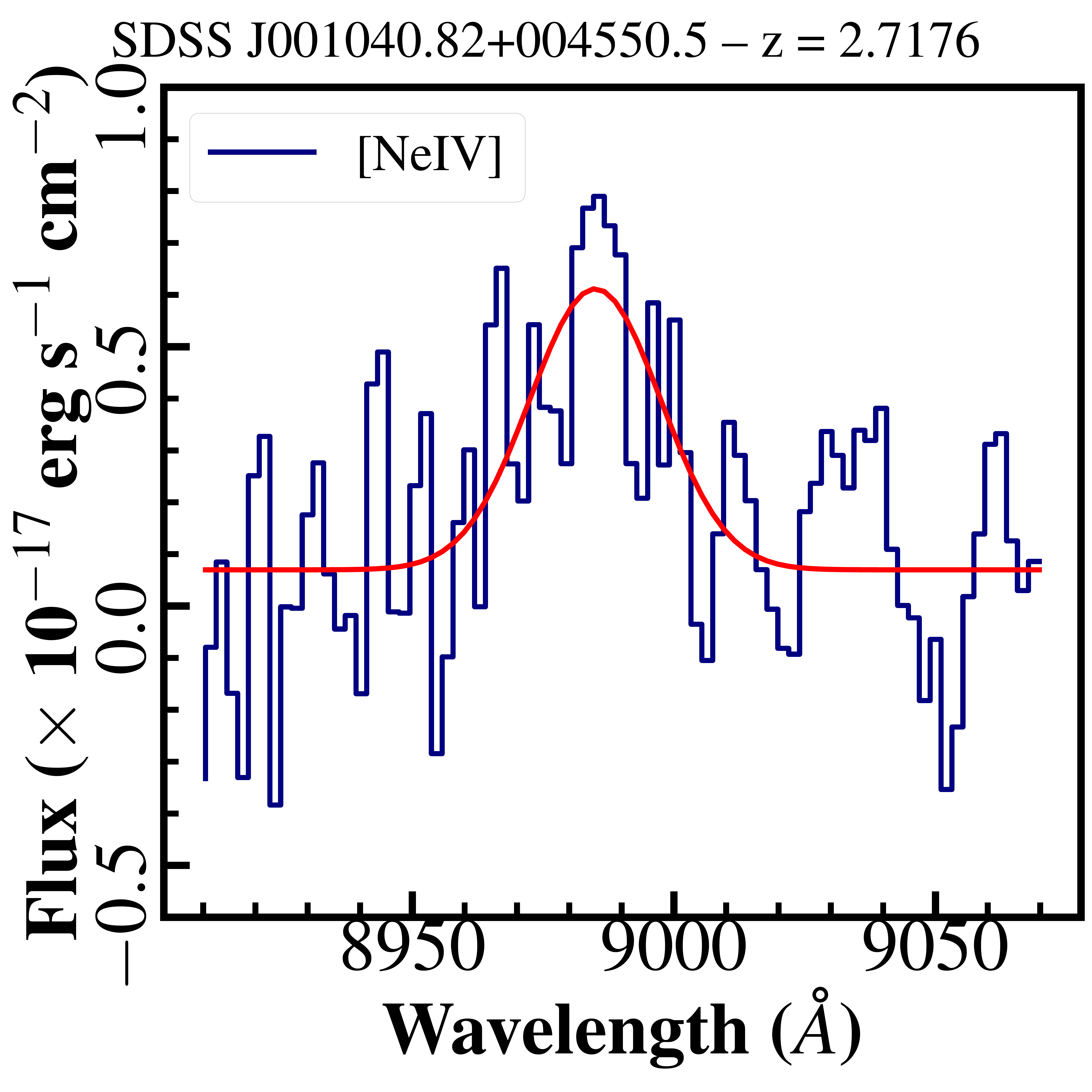}}
	\subfloat[]{
		\includegraphics[width=\columnwidth,height=2.0in,keepaspectratio]{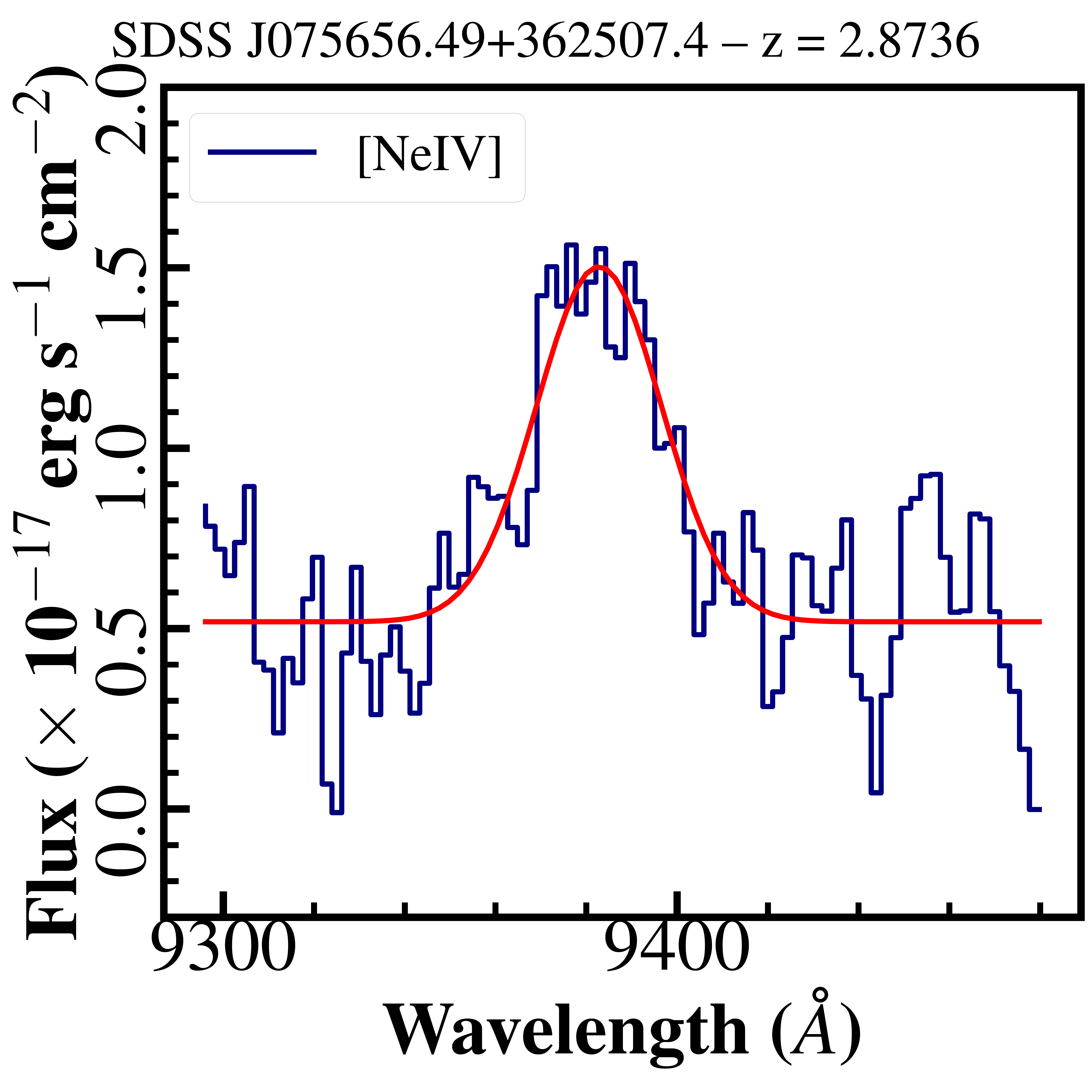}}
	\subfloat[]{
		\includegraphics[width=\columnwidth,height=2.0in,keepaspectratio]{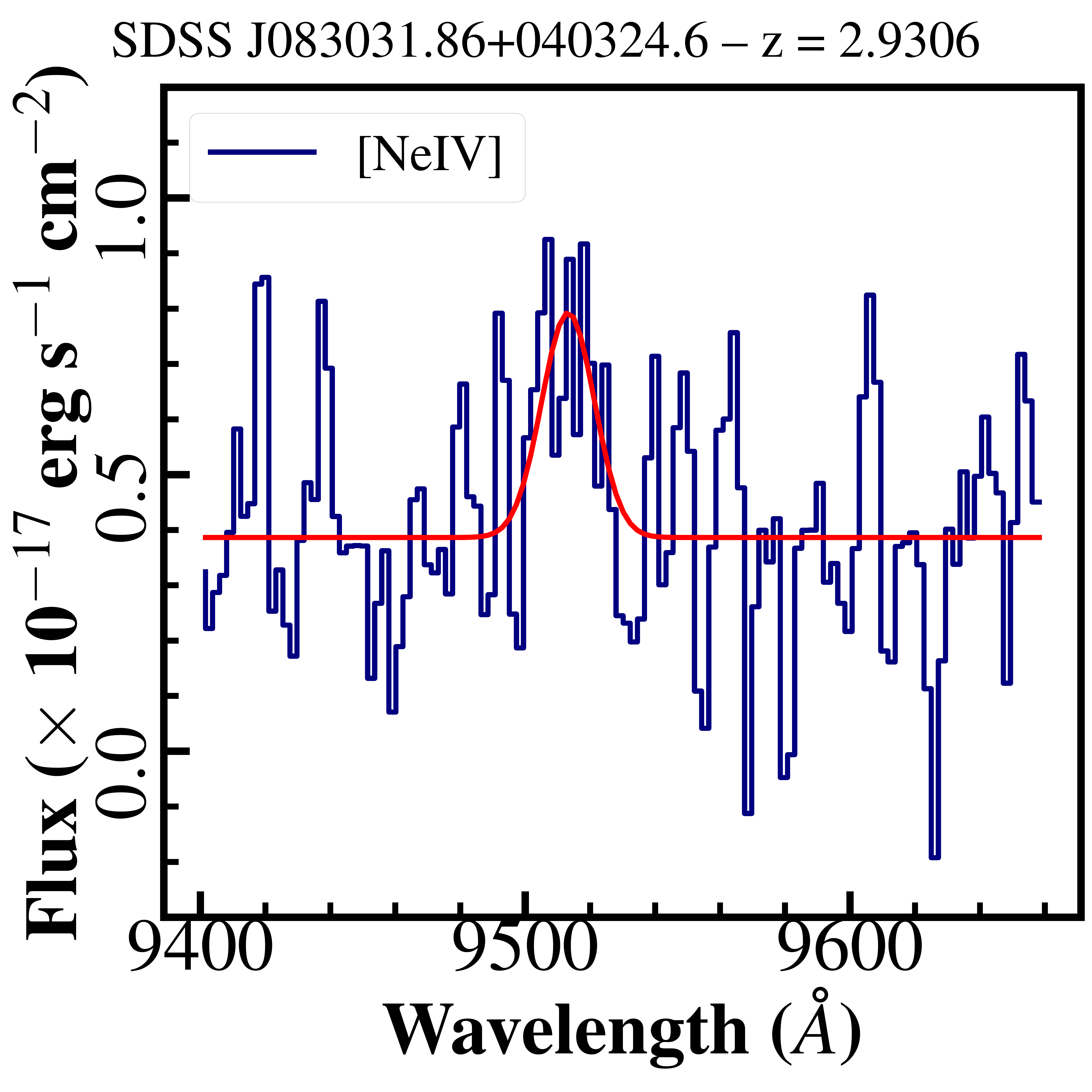}}

	\caption{Spectra of the emission line \ion{[Ne}{IV]} $\lambda$ 2422 detected in some type II quasars.}
	\label{neiv2422}
\end{figure*}

\begin{figure*}
		\subfloat[]{
		\includegraphics[width=\columnwidth,height=2.0in,keepaspectratio]{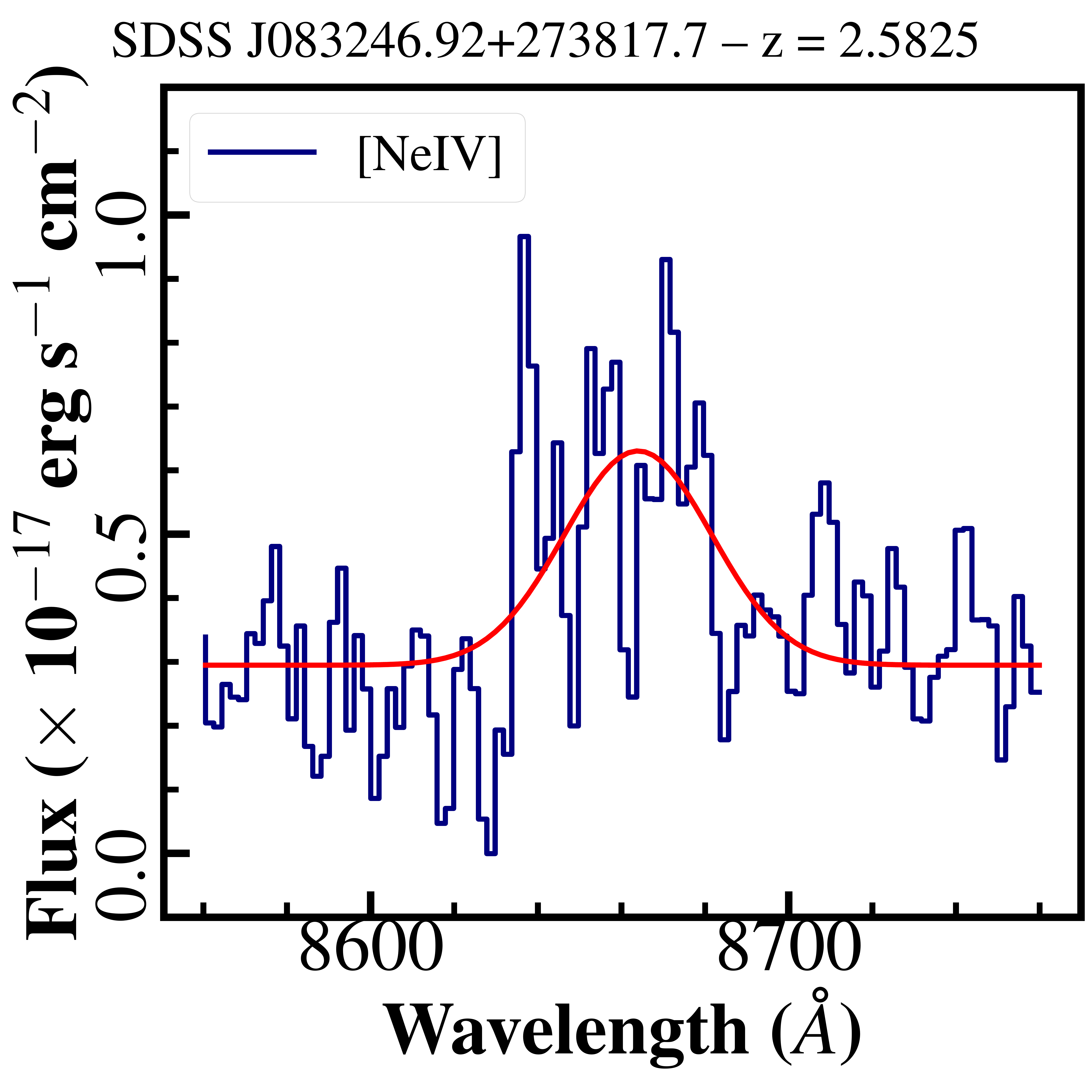}}
	\subfloat[]{
		\includegraphics[width=\columnwidth,height=2.0in,keepaspectratio]{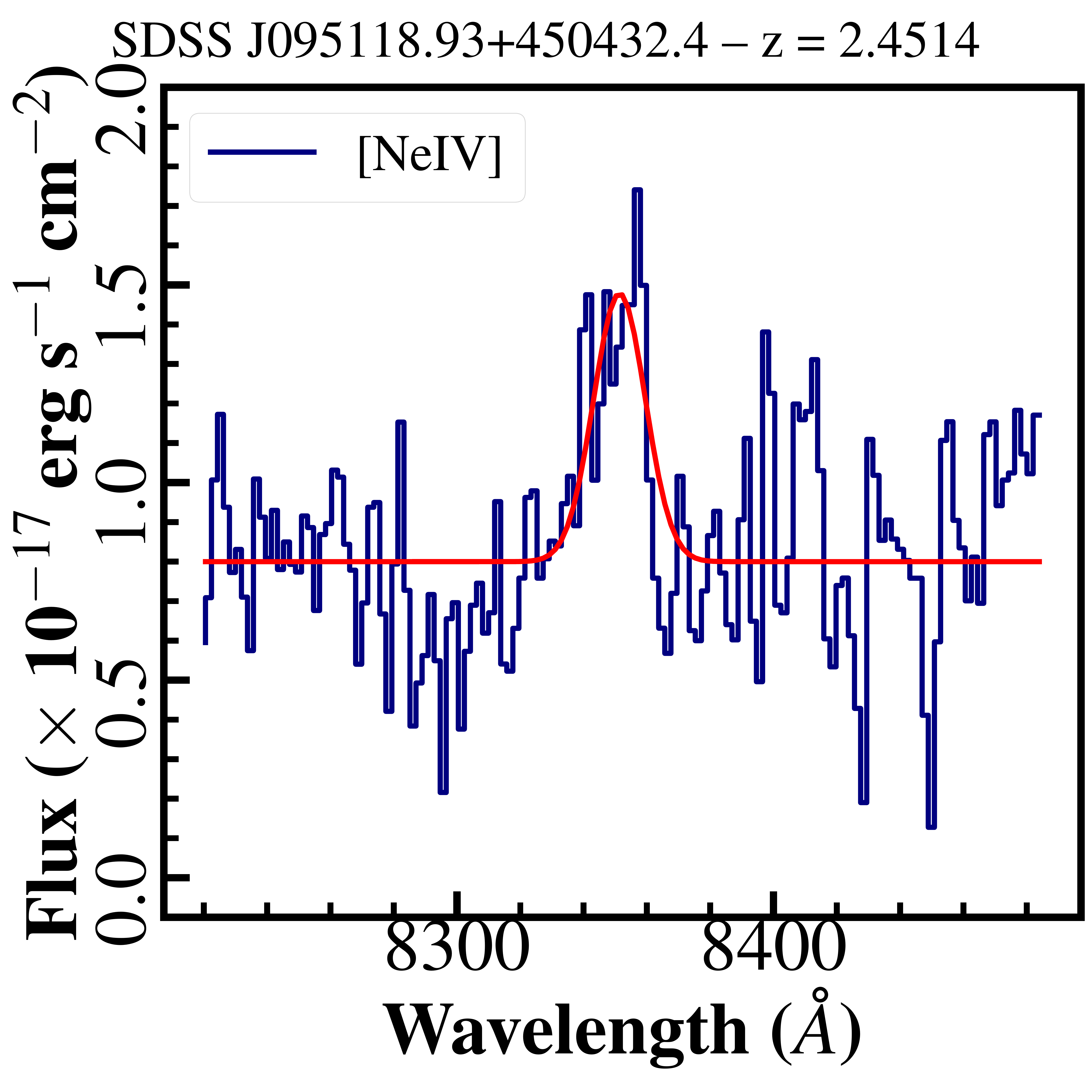}}
	\subfloat[]{
		\includegraphics[width=\columnwidth,height=2.0in,keepaspectratio]{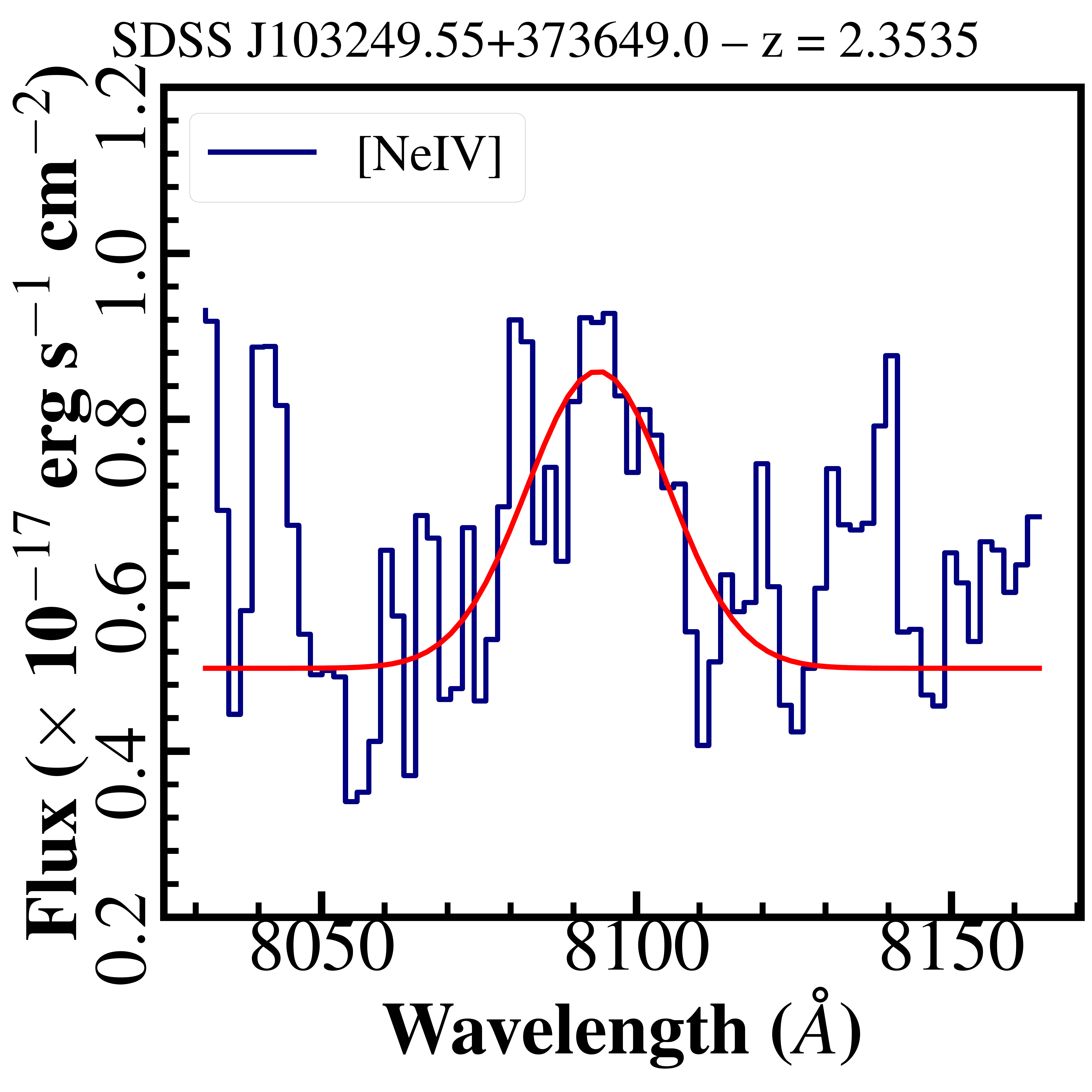}}
	\quad
	\subfloat[]{
		\includegraphics[width=\columnwidth,height=2.0in,keepaspectratio]{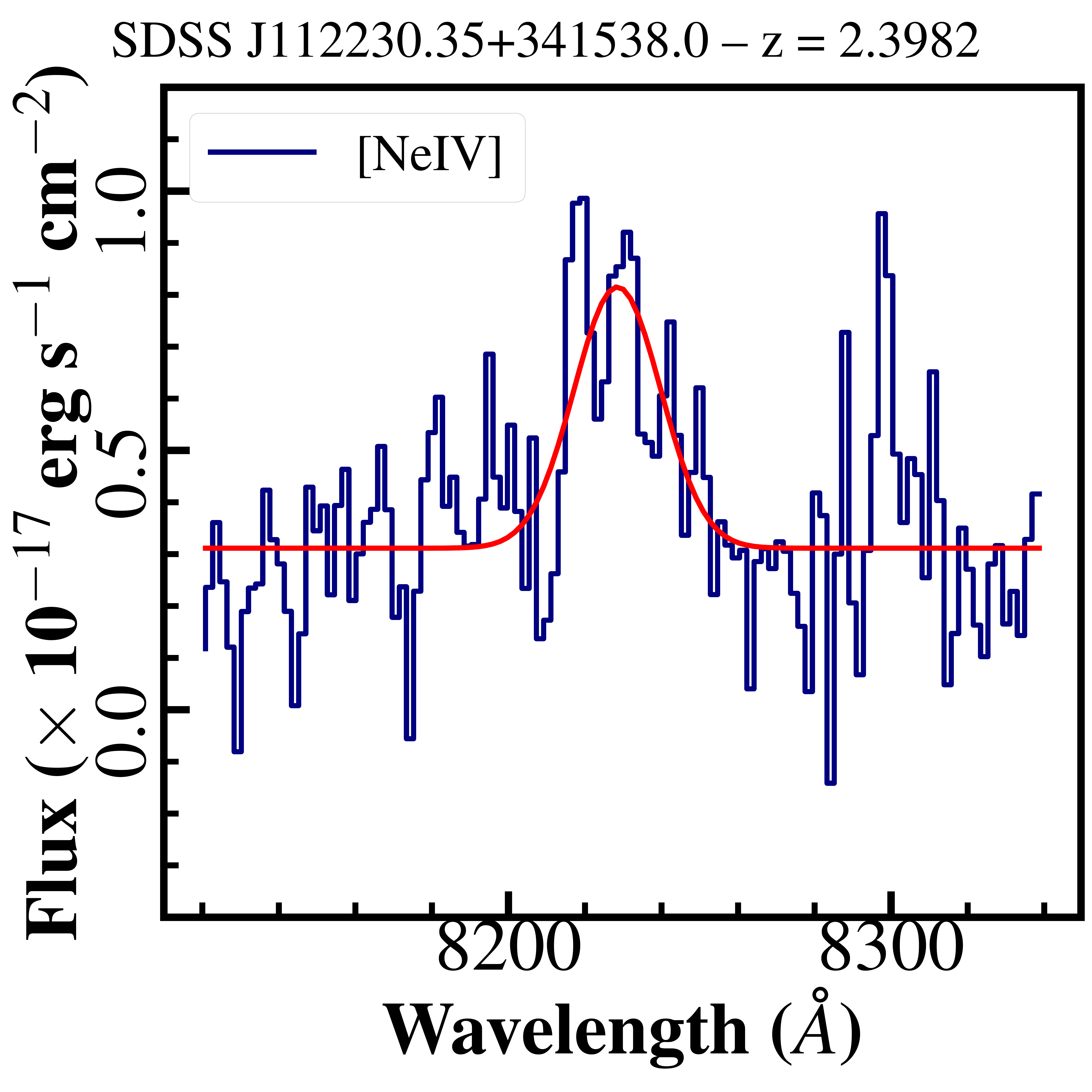}}
	\subfloat[]{
		\includegraphics[width=\columnwidth,height=2.0in,keepaspectratio]{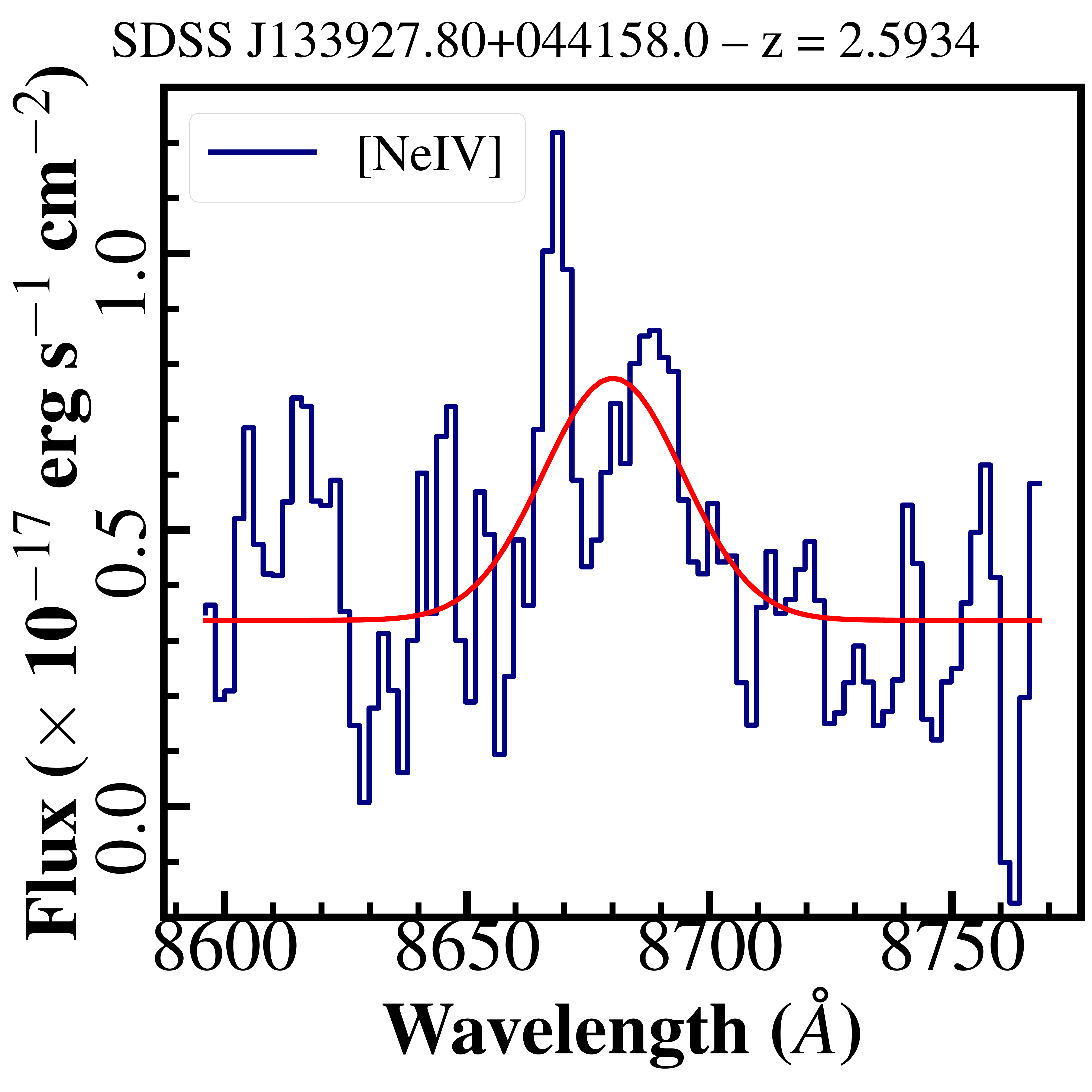}}
	\subfloat[]{
		\includegraphics[width=\columnwidth,height=2.0in,keepaspectratio]{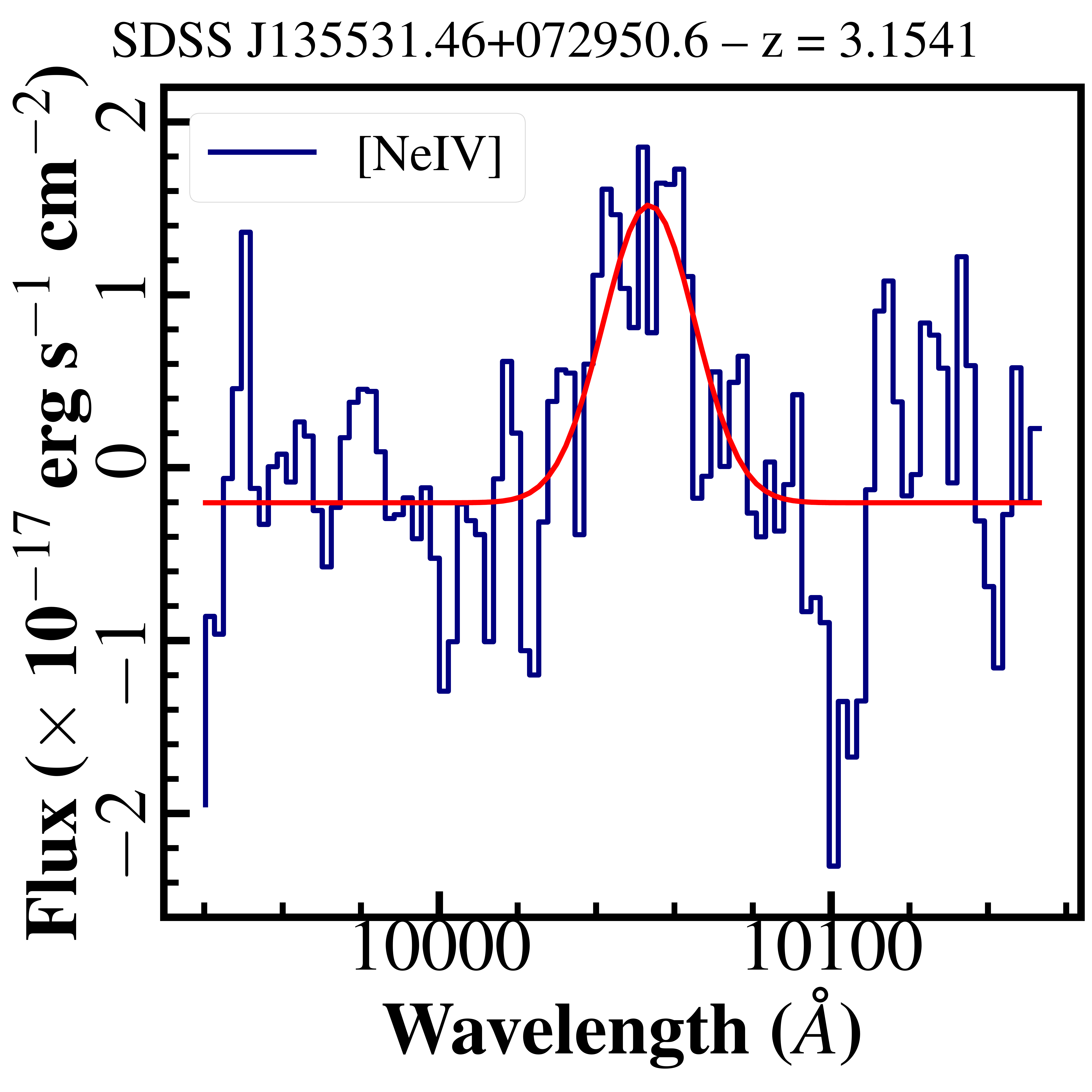}}
	\quad
	\subfloat[]{
		\includegraphics[width=\columnwidth,height=2.0in,keepaspectratio]{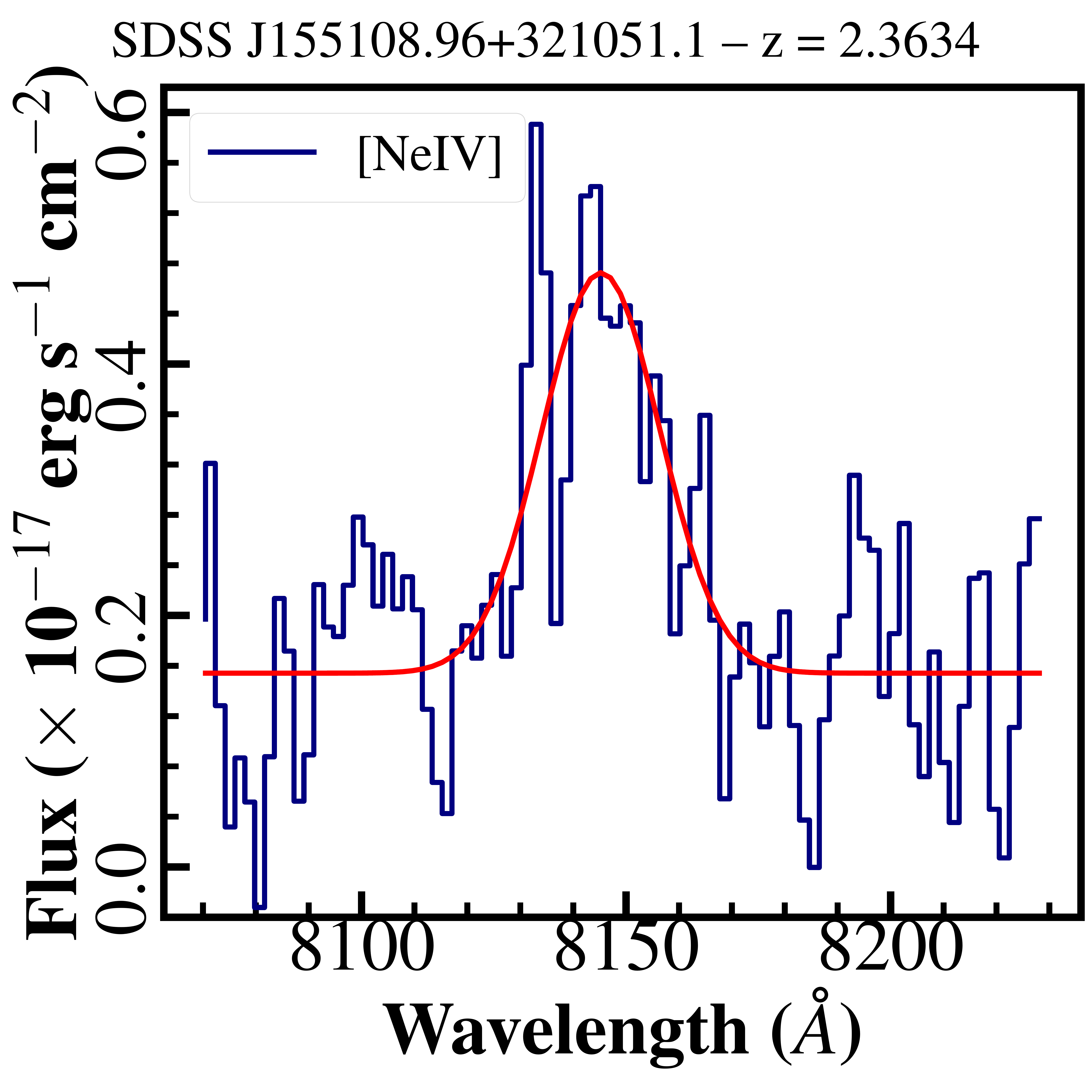}}
	\subfloat[]{
		\includegraphics[width=\columnwidth,height=2.0in,keepaspectratio]{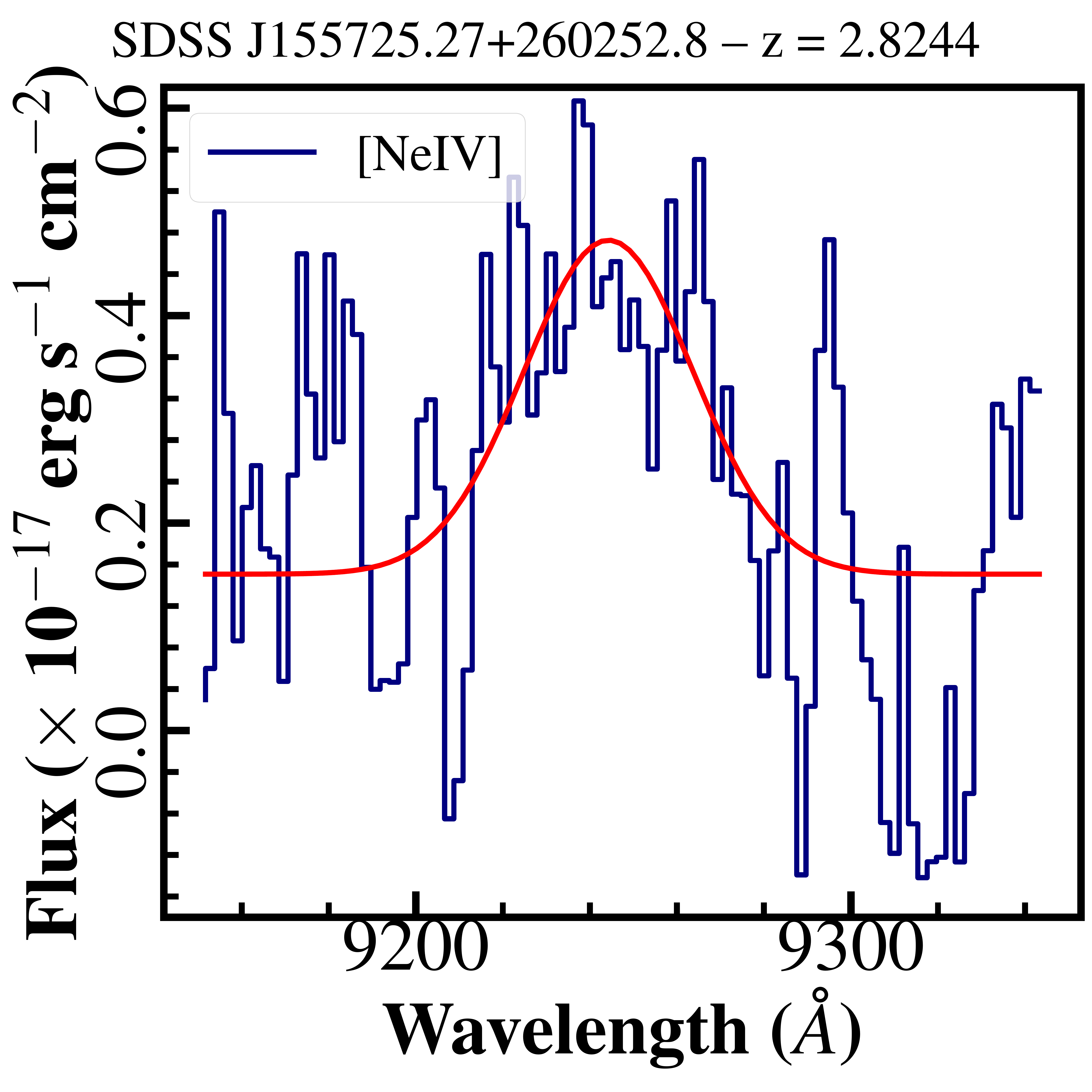}}
	\subfloat[]{
		\includegraphics[width=\columnwidth,height=2.0in,keepaspectratio]{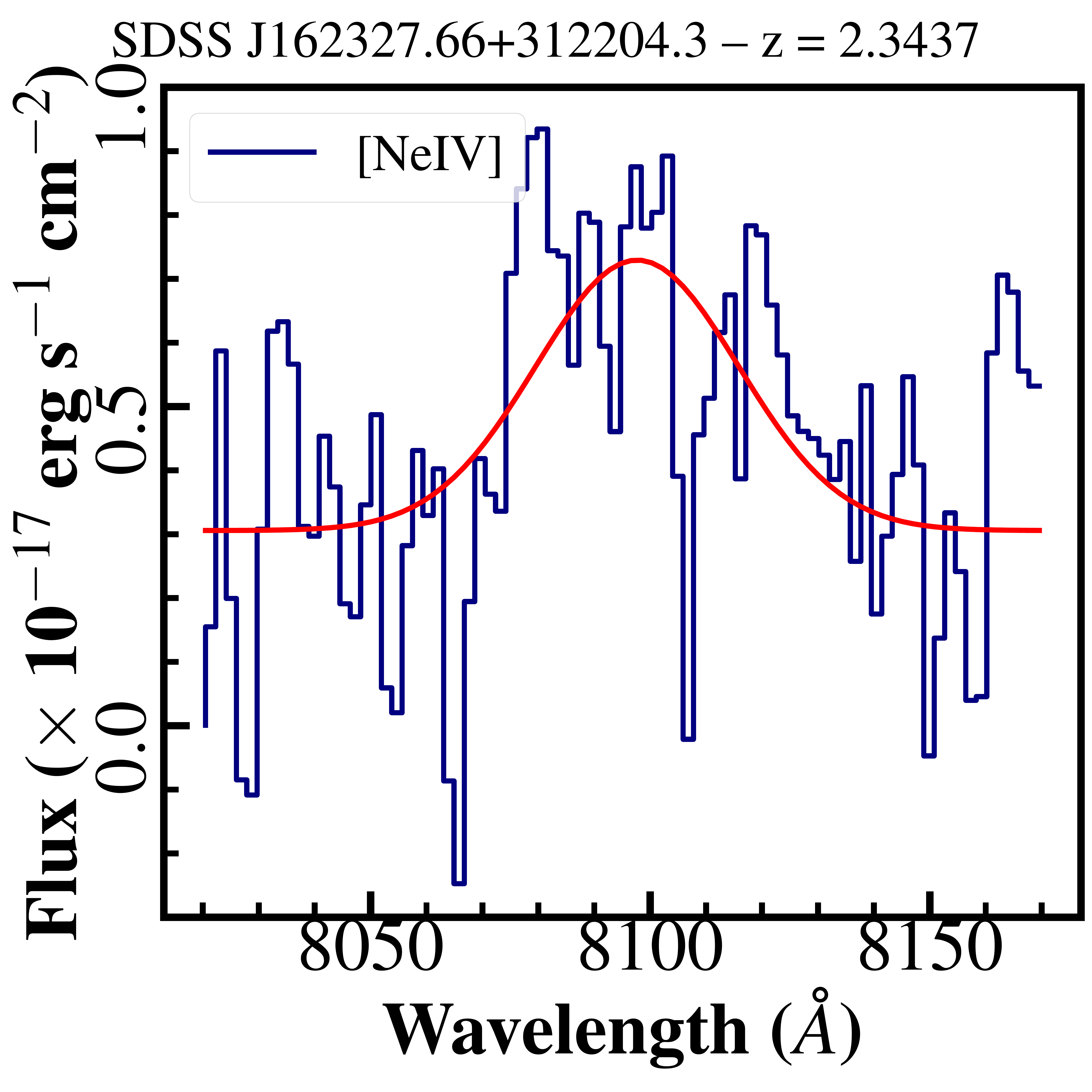}}
	\quad
	\subfloat[]{
		\includegraphics[width=\columnwidth,height=2.0in,keepaspectratio]{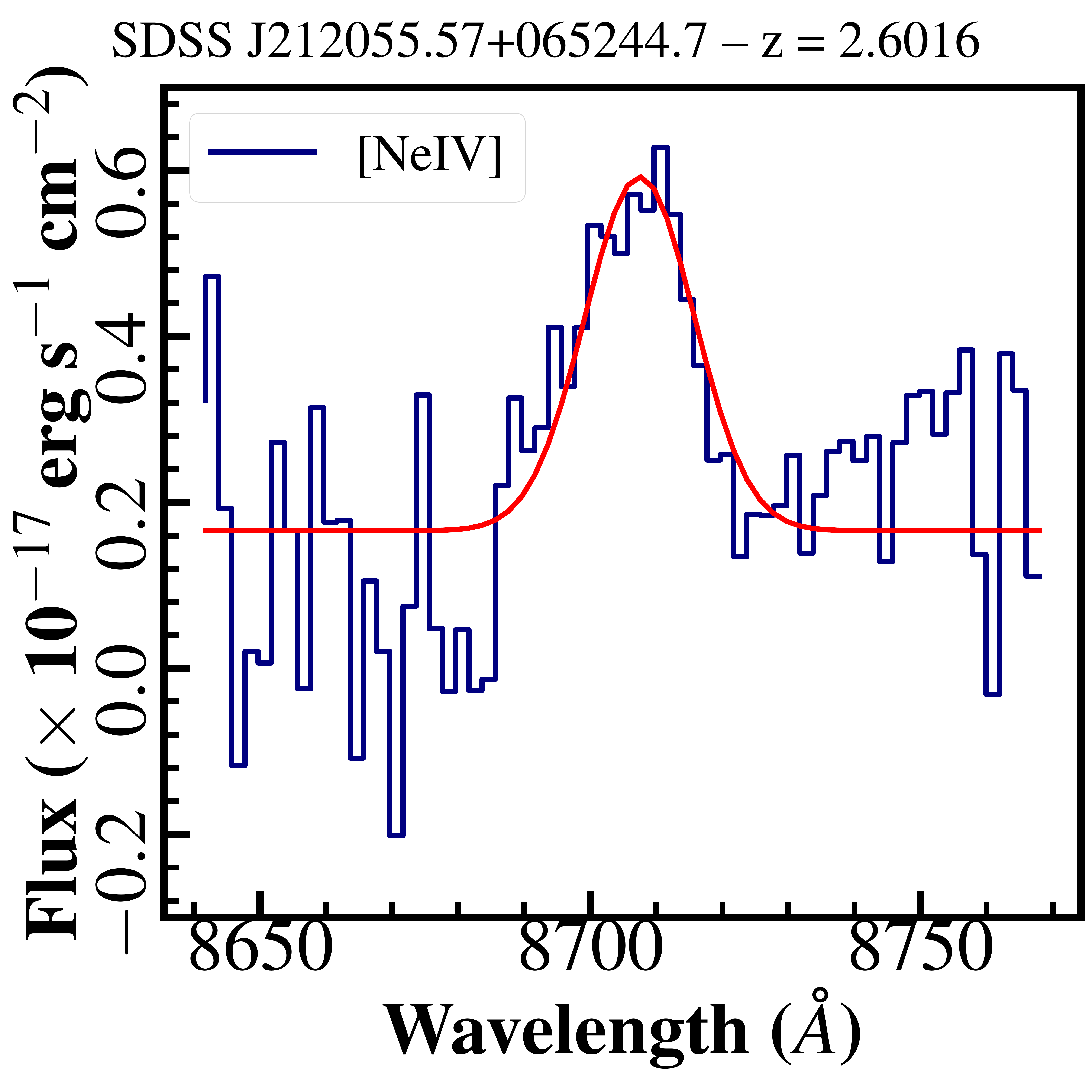}}
	\subfloat[]{
		\includegraphics[width=\columnwidth,height=2.0in,keepaspectratio]{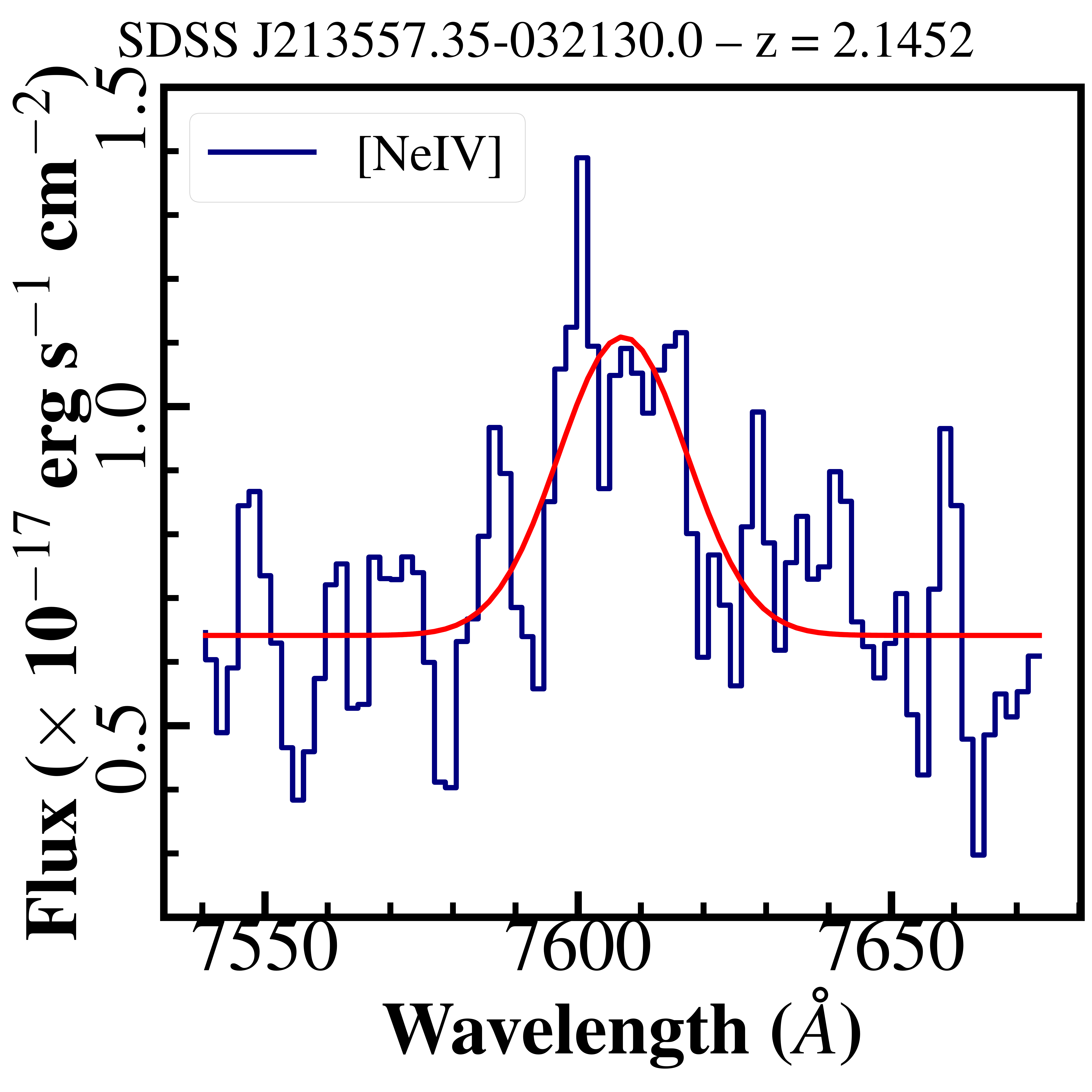}}
	\subfloat[]{
		\includegraphics[width=\columnwidth,height=2.0in,keepaspectratio]{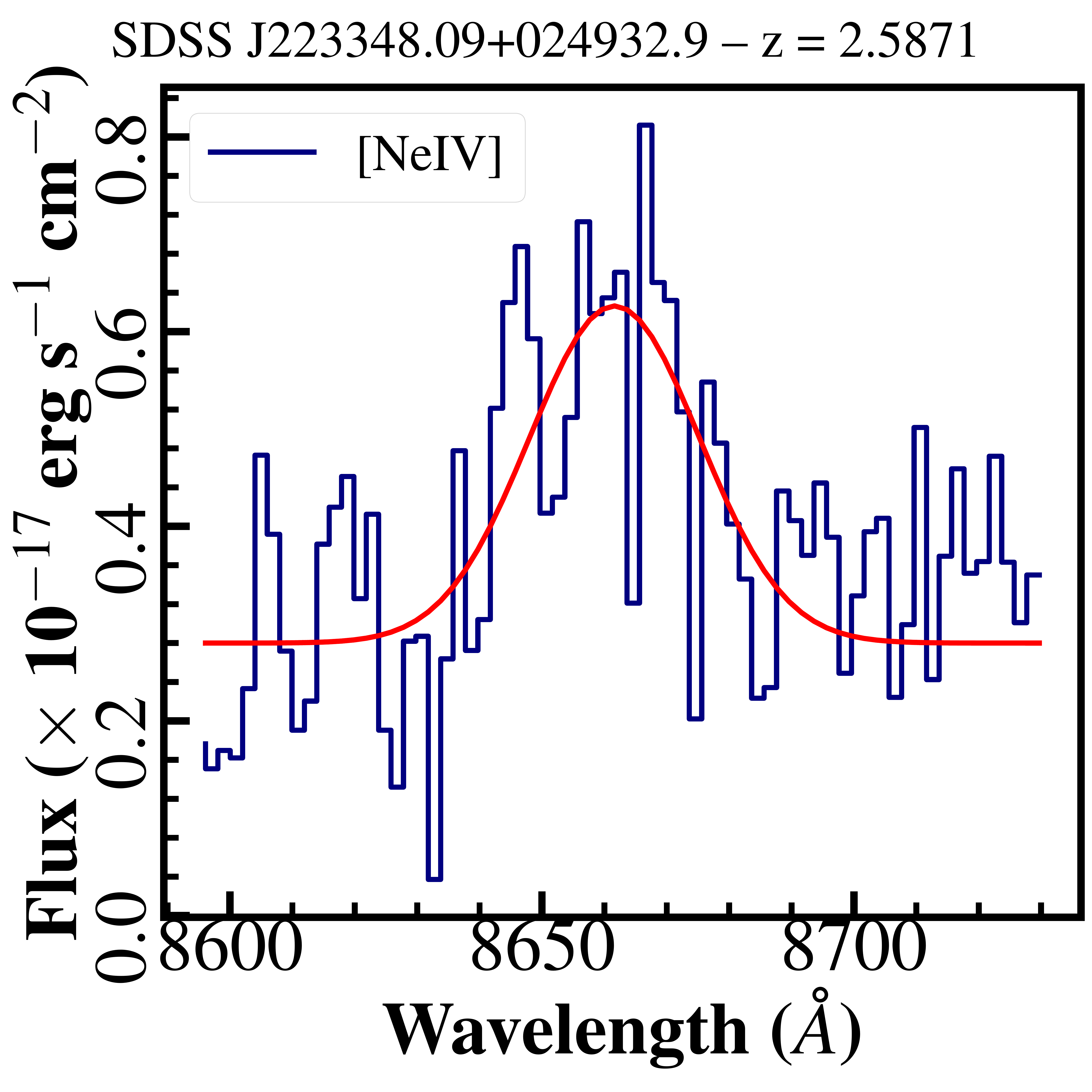}}

	\caption{(continued).}
	\label{neiv2422_02}
\end{figure*}

\begin{figure*}

		\includegraphics[width=\columnwidth,height=1.60in,keepaspectratio]{Fig/legend.pdf}
		\includegraphics[width=\columnwidth,height=1.5in,keepaspectratio]{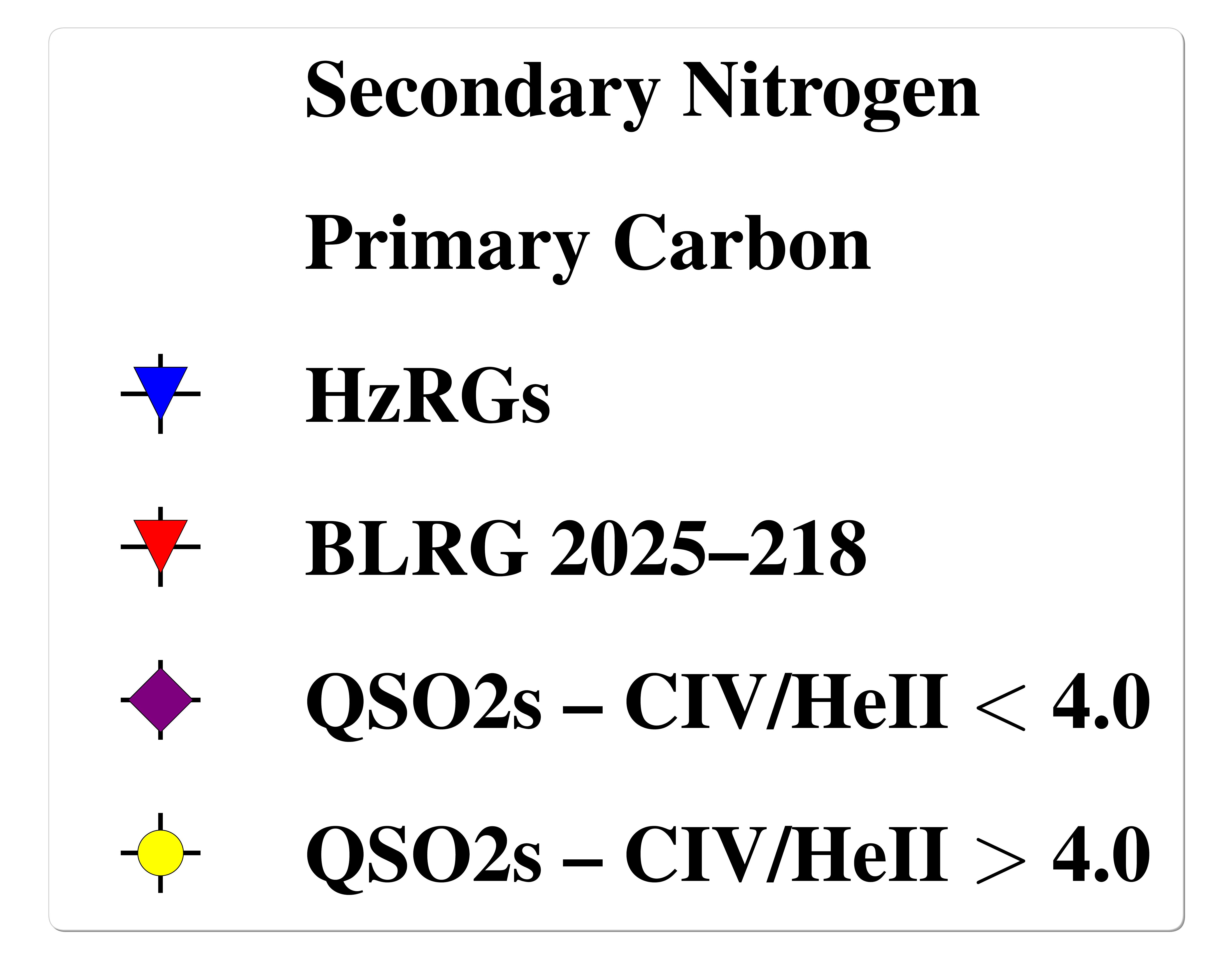}
	\quad
		\includegraphics[width=7.0in,height=7.0in]{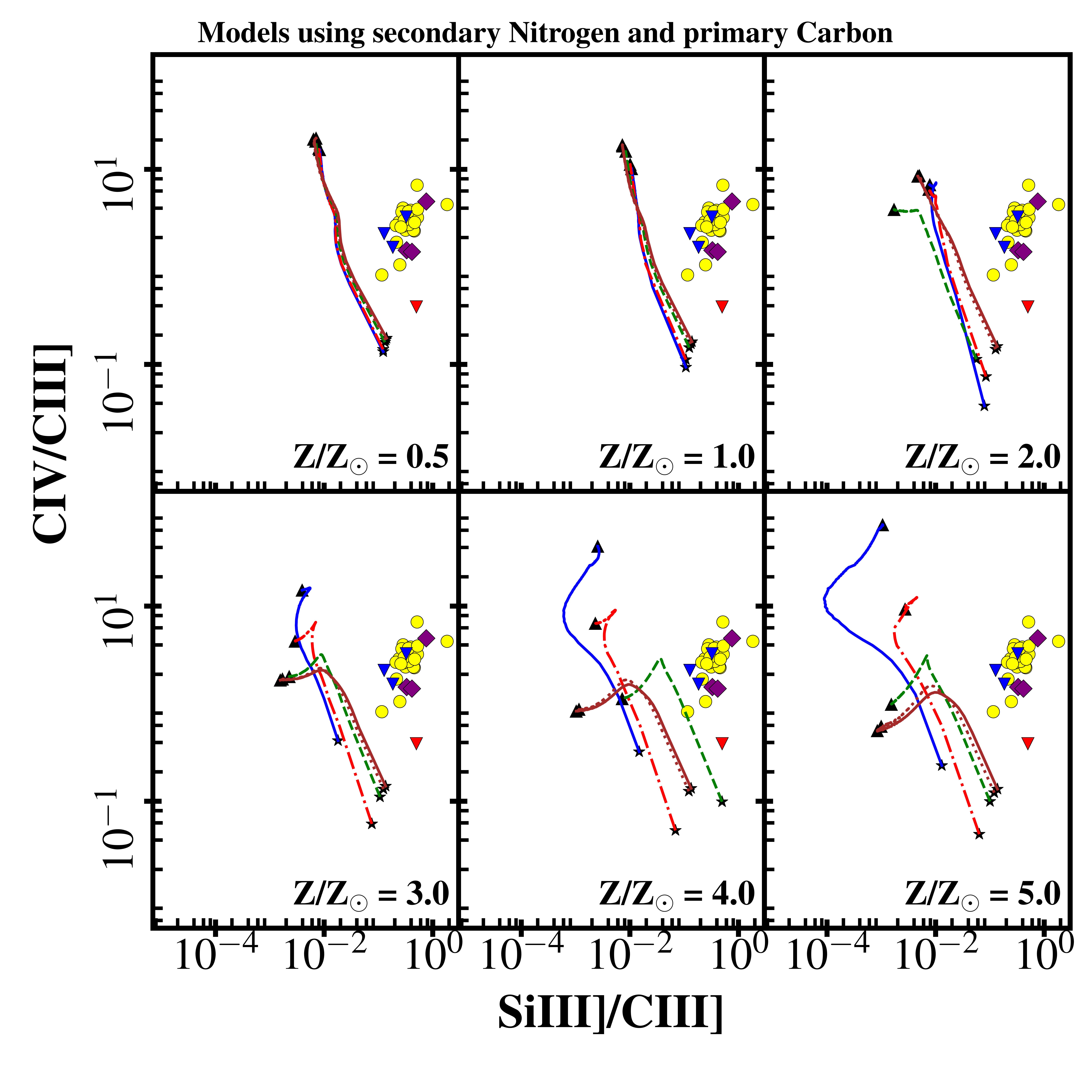}

	\caption{Comparison of the observed emission line ratios from the SDSS type II quasars divided in objects with \ion{C}{IV}/\ion{He}{II} $<$ 4 and \ion{C}{IV}/\ion{He}{II} $>$ 4 (purple filled diamond and yellow filled circles, respectively), Keck II HzRGs from \citet{Ve2001} (blue filled triangles) and BLRG 2025--218 (red filled triangle) from \citet{Hu6} with photoionization models using ionizing continuum power law index $\alpha$ = -1.0. Each diagram presents a different gas metallicity, i.e., Z/Z$_{\odot}$ = 0.5, 1.0, 2.0, 3.0, 4.0, 5.0. Curves with different colors represent the hydrogen gas density (n$_{H}$). At the end of each sequence, a solid black triangle corresponds to the lowest ionization parameter (U = 2$\times$10$^{-3}$) while the solid black star corresponds to the maximum value of the ionization parameter (U = 1.0).}
	\label{SiIII}
\end{figure*}

\begin{figure*}

		\includegraphics[width=\columnwidth,height=1.60in,keepaspectratio]{Fig/legend.pdf}
		\includegraphics[width=\columnwidth,height=1.5in,keepaspectratio]{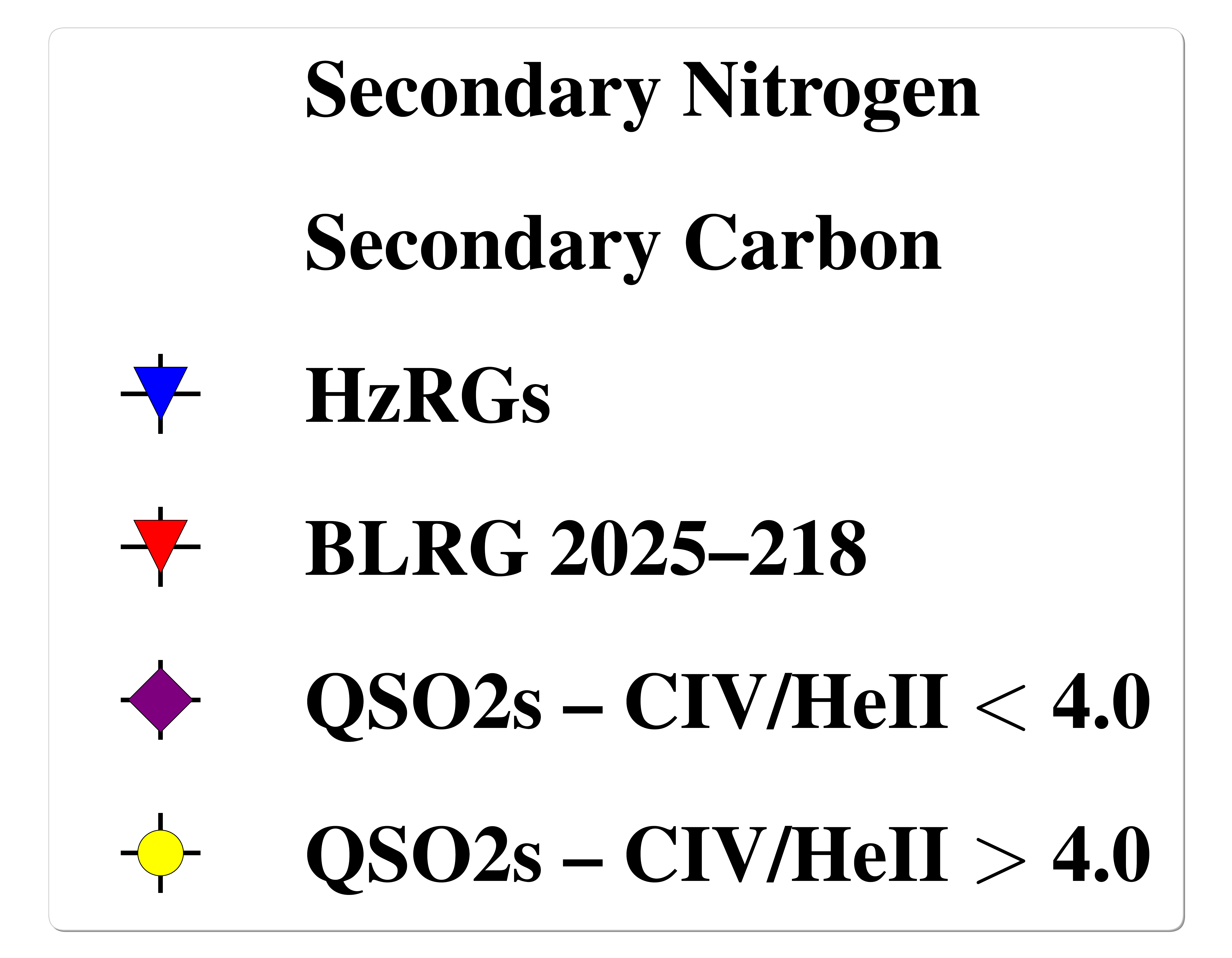}
	\quad
		\includegraphics[width=7.0in,height=7.0in]{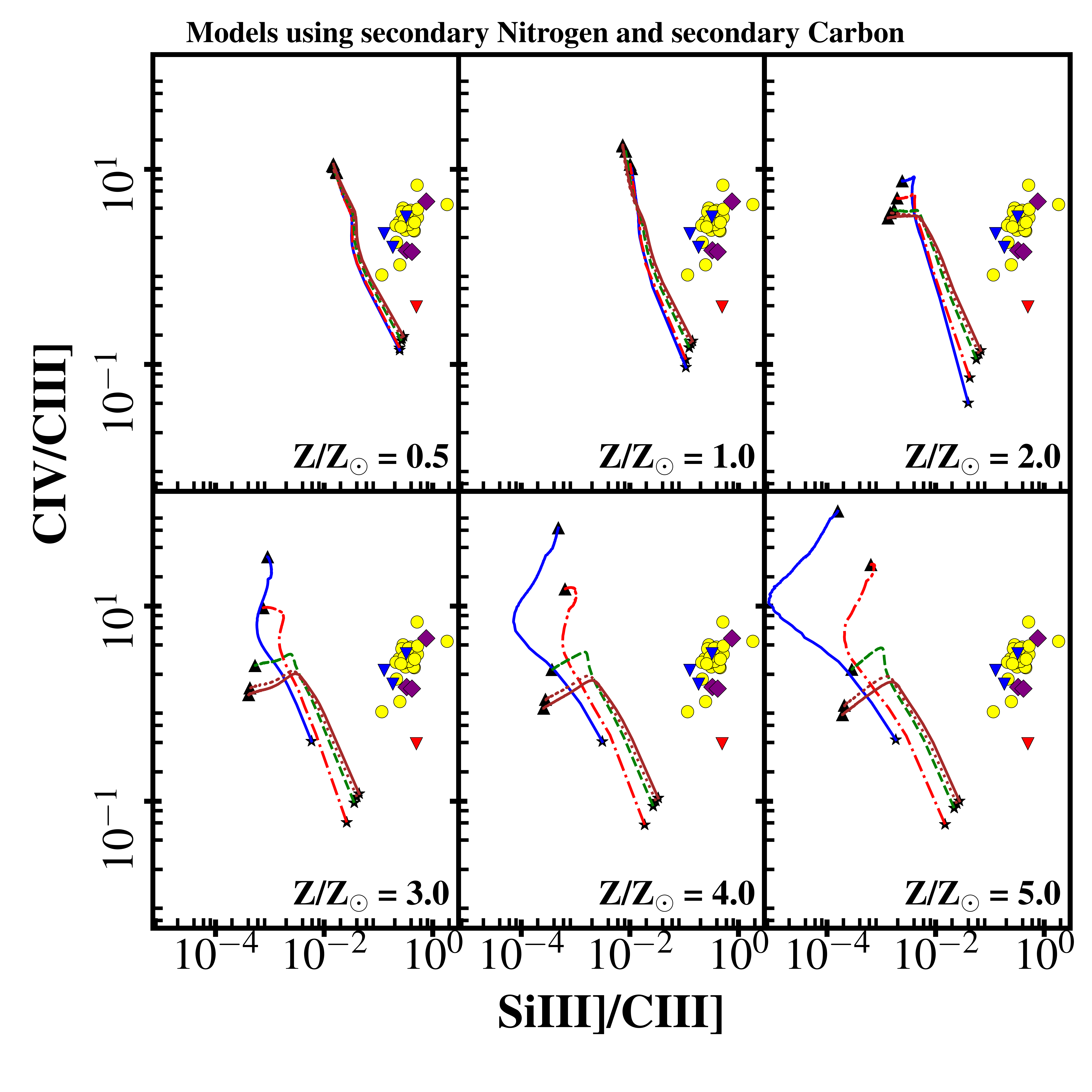}

	\caption{Comparison of the observed emission line ratios from the SDSS type II quasars divided in objects with \ion{C}{IV}/\ion{He}{II} $<$ 4 and \ion{C}{IV}/\ion{He}{II} $>$ 4 (purple filled diamond and yellow filled circles, respectively), Keck II HzRGs from \citet{Ve2001} (blue filled triangles) and BLRG 2025--218 (red filled triangle) from \citet{Hu6} with photoionization models using ionizing continuum power law index $\alpha$ = -1.0. Carbon is a secondary element in which its abundance is proportional to the square of the metallicity. Each diagram presents a different gas metallicity, i.e., Z/Z$_{\odot}$ = 0.5, 1.0, 2.0, 3.0, 4.0, 5.0. Curves with different colors represent the hydrogen gas density (n$_{H}$). At the end of each sequence, a solid black triangle corresponds to the lowest ionization parameter (U = 2$\times$10$^{-3}$) while the solid black star corresponds to the maximum value of the ionization parameter (U = 1.0).}
	\label{SiIIIs}
\end{figure*}

\begin{figure*}

		\includegraphics[width=\columnwidth,height=1.5in,keepaspectratio]{Fig/legend.pdf}
		\includegraphics[width=\columnwidth,height=1.5in,keepaspectratio]{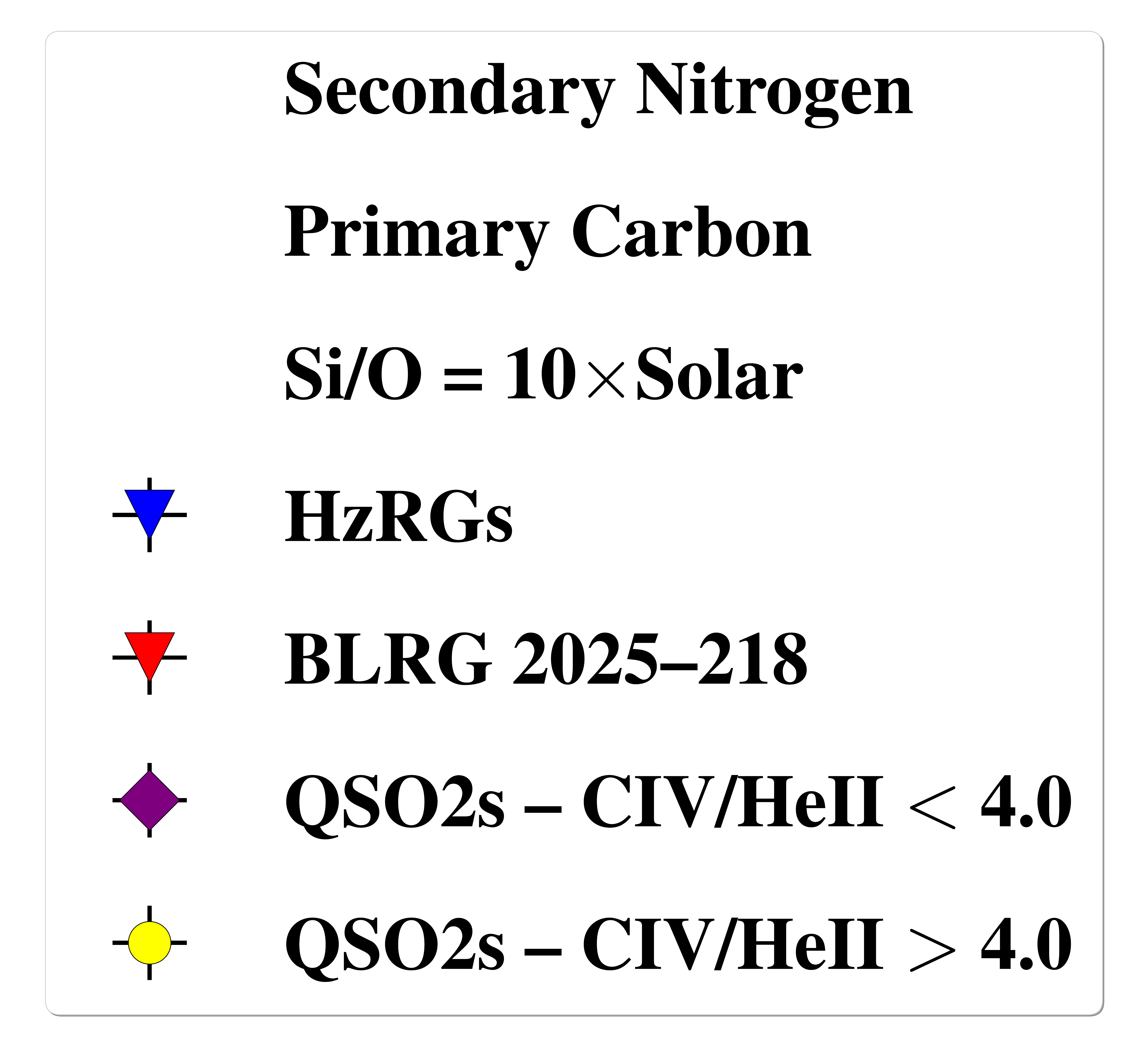}
	\quad
		\includegraphics[width=7.0in,height=7.0in]{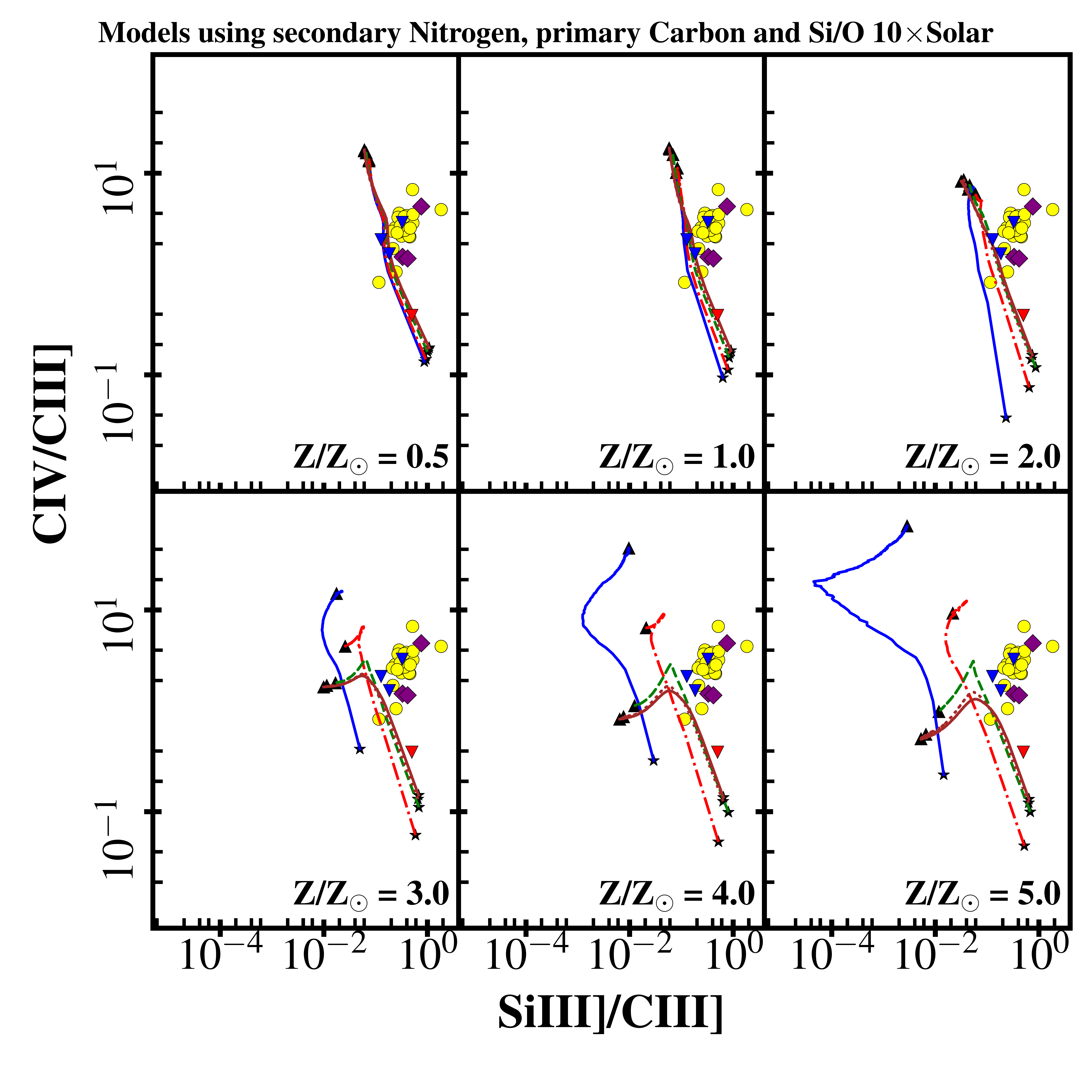}
	\caption{Comparison of the observed emission line ratios from the SDSS type II quasars divided in objects with \ion{C}{IV}/\ion{He}{II} $<$ 4 and \ion{C}{IV}/\ion{He}{II} $>$ 4 (purple filled diamond and yellow filled circles, respectively), Keck II HzRGs from \citet{Ve2001} (blue filled triangles) and BLRG 2025--218 (red filled triangle) from \citet{Hu6} with photoionization models using ionizing continuum power law index $\alpha$ = -1.0. Each diagram presents a different gas metallicity, i.e., Z/Z$_{\odot}$ = 0.5, 1.0, 2.0, 3.0, 4.0, 5.0. In addition, the abundance of Si was increased by a factor of 10, resulting in a Si/O abundance ratio ten times its Solar value. Curves with different colors represent the hydrogen gas density (n$_{H}$). At the end of each sequence, a solid black triangle corresponds to the lowest ionization parameter (U = 2$\times$10$^{-3}$) while the solid black star corresponds to the maximum value of the ionization parameter (U = 1.0).}
	\label{SiIIIx10}
\end{figure*}

\begin{figure*}
		\includegraphics[width=\columnwidth,height=1.5in,keepaspectratio]{Fig/legend.pdf}
		\includegraphics[width=\columnwidth,height=1.5in,keepaspectratio]{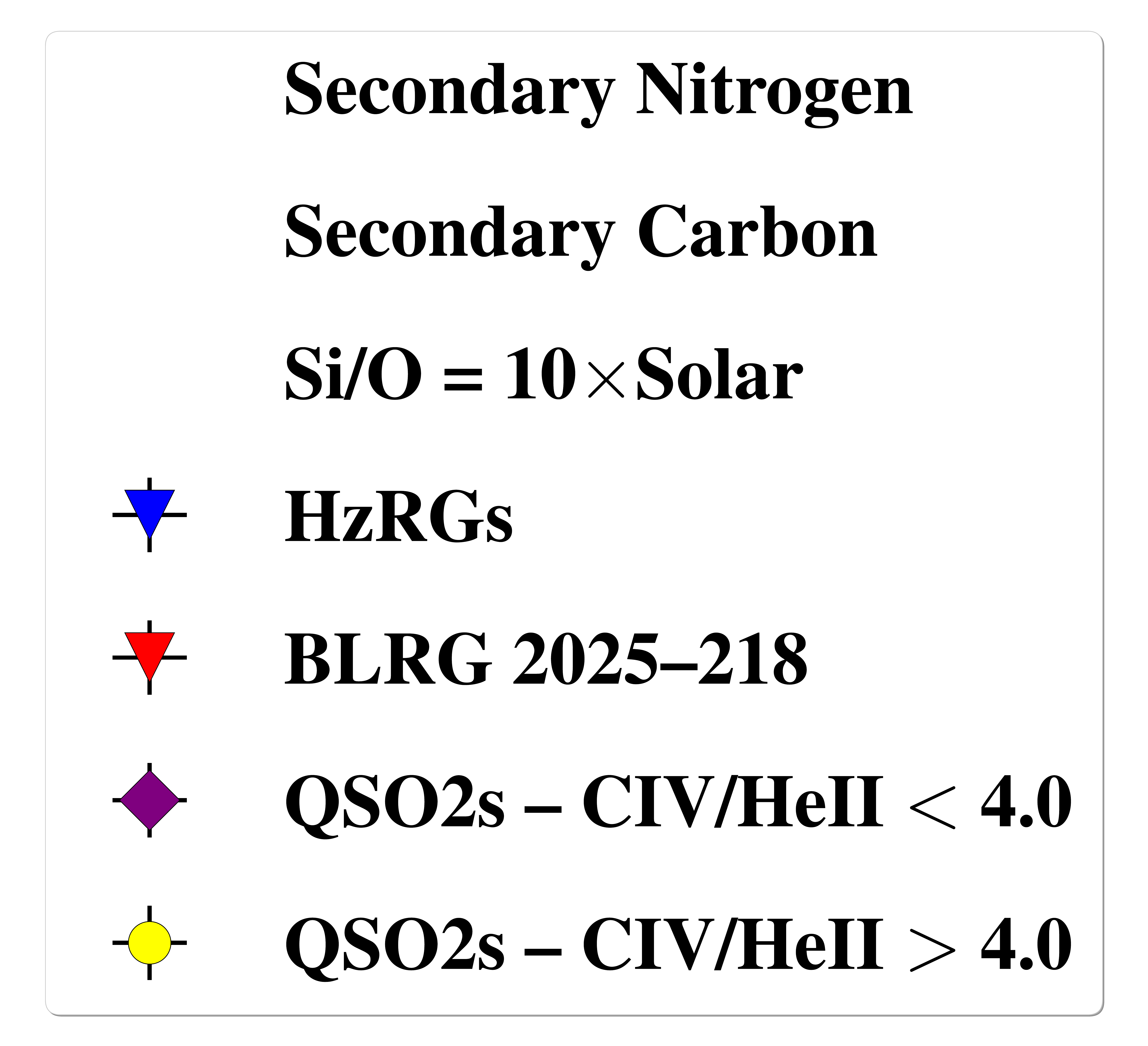}
	\quad
		\includegraphics[width=7.0in,height=7.0in]{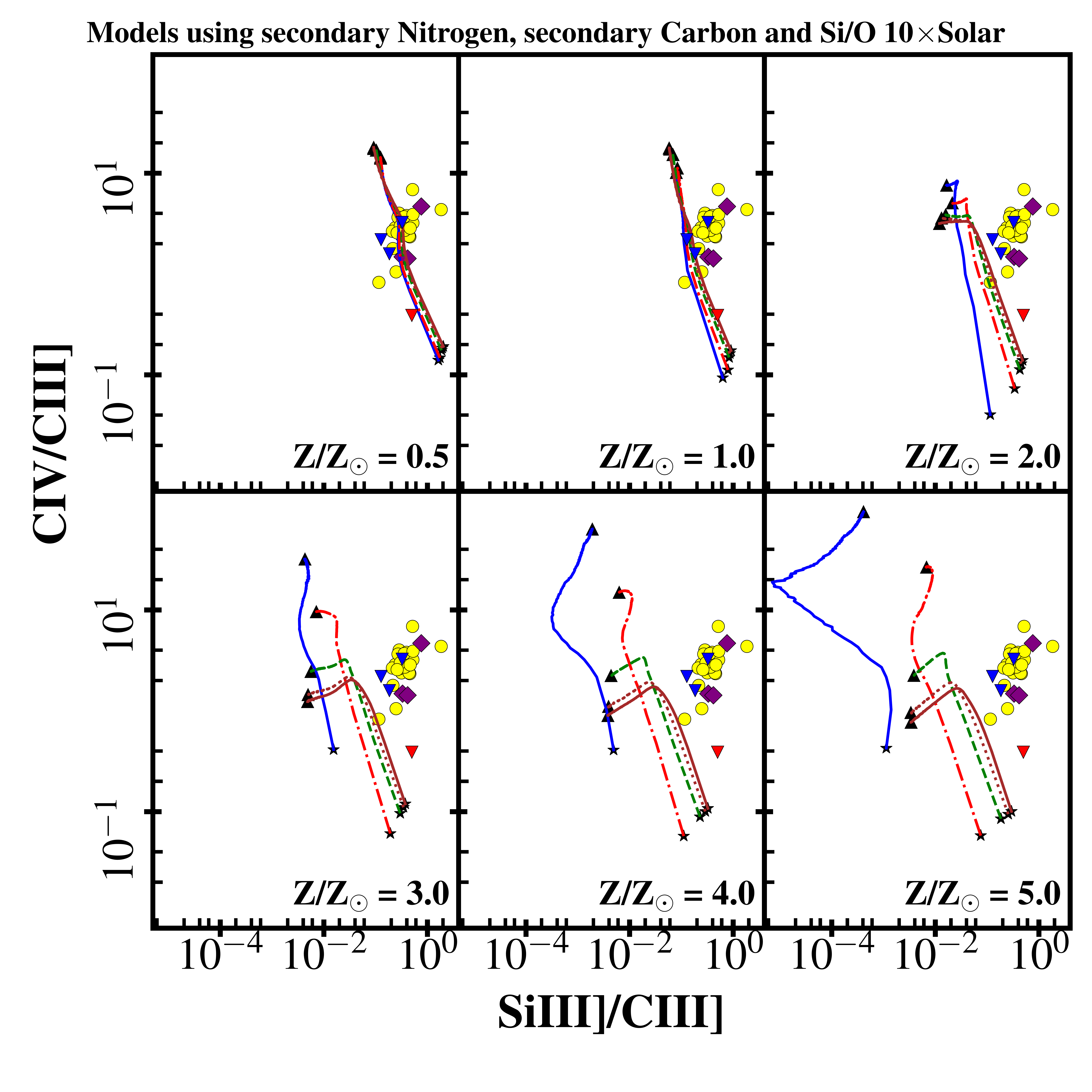}

	\caption{Comparison of the observed emission line ratios from the SDSS type II quasars divided in objects with \ion{C}{IV}/\ion{He}{II} $<$ 4 and \ion{C}{IV}/\ion{He}{II} $>$ 4 (purple filled diamond and yellow filled circles, respectively), Keck II HzRGs from \citet{Ve2001} (blue filled triangles) and BLRG 2025--218 (red filled triangle) from \citet{Hu6} with photoionization models using ionizing continuum power law index $\alpha$ = -1.0. Each diagram presents a different gas metallicity, i.e., Z/Z$_{\odot}$ = 0.5, 1.0, 2.0, 3.0, 4.0, 5.0. Carbon is a secondary element in which its abundance is proportional to the square of the metallicity and at each of the gas metallicities shown in this Figure, the abundance of Si was increased by a factor of 10, resulting in a Si/O abundance ratio ten times its Solar value. Curves with different colors represent the hydrogen gas density (n$_{H}$). At the end of each sequence, a solid black triangle corresponds to the lowest ionization parameter (U = 2$\times$10$^{-3}$) while the solid black star corresponds to the maximum value of the ionization parameter (U = 1.0).}
	\label{SiIIIx10s}
\end{figure*}

\begin{figure*}

		\includegraphics[width=\columnwidth,height=1.60in,keepaspectratio]{Fig/legend.pdf}
		\includegraphics[width=\columnwidth,height=1.5in,keepaspectratio]{Fig/legend_03.pdf}
	\quad
		\includegraphics[width=7.0in,height=7.0in]{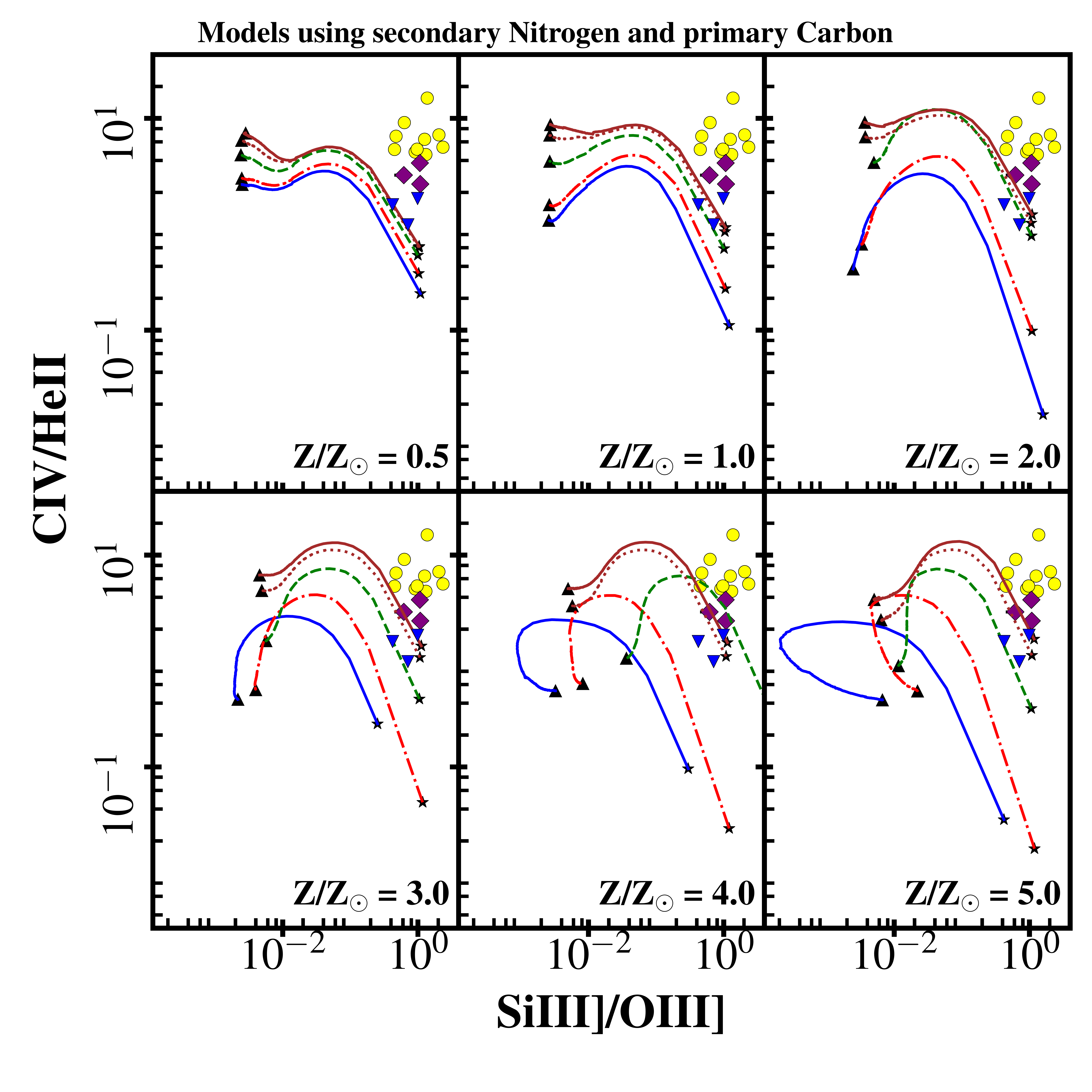}

	\caption{Comparison of the observed emission line ratios from the SDSS type II quasars divided in objects with \ion{C}{IV}/\ion{He}{II} $<$ 4 and \ion{C}{IV}/\ion{He}{II} $>$ 4 (purple filled diamond and yellow filled circles, respectively), and Keck II HzRGs from \citet{Ve2001} (blue filled triangles) with photoionization models using ionizing continuum power law index $\alpha$ = -1.0. Each diagram presents a different gas metallicity, i.e., Z/Z$_{\odot}$ = 0.5, 1.0, 2.0, 3.0, 4.0, 5.0. Curves with different colors represent the hydrogen gas density (n$_{H}$). At the end of each sequence, a solid black triangle corresponds to the lowest ionization parameter (U = 2$\times$10$^{-3}$) while the solid black star corresponds to the maximum value of the ionization parameter (U = 1.0).}
	\label{OIII}
\end{figure*}

\begin{figure*}

		\includegraphics[width=\columnwidth,height=1.50in,keepaspectratio]{Fig/legend.pdf}
		\includegraphics[width=\columnwidth,height=1.5in,keepaspectratio]{Fig/legend_04.pdf}
	\quad
		\includegraphics[width=7.0in,height=7.0in]{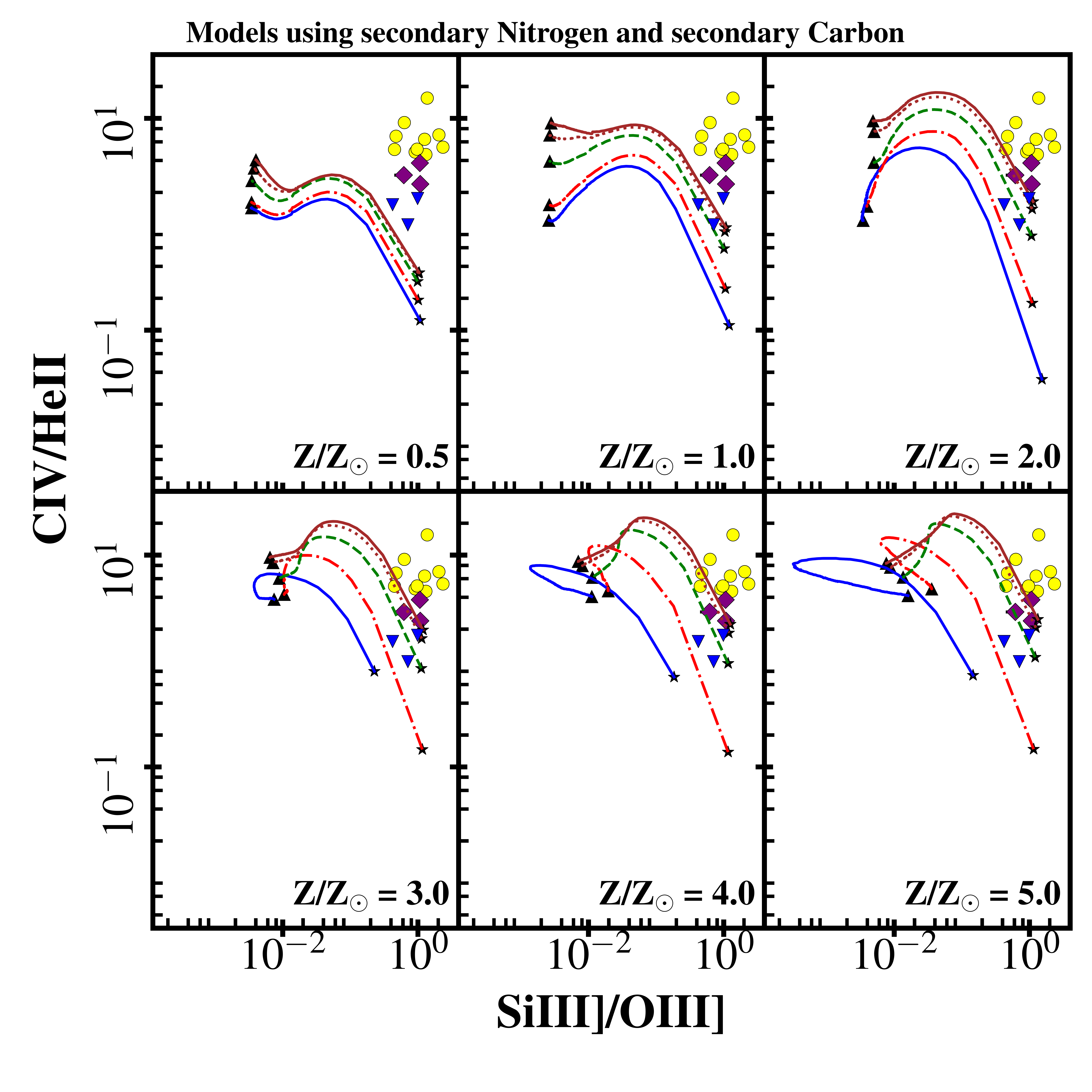}

	\caption{Comparison of the observed emission line ratios from the SDSS type II quasars divided in objects with \ion{C}{IV}/\ion{He}{II} $<$ 4 and \ion{C}{IV}/\ion{He}{II} $>$ 4 (purple filled diamond and yellow filled circles, respectively), and Keck II HzRGs from \citet{Ve2001} (blue filled triangles) with photoionization models using ionizing continuum power law index $\alpha$ = -1.0. Carbon is a secondary element in which its abundance is proportional to the square of the metallicity. Each diagram presents a different gas metallicity, i.e., Z/Z$_{\odot}$ = 0.5, 1.0, 2.0, 3.0, 4.0, 5.0. Curves with different colors represent the hydrogen gas density (n$_{H}$). At the end of each sequence, a solid black triangle corresponds to the lowest ionization parameter (U = 2$\times$10$^{-3}$) while the solid black star corresponds to the maximum value of the ionization parameter (U = 1.0).}
	\label{OIIIs}
\end{figure*}

\begin{figure*}

		\includegraphics[width=\columnwidth,height=1.60in,keepaspectratio]{Fig/legend.pdf}
		\includegraphics[width=\columnwidth,height=1.6in,keepaspectratio]{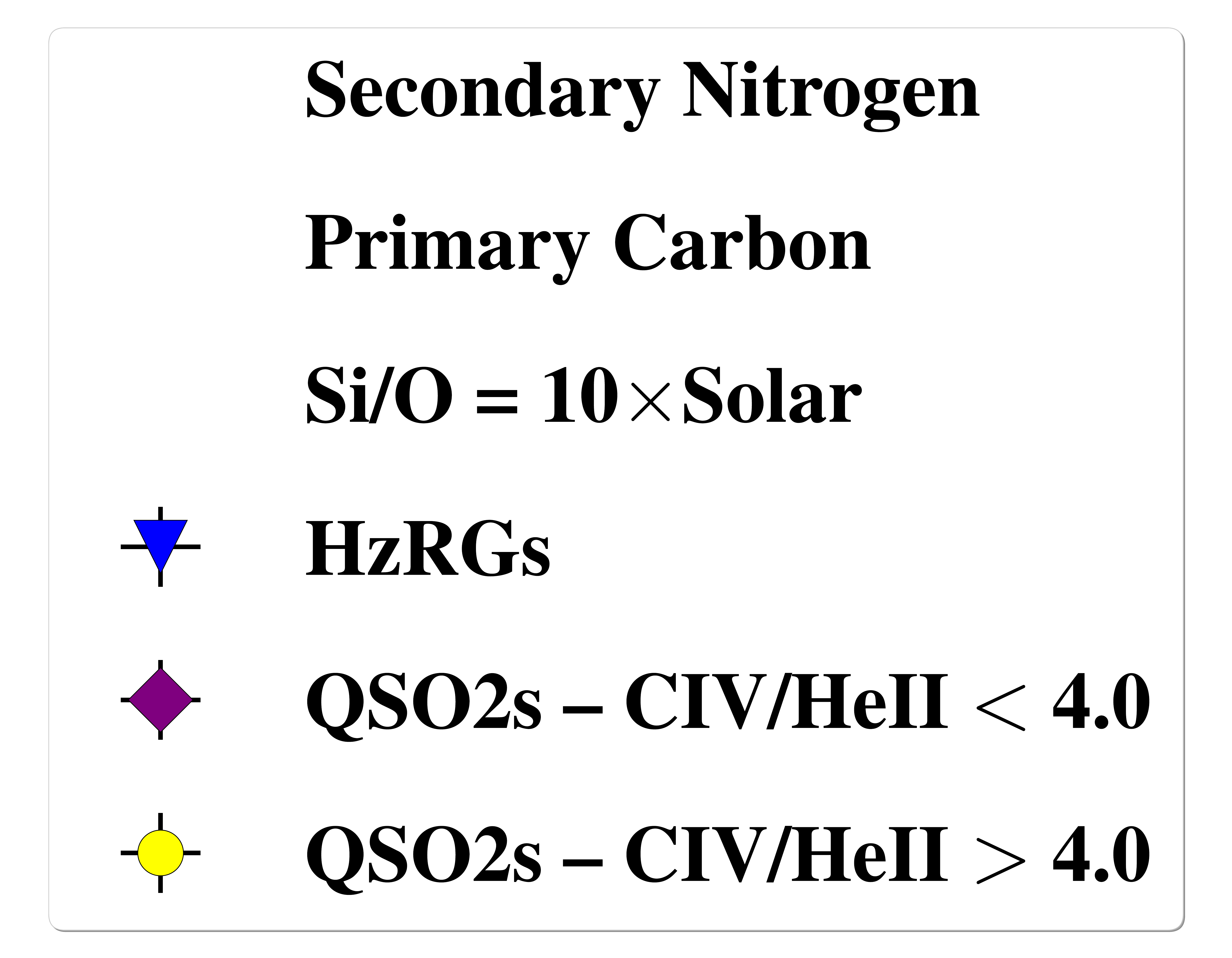}
	\quad
		\includegraphics[width=7.0in,height=7.0in]{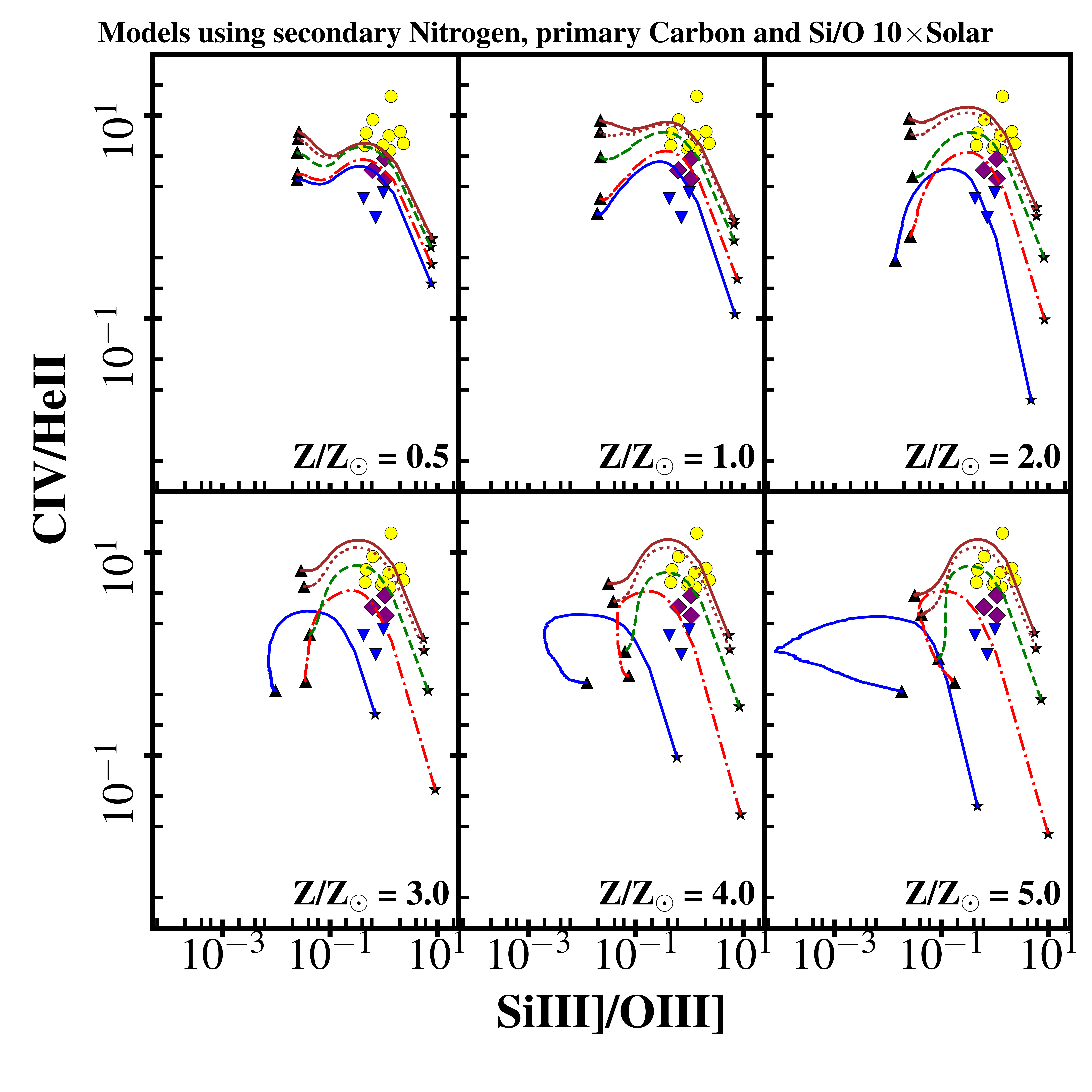}

	\caption{Comparison of the observed emission line ratios from the SDSS type II quasars divided in objects with \ion{C}{IV}/\ion{He}{II} $<$ 4 and \ion{C}{IV}/\ion{He}{II} $>$ 4 (purple filled diamond and yellow filled circles, respectively), and Keck II HzRGs from \citet{Ve2001} (blue filled triangles) with photoionization models using ionizing continuum power law index $\alpha$ = -1.0. Each diagram presents a different gas metallicity, i.e., Z/Z$_{\odot}$ = 0.5, 1.0, 2.0, 3.0, 4.0, 5.0. In addition, at each of the gas metallicities shown in this Figure, the abundance of Si was increased by a factor of 10, resulting in a Si/O abundance ratio ten times its Solar value. Curves with different colors represent the hydrogen gas density (n$_{H}$). At the end of each sequence, a solid black triangle corresponds to the lowest ionization parameter (U = 2$\times$10$^{-3}$) while the solid black star corresponds to the maximum value of the ionization parameter (U = 1.0).}
	\label{OIIIx10}
\end{figure*}

\begin{figure*}
		\includegraphics[width=\columnwidth,height=1.5in,keepaspectratio]{Fig/legend.pdf}
		\includegraphics[width=\columnwidth,height=1.5in,keepaspectratio]{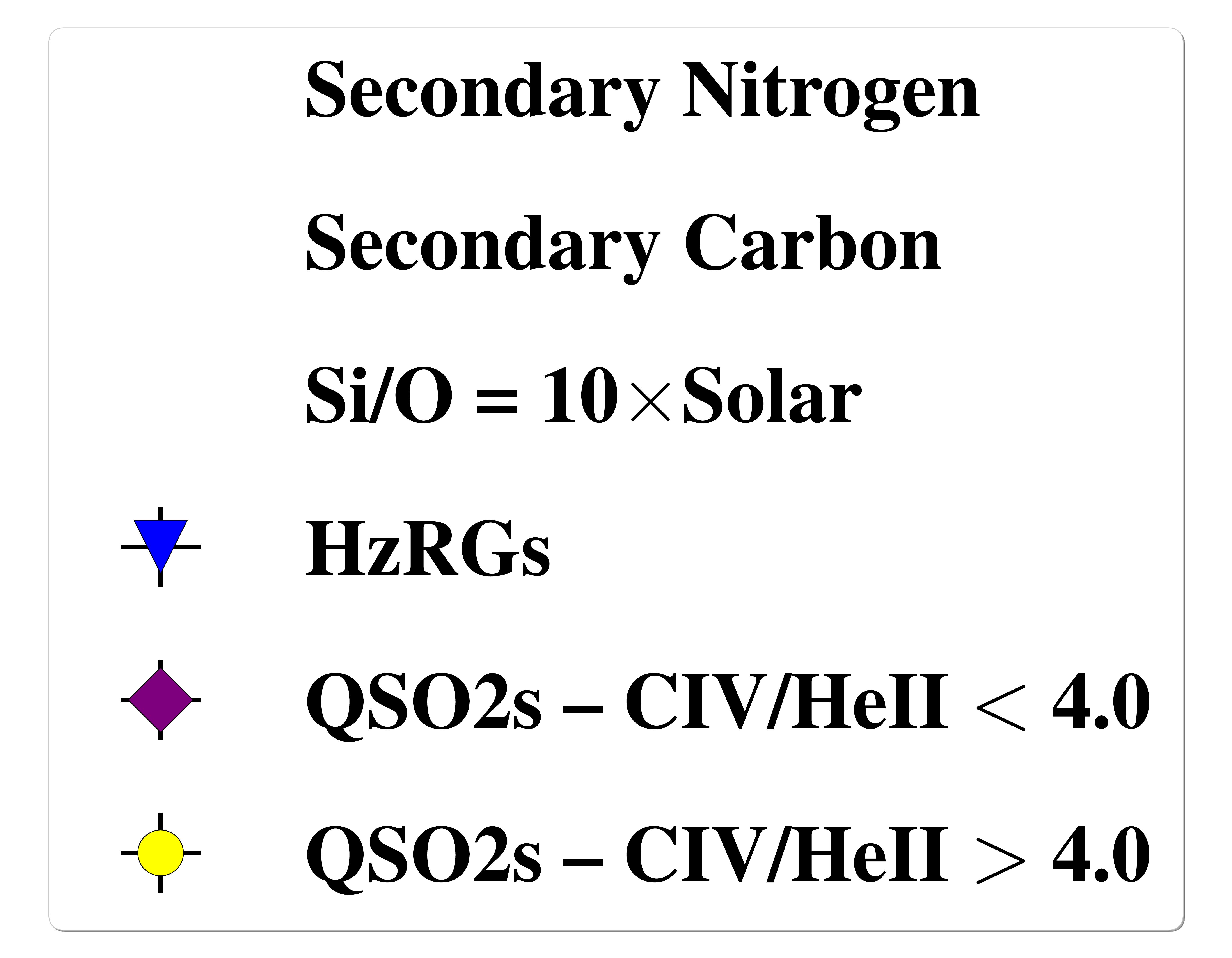}
	\quad
		\includegraphics[width=7.0in,height=7.0in]{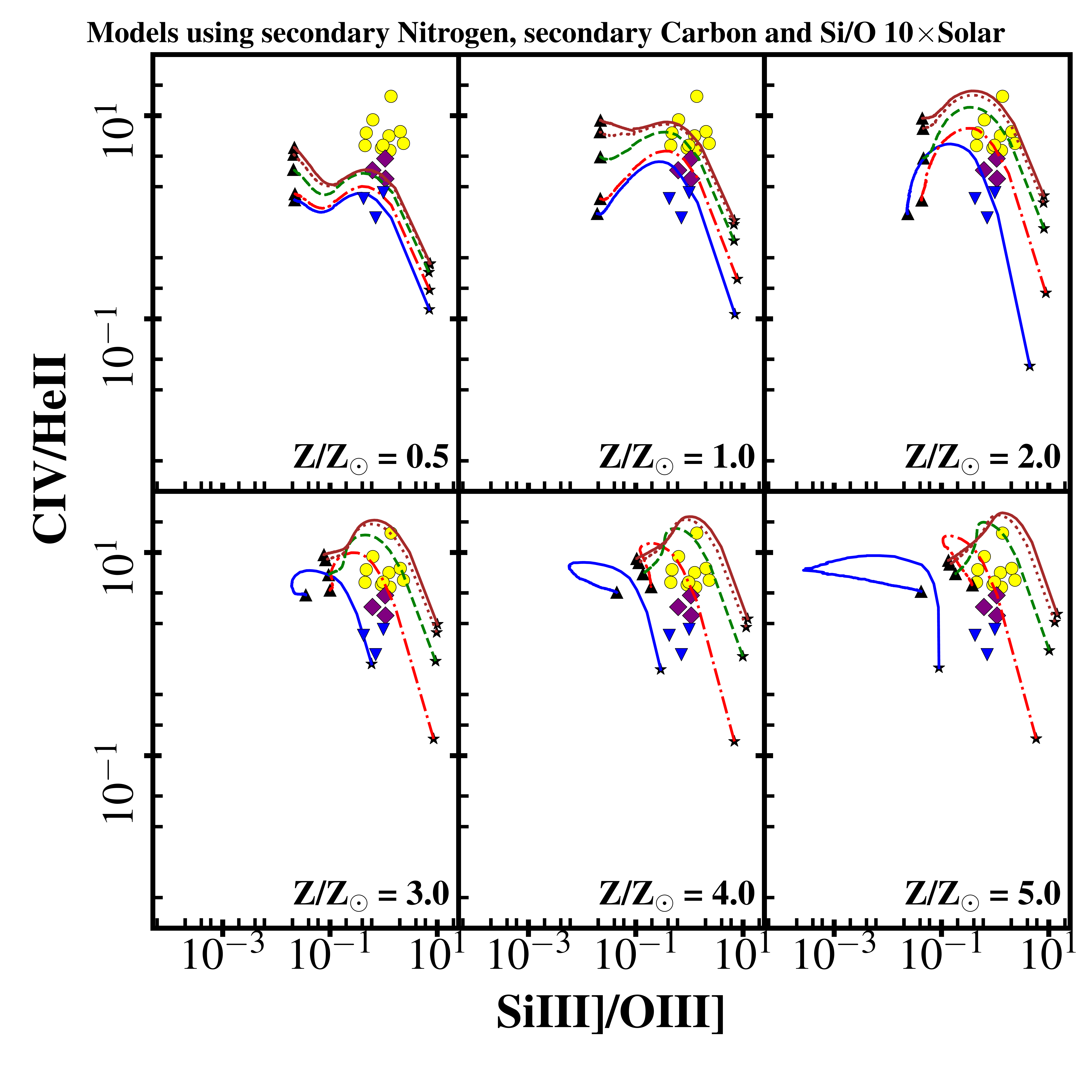}

	\caption{Comparison of the observed emission line ratios from the SDSS type II quasars divided in objects with \ion{C}{IV}/\ion{He}{II} $<$ 4 and \ion{C}{IV}/\ion{He}{II} $>$ 4 (purple filled diamond and yellow filled circles, respectively), and Keck II HzRGs from \citet{Ve2001} (blue filled triangles) with photoionization models using ionizing continuum power law index $\alpha$ = -1.0. Each diagram presents a different gas metallicity, i.e., Z/Z$_{\odot}$ = 0.5, 1.0, 2.0, 3.0, 4.0, 5.0. Carbon is a secondary element in which its abundance is proportional to the square of the metallicity and at each of the gas metallicities shown in this Figure, the abundance of Si was increased by a factor of 10, resulting in a Si/O abundance ratio ten times its Solar value. Curves with different colors represent the hydrogen gas density (n$_{H}$). At the end of each sequence, a solid black triangle corresponds to the lowest ionization parameter (U = 2$\times$10$^{-3}$) while the solid black star corresponds to the maximum value of the ionization parameter (U = 1.0).}
	\label{OIIIx10s}
\end{figure*}

\begin{figure*}

		\includegraphics[width=\columnwidth,height=1.50in,keepaspectratio]{Fig/legend.pdf}
		\includegraphics[width=\columnwidth,height=1.50in,keepaspectratio]{Fig/legend_03.pdf}
	\quad
		\includegraphics[width=7.0in,height=7.0in]{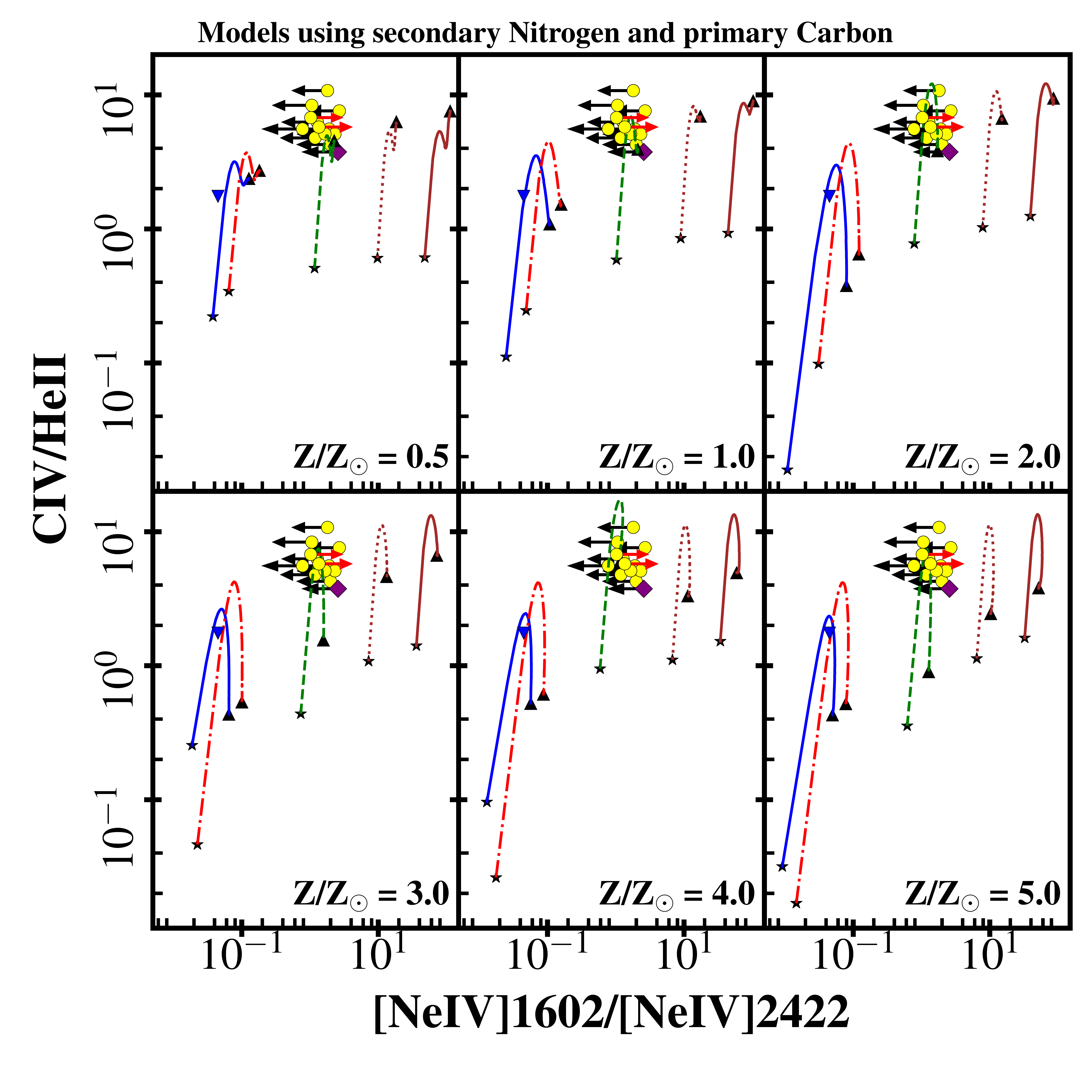}

	\caption{Comparison of the observed emission line ratios from the SDSS type II quasars divided in objects with \ion{C}{IV}/\ion{He}{II} $<$ 4 and \ion{C}{IV}/\ion{He}{II} $>$ 4 (purple filled diamond and yellow filled circles, respectively), and Keck II HzRGs from \citet{Ve2001} (blue filled triangles) with photoionization models using ionizing continuum power law index $\alpha$ = -1.0. Each diagram presents a different gas metallicity, i.e., Z/Z$_{\odot}$ = 0.5, 1.0, 2.0, 3.0, 4.0, 5.0. The black arrows represent the estimated 3$\sigma$ upper limit for the emission line \ion{[Ne}{IV]} $\lambda$ 1602 while the red arrows represent the estimated 3$\sigma$ upper limit for the emission line \ion{[Ne}{IV]} $\lambda$ 2422. Curves with different colors represent the hydrogen gas density (n$_{H}$). At the end of each sequence, a solid black triangle corresponds to the lowest ionization parameter (U = 2$\times$10$^{-3}$) while the solid black star corresponds to the maximum value of the ionization parameter (U = 1.0).}
	\label{neiv01}
\end{figure*}

\begin{figure*}
		\includegraphics[width=\columnwidth,height=1.50in,keepaspectratio]{Fig/legend.pdf}
		\includegraphics[width=\columnwidth,height=1.5in,keepaspectratio]{Fig/legend_04.pdf}
	\quad
		\includegraphics[width=7.0in,height=7.0in]{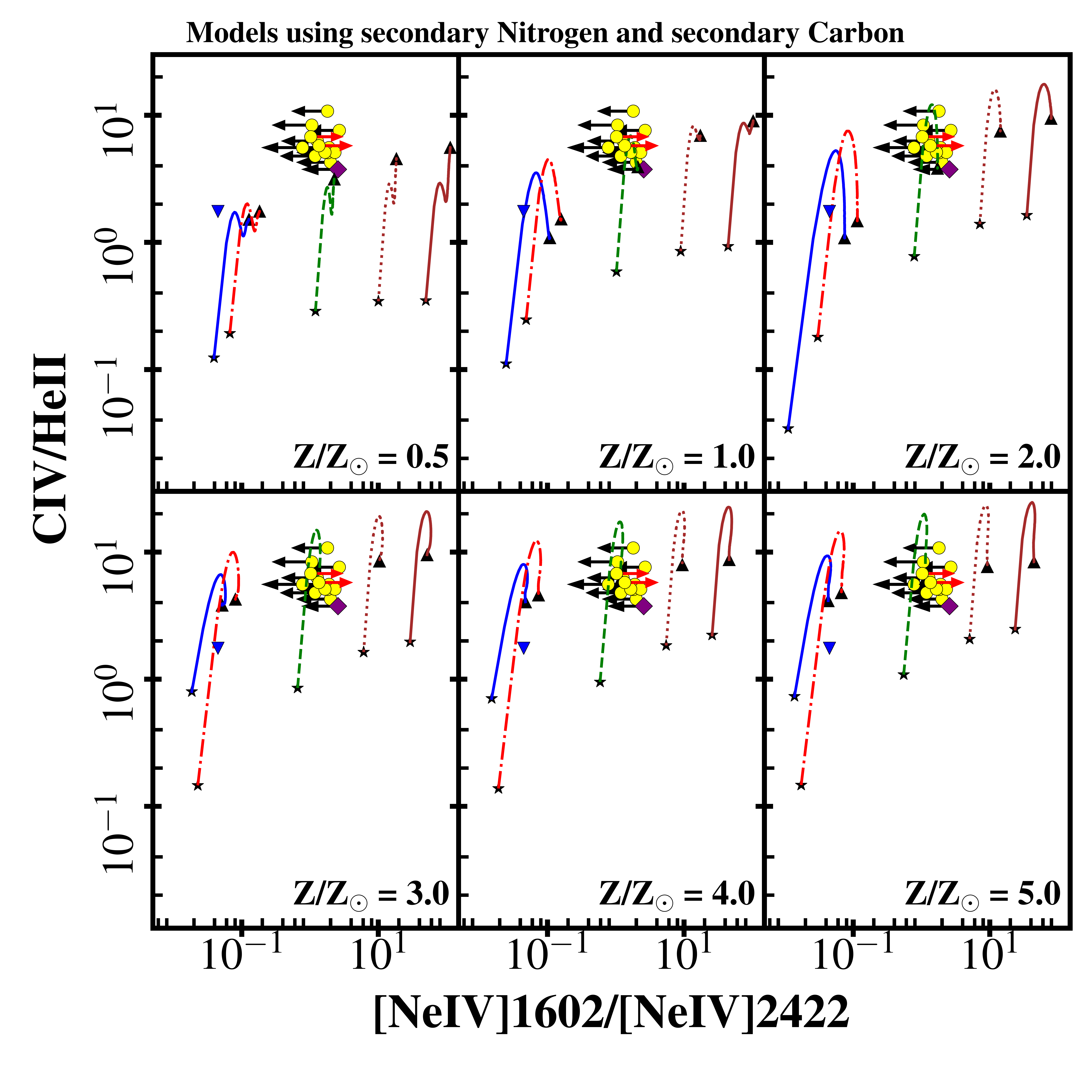}

	\caption{Comparison of the observed emission line ratios from the SDSS type II quasars divided in objects with \ion{C}{IV}/\ion{He}{II} $<$ 4 and \ion{C}{IV}/\ion{He}{II} $>$ 4 (purple filled diamond and yellow filled circles, respectively), and Keck II HzRGs from \citet{Ve2001} (blue filled triangles) with photoionization models using ionizing continuum power law index $\alpha$ = -1.0. Carbon is a secondary element in which its abundance is proportional to the square of the metallicity. Each diagram presents a different gas metallicity, i.e., Z/Z$_{\odot}$ = 0.5, 1.0, 2.0, 3.0, 4.0, 5.0.The black arrows represent the estimated 3$\sigma$ upper limit for the emission line \ion{[Ne}{IV]} $\lambda$ 1602 while the red arrows represent the estimated 3$\sigma$ upper limit for the emission line \ion{[Ne}{IV]} $\lambda$ 2422. Curves with different colors represent the hydrogen gas density (n$_{H}$). At the end of each sequence, a solid black triangle corresponds to the lowest ionization parameter (U = 2$\times$10$^{-3}$) while the solid black star corresponds to the maximum value of the ionization parameter (U = 1.0).}
	\label{neiv01s}
\end{figure*}

\begin{figure*}

		\includegraphics[width=\columnwidth,height=1.50in,keepaspectratio]{Fig/legend.pdf}
		\includegraphics[width=\columnwidth,height=1.5in,keepaspectratio]{Fig/legend_03.pdf}
	\quad
		\includegraphics[width=7.0in,height=7.0in]{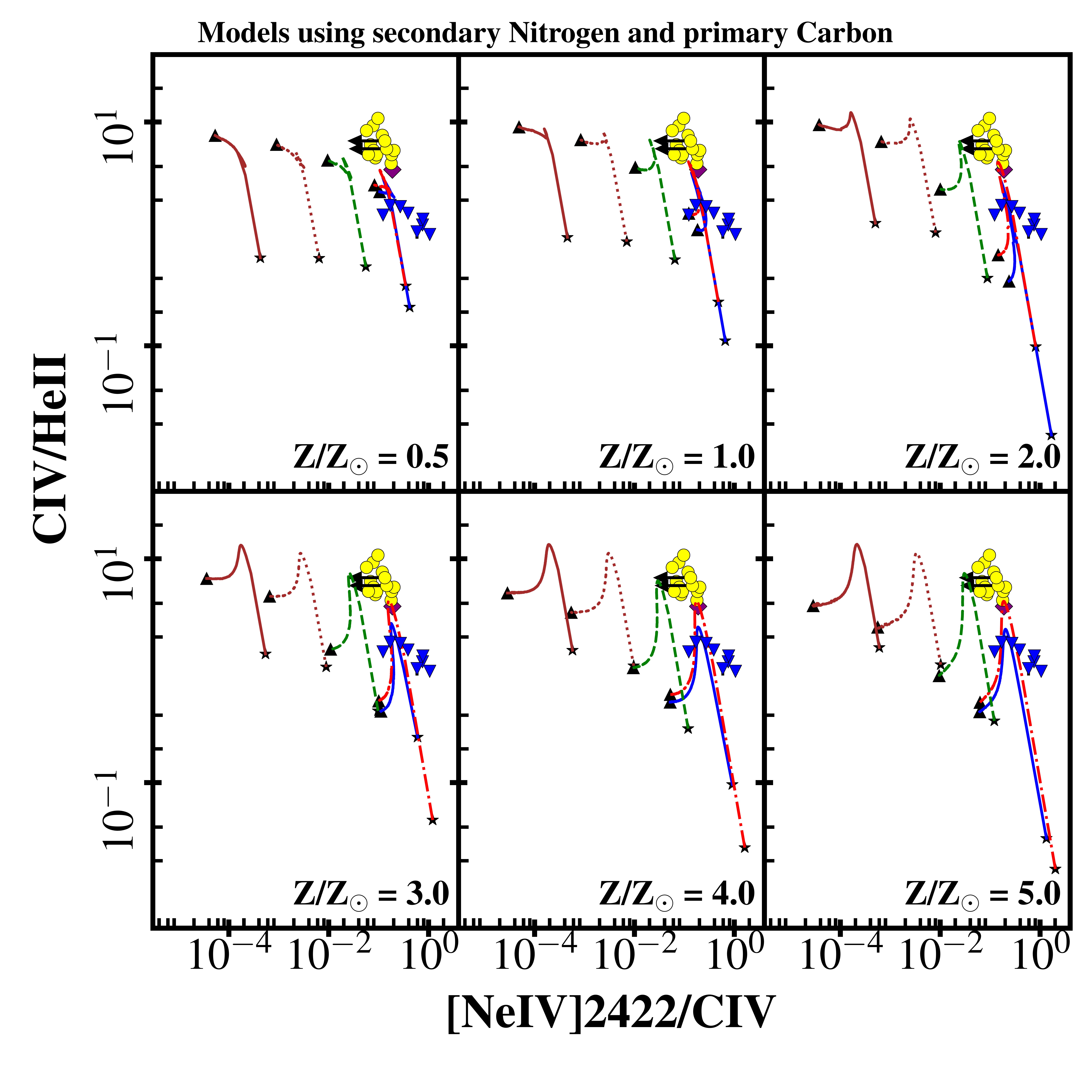}

	\caption{Comparison of the observed emission line ratios from the SDSS type II quasars divided in objects with \ion{C}{IV}/\ion{He}{II} $<$ 4 and \ion{C}{IV}/\ion{He}{II} $>$ 4 (purple filled diamond and yellow filled circles, respectively), and Keck II HzRGs from \citet{Ve2001} (blue filled triangles) with photoionization models using ionizing continuum power law index $\alpha$ = -1.0. Each diagram presents a different gas metallicity, i.e., Z/Z$_{\odot}$ = 0.5, 1.0, 2.0, 3.0, 4.0, 5.0. Curves with different colors represent the hydrogen gas density (n$_{H}$). At the end of each sequence, a solid black triangle corresponds to the lowest ionization parameter (U = 2$\times$10$^{-3}$) while the solid black star corresponds to the maximum value of the ionization parameter (U = 1.0).}
	\label{neiv02}
\end{figure*}

\begin{figure*}
		\includegraphics[width=\columnwidth,height=1.50in,keepaspectratio]{Fig/legend.pdf}
		\includegraphics[width=\columnwidth,height=1.5in,keepaspectratio]{Fig/legend_04.pdf}
	\quad
		\includegraphics[width=7.0in,height=7.0in]{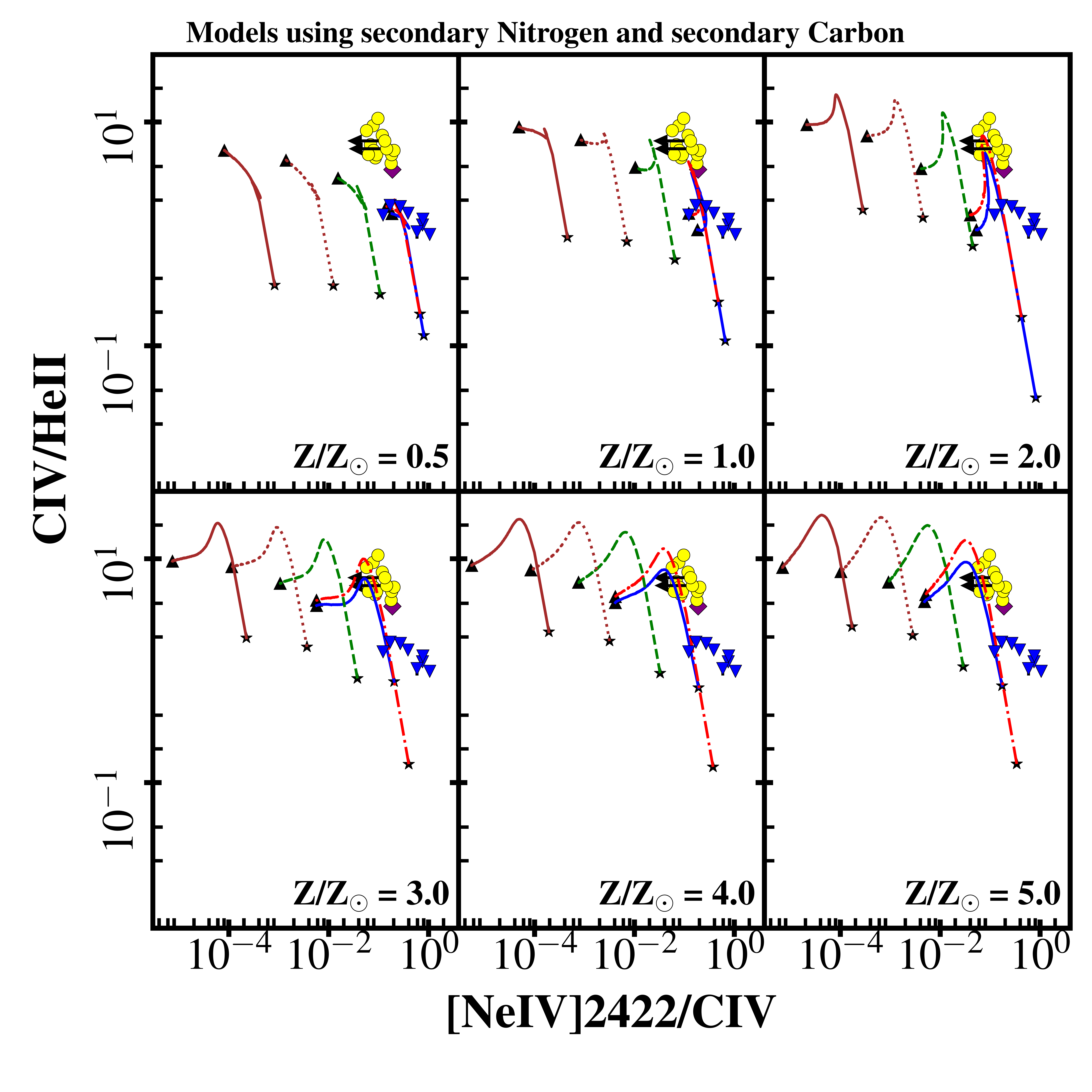}

	\caption{Comparison of the observed emission line ratios from the SDSS type II quasars divided in objects with \ion{C}{IV}/\ion{He}{II} $<$ 4 and \ion{C}{IV}/\ion{He}{II} $>$ 4 (purple filled diamond and yellow filled circles, respectively), and Keck II HzRGs from \citet{Ve2001} (blue filled triangles) with photoionization models using ionizing continuum power law index $\alpha$ = -1.0. Carbon is a secondary element in which its abundance is proportional to the square of the metallicity. Each diagram presents a different gas metallicity, i.e., Z/Z$_{\odot}$ = 0.5, 1.0, 2.0, 3.0, 4.0, 5.0. Curves with different colors represent the hydrogen gas density (n$_{H}$). At the end of each sequence, a solid black triangle corresponds to the lowest ionization parameter (U = 2$\times$10$^{-3}$) while the solid black star corresponds to the maximum value of the ionization parameter (U = 1.0).}
	\label{neiv02s}
\end{figure*}

\begin{figure*}
		\includegraphics[width=\columnwidth,height=1.50in,keepaspectratio]{Fig/legend.pdf}
		\includegraphics[width=\columnwidth,height=1.5in,keepaspectratio]{Fig/legend_03.pdf}
	\quad
		\includegraphics[width=7.0in,height=7.0in]{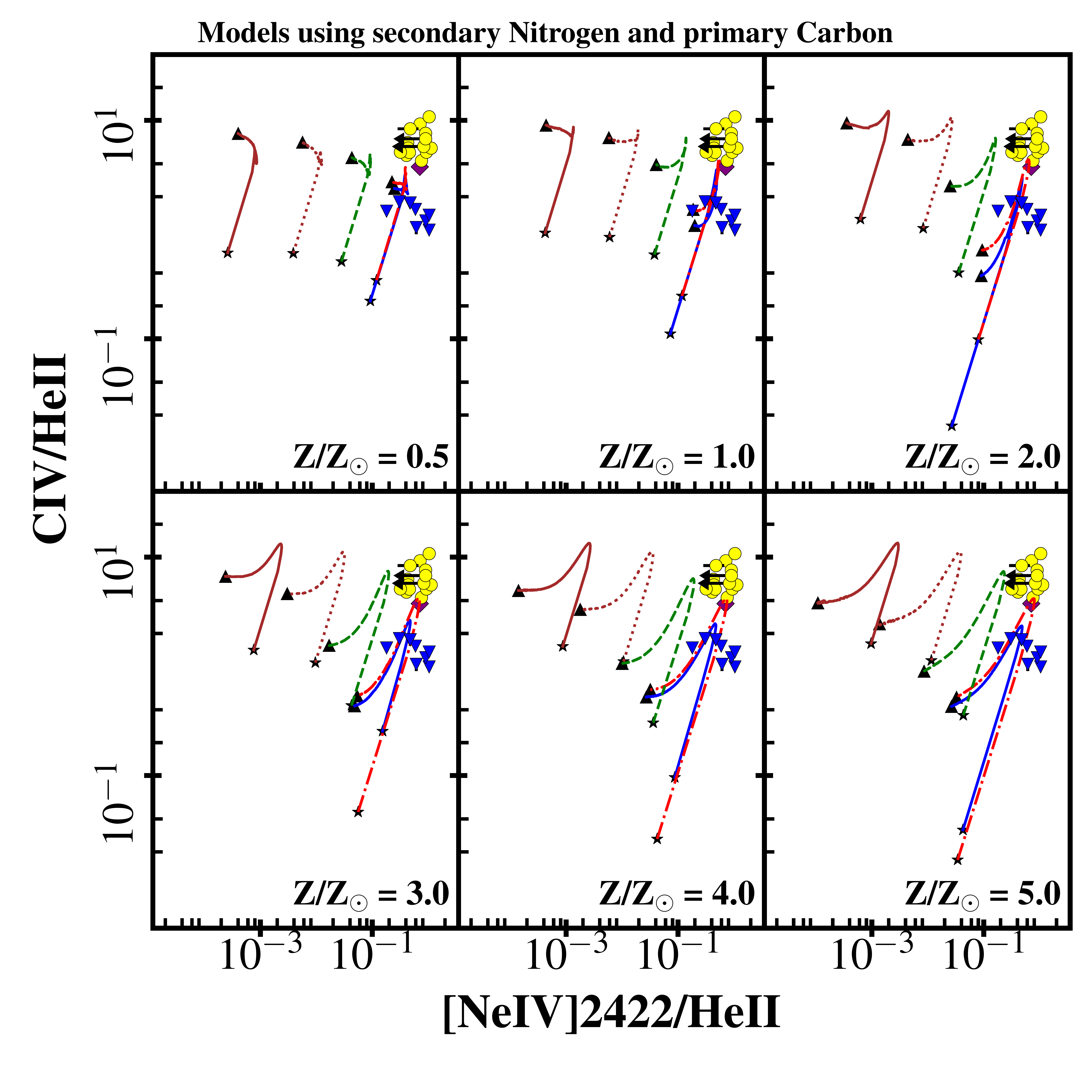}

	\caption{Comparison of the observed emission line ratios from the SDSS type II quasars divided in objects with \ion{C}{IV}/\ion{He}{II} $<$ 4 and \ion{C}{IV}/\ion{He}{II} $>$ 4 (purple filled diamond and yellow filled circles, respectively), and Keck II HzRGs from \citet{Ve2001} (blue filled triangles) with photoionization models using ionizing continuum power law index $\alpha$ = -1.0. Each diagram presents a different gas metallicity, i.e., Z/Z$_{\odot}$ = 0.5, 1.0, 2.0, 3.0, 4.0, 5.0. Curves with different colors represent the hydrogen gas density (n$_{H}$). At the end of each sequence, a solid black triangle corresponds to the lowest ionization parameter (U = 2$\times$10$^{-3}$) while the solid black star corresponds to the maximum value of the ionization parameter (U = 1.0).}
	\label{neiv03}
\end{figure*}

\begin{figure*}
		\includegraphics[width=\columnwidth,height=1.50in,keepaspectratio]{Fig/legend.pdf}
		\includegraphics[width=\columnwidth,height=1.5in,keepaspectratio]{Fig/legend_04.pdf}
	\quad
		\includegraphics[width=7.0in,height=7.0in]{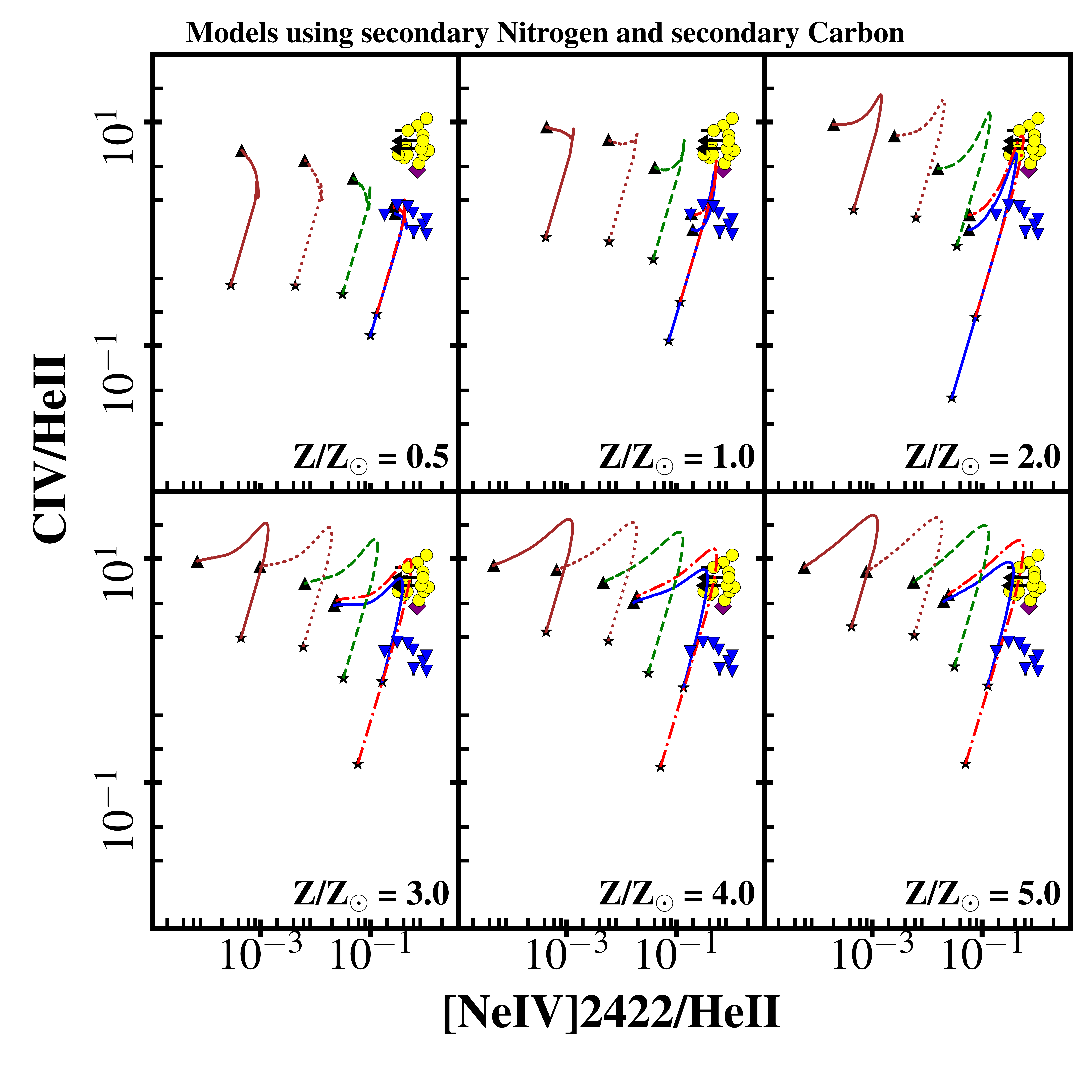}

	\caption{Comparison of the observed emission line ratios from the SDSS type II quasars divided in objects with \ion{C}{IV}/\ion{He}{II} $<$ 4 and \ion{C}{IV}/\ion{He}{II} $>$ 4 (purple filled diamond and yellow filled circles, respectively), and Keck II HzRGs from \citet{Ve2001} (blue filled triangles) with photoionization models using ionizing continuum power law index $\alpha$ = -1.0. Each diagram presents a different gas metallicity, i.e., Z/Z$_{\odot}$ = 0.5, 1.0, 2.0, 3.0, 4.0, 5.0. Carbon is a secondary element in which its abundance is proportional to the square of the metallicity. Curves with different colors represent the hydrogen gas density (n$_{H}$). At the end of each sequence, a solid black triangle corresponds to the lowest ionization parameter (U = 2$\times$10$^{-3}$) while the solid black star corresponds to the maximum value of the ionization parameter (U = 1.0).}
	\label{neiv03s}
\end{figure*}

\begin{figure*}
		\includegraphics[width=\columnwidth,height=1.6in,keepaspectratio]{Fig/legend.pdf}
		\includegraphics[width=\columnwidth,height=1.6in,keepaspectratio]{Fig/legend_01.pdf}
		\quad
		\includegraphics[width=7.0in,height=7.0in]{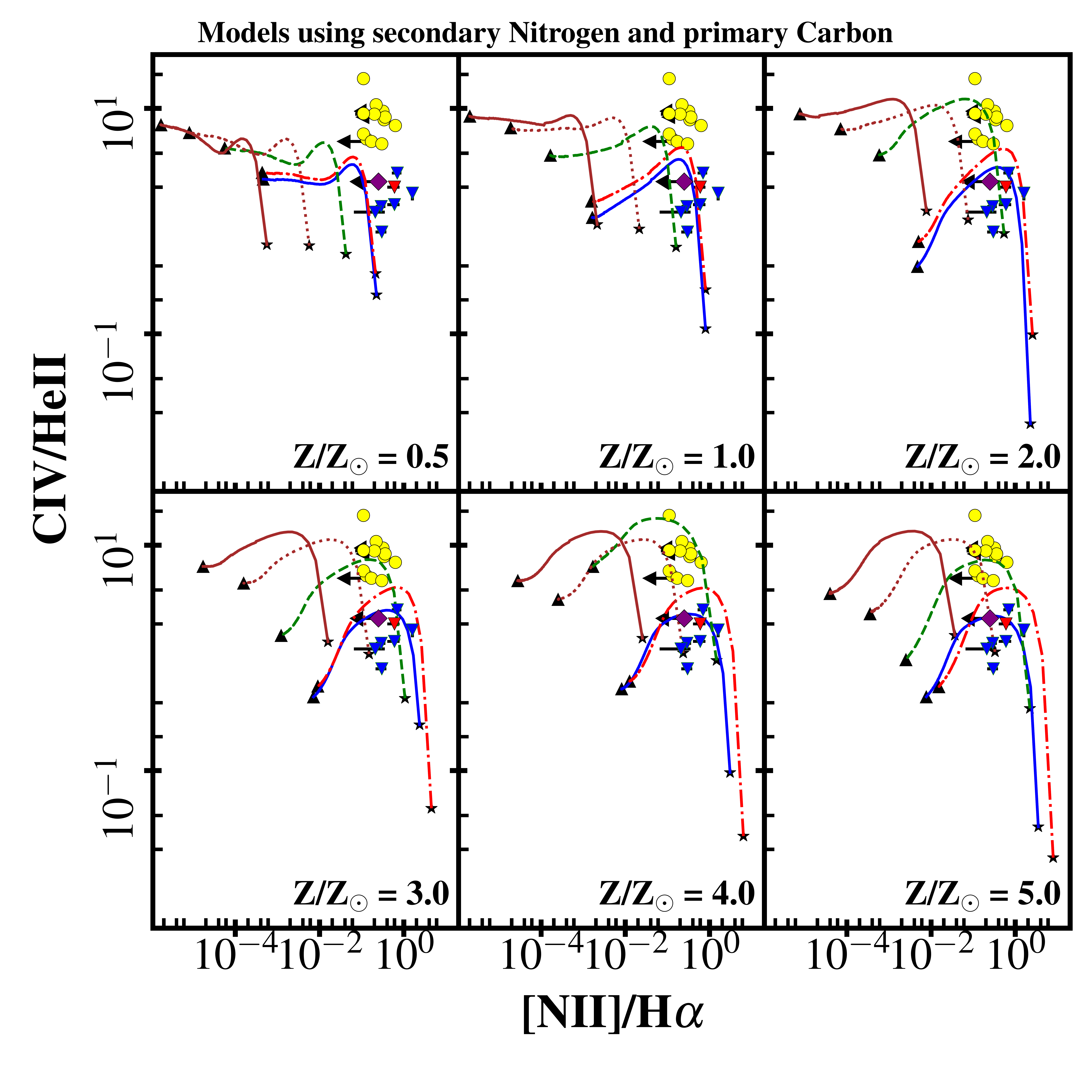}

	\caption{Comparison of the observed emission line ratios of type II quasars from \citet{greene2014}, divided in objects with \ion{C}{IV}/\ion{He}{II} $<$ 4 and \ion{C}{IV}/\ion{He}{II} $>$ 4 (purple filled diamond and yellow filled circles, respectively), and HzRGs from \citet{Hu4} (blue and red filled triangles) with photoionization models using ionizing continuum power law index $\alpha$ = -1.0. Each diagram presents a different gas metallicity, i.e., Z/Z$_{\odot}$ = 0.5, 1.0, 2.0, 3.0, 4.0, 5.0. Curves with different colors represent the hydrogen gas density (n$_{H}$). At the end of each sequence, a solid black triangle corresponds to the lowest ionization parameter (U = 2$\times$10$^{-3}$) while the solid black star corresponds to the maximum value of the ionization parameter (U = 1.0).}
	\label{nii02}
\end{figure*}

\begin{figure*}
		\includegraphics[width=\columnwidth,height=1.6in,keepaspectratio]{Fig/legend.pdf}
		\includegraphics[width=\columnwidth,height=1.6in,keepaspectratio]{Fig/legend_02.pdf}
		\quad
		\includegraphics[width=7.0in,height=7.0in]{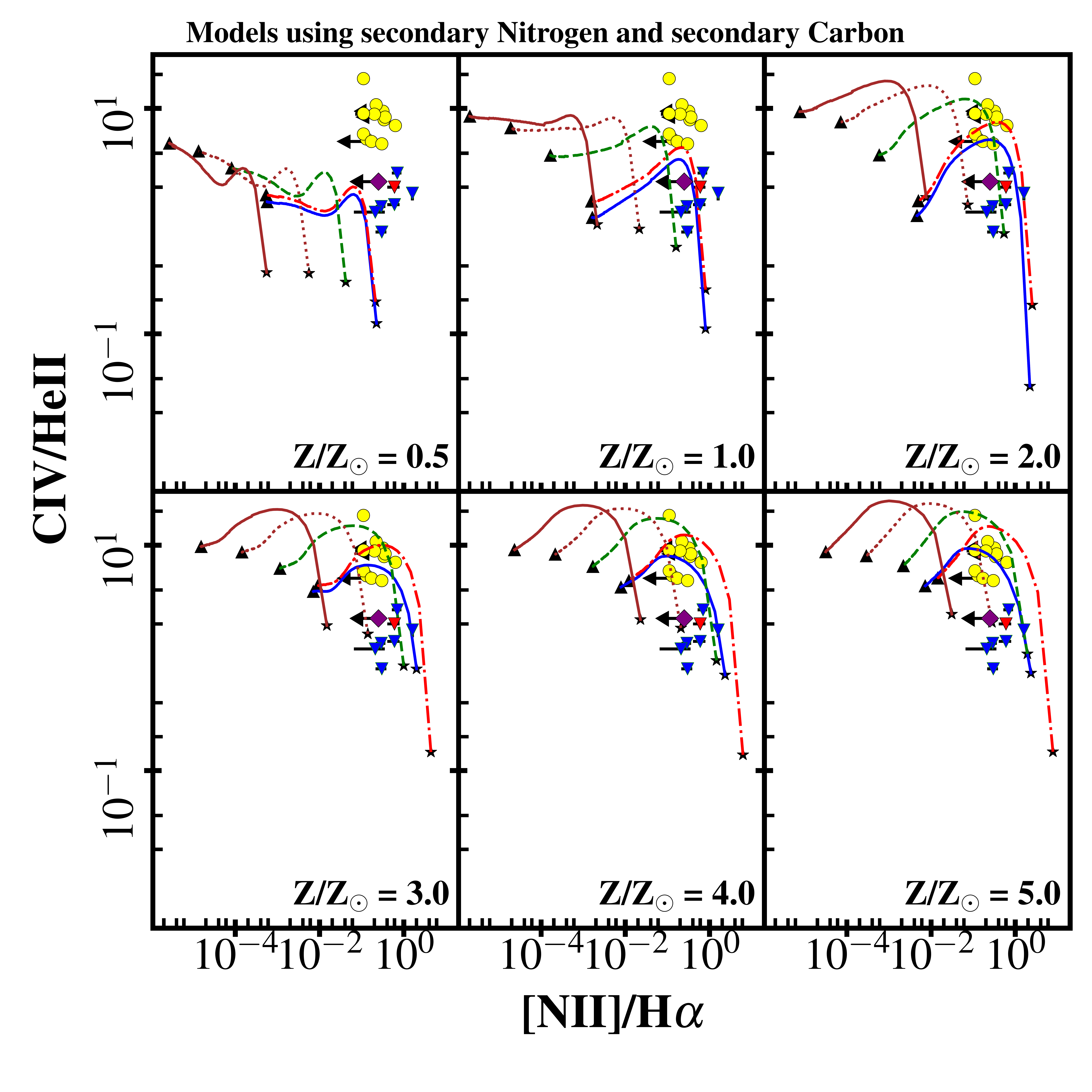}

	\caption{Comparison of the observed emission line ratios of type II quasars from \citet{greene2014}, divided in objects with \ion{C}{IV}/\ion{He}{II} $<$ 4 and \ion{C}{IV}/\ion{He}{II} $>$ 4 (purple filled diamond and yellow filled circles, respectively), and HzRGs from \citet{Hu4} (blue and red filled triangles) with photoionization models using ionizing continuum power law index $\alpha$ = -1.0. Each diagram presents a different gas metallicity, i.e., Z/Z$_{\odot}$ = 0.5, 1.0, 2.0, 3.0, 4.0, 5.0. Carbon is a secondary element in which its abundance is proportional to the square of the metallicity. Curves with different colors represent the hydrogen gas density (n$_{H}$). At the end of each sequence, a solid black triangle corresponds to the lowest ionization parameter (U = 2$\times$10$^{-3}$) while the solid black star corresponds to the maximum value of the ionization parameter (U = 1.0).}
	\label{nii02s}
\end{figure*}

\begin{figure*}
		\includegraphics[width=\columnwidth,height=1.6in,keepaspectratio]{Fig/legend.pdf}
		\includegraphics[width=\columnwidth,height=1.6in,keepaspectratio]{Fig/legend_01.pdf}
		\quad
		\includegraphics[width=7.0in,height=7.0in]{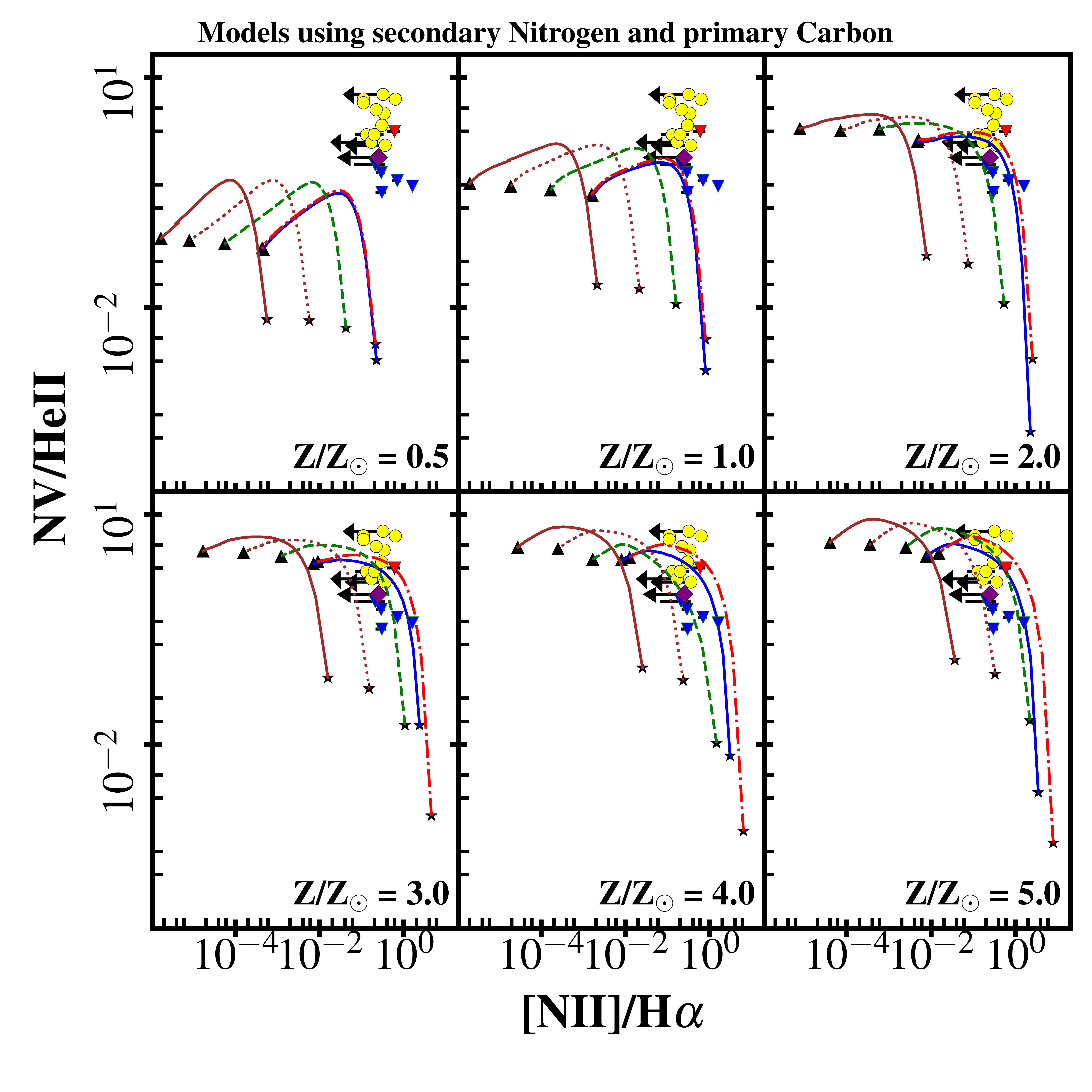}

		\caption{Comparison of the observed emission line ratios of type II quasars from \citet{greene2014}, divided in objects with \ion{C}{IV}/\ion{He}{II} $<$ 4 and \ion{C}{IV}/\ion{He}{II} $>$ 4 (purple filled diamond and yellow filled circles, respectively), and HzRGs from \citet{Hu4} (blue and red filled triangles) with photoionization models using ionizing continuum power law index $\alpha$ = -1.0. Each diagram presents a different gas metallicity, i.e., Z/Z$_{\odot}$ = 0.5, 1.0, 2.0, 3.0, 4.0, 5.0. Curves with different colors represent the hydrogen gas density (n$_{H}$). At the end of each sequence, a solid black triangle corresponds to the lowest ionization parameter (U = 2$\times$10$^{-3}$) while the solid black star corresponds to the maximum value of the ionization parameter (U = 1.0).}
		\label{nii03}
\end{figure*}

\begin{figure*}
		\includegraphics[width=\columnwidth,height=1.6in,keepaspectratio]{Fig/legend.pdf}
		\includegraphics[width=\columnwidth,height=1.6in,keepaspectratio]{Fig/legend_02.pdf}
		\quad
		\includegraphics[width=7.0in,height=7.0in]{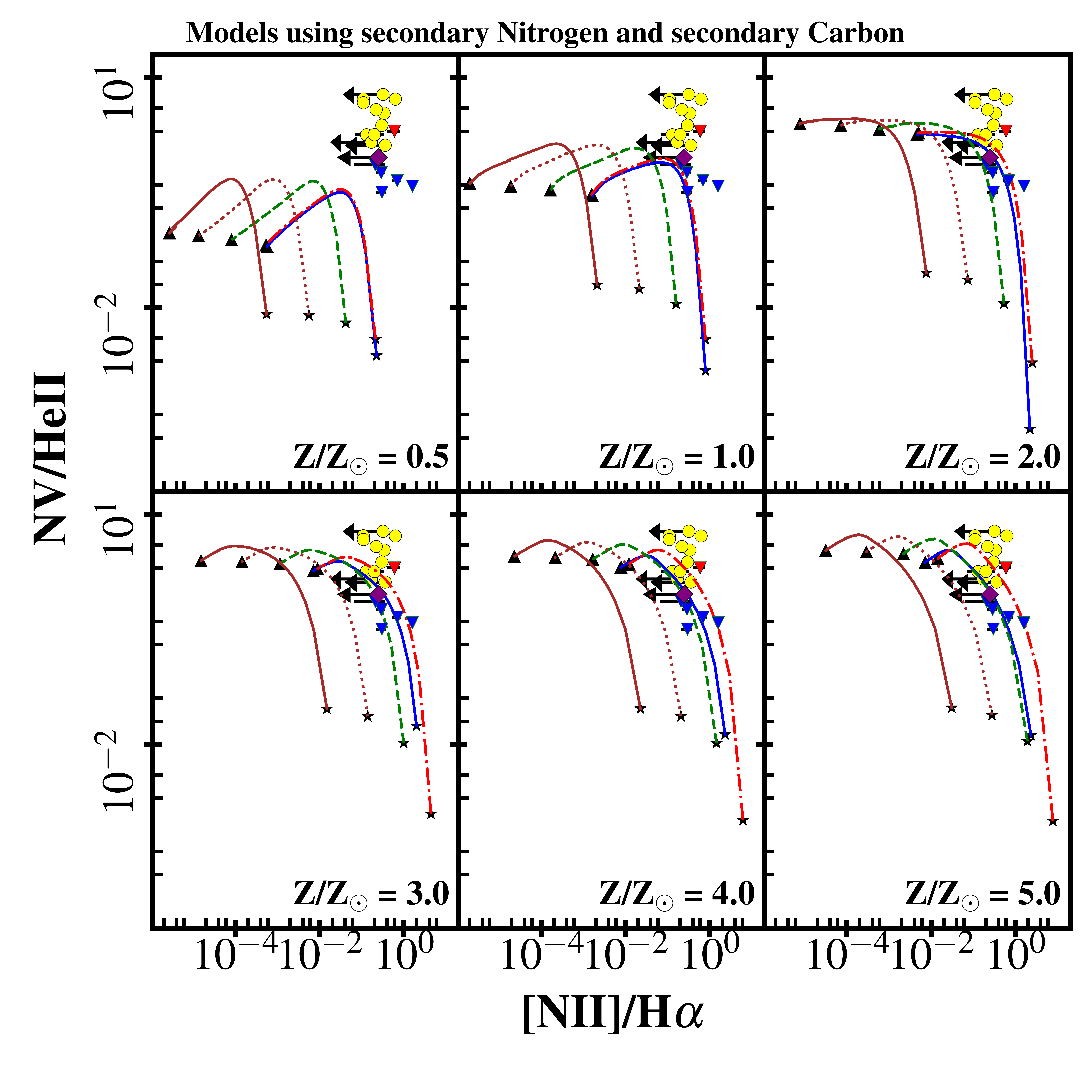}

	\caption{Comparison of the observed emission line ratios of type II quasars from \citet{greene2014}, divided in objects with \ion{C}{IV}/\ion{He}{II} $<$ 4 and \ion{C}{IV}/\ion{He}{II} $>$ 4 (purple filled diamond and yellow filled circles, respectively), and HzRGs from \citet{Hu4} (blue and red filled triangles) with photoionization models using ionizing continuum power law index $\alpha$ = -1.0. Each diagram presents a different gas metallicity, i.e., Z/Z$_{\odot}$ = 0.5, 1.0, 2.0, 3.0, 4.0, 5.0. Carbon is a secondary element in which its abundance is proportional to the square of the metallicity. Curves with different colors represent the hydrogen gas density (n$_{H}$). At the end of each sequence, a solid black triangle corresponds to the lowest ionization parameter (U = 2$\times$10$^{-3}$) while the solid black star corresponds to the maximum value of the ionization parameter (U = 1.0).}
	\label{nii03s}
\end{figure*}

\begin{table*}
	\caption{List of objects separated according to the selection criteria presented in \S \ref{new-crit}.}
	\label{class_tab}
	\begin{tabular}{cccccccc}
		\hline
		\hline
		SDSS name           & z    & NV/HeII         & CIV/HeII        & {[}NeIV{]}2422/HeII & {[}NeIV{]}2422/CIV & {[}NeIV{]}1602/{[}NeIV{]}2422 & FWHM$_{CIV}$ (km s$^{-1}$)   \\
		\hline
		\multicolumn{8}{c}{Class 1}                                                                                                                               \\
		\hline
		J213557.35-032130.0 & 2.15 & 1.51 $\pm$ 0.11 & 3.75 $\pm$ 0.29 & 0.71 $\pm$ 0.20     & 0.19 $\pm$ 0.02    & $<$ 2.57                            & 846 $\pm$  115 \\
		\hline
		\multicolumn{8}{c}{Class 2}                                                                                                                                \\
		\hline
		J161059.96+260321.5 & 2.28 & 1.19 $\pm$ 0.19 & 3.85 $\pm$ 0.38 & --                  & --                 & --                            & 1034 $\pm$ 87  \\
		J001814.72+023258.8 & 2.90 &                 & 3.45 $\pm$ 0.54 & --                  & --                 & --                            & 557 $\pm$ 47   \\
		J074725.50+160053.7 & 2.83 & 0.53 $\pm$ 0.09 & 3.16 $\pm$ 0.19 & --                  & --                 & --                            & 1300 $\pm$ 67  \\
		J080826.02+435652.1 & 2.85 & --              & 3.44 $\pm$ 0.55 & --                  & --                 & --                            & 727 $\pm$ 62   \\
		J081257.15+181916.8 & 2.38 & 0.91 $\pm$ 0.35 & 2.24 $\pm$ 0.33 & --                  & --                 & --                            & 532 $\pm$ 38   \\
		J094308.14+380923.0 & 3.12 & 1.64 $\pm$ 0.17 & 2.12 $\pm$ 0.27 & --                  & --                 & --                            & 545 $\pm$ 116  \\
		J101448.79+451444.7 & 2.28 & 2.16 $\pm$ 0.06 & 3.88 $\pm$ 0.49 & --                  & --                 & --                            & 680 $\pm$ 69   \\
		J113351.03+400851.1 & 3.00 & 1.11 $\pm$ 0.08 & 2.91 $\pm$ 0.20 & --                  & --                 & --                            & 574 $\pm$ 46   \\
		J114703.82+382727.2 & 2.26 & 2.94 $\pm$ 0.47 & 3.10 $\pm$ 0.65 & --                  & --                 & --                            & 620 $\pm$ 111  \\
		J122214.45+374837.9 & 3.16 & 0.98 $\pm$ 0.13 & 3.48 $\pm$ 0.44 & --                  & --                 & --                            & 1452 $\pm$ 41  \\
		J124302.62+421245.3 & 2.41 & 0.97 $\pm$ 0.29 & 2.39 $\pm$ 0.22 & --                  & --                 & --                            & 1375 $\pm$ 76  \\
		J141853.65+034005.2 & 3.19 & --              & 3.79 $\pm$ 0.34 & --                  & --                 & --                            & 619 $\pm$ 74   \\
		J152051.00+252830.1 & 3.77 & 1.38 $\pm$ 0.27 & 3.17 $\pm$ 0.47 & --                  & --                 & --                            & 1015 $\pm$ 86  \\
		J161353.27+201526.5 & 2.97 & --              & 3.64 $\pm$ 0.36 & --                  & --                 & --                            & 982 $\pm$ 56   \\
		\hline
		\multicolumn{8}{c}{Class 3}                                                                                                                                \\
		\hline
		J001040.82+004550.5 & 2.72 & 2.09 $\pm$ 0.28 & 4.26 $\pm$ 0.16 & 0.76 $\pm$ 0.15     & 0.18 $\pm$ 0.09    & $<$ 1.99                      & 1362 $\pm$ 98  \\
		J155108.96+321051.1 & 2.36 & 0.88 $\pm$ 0.36 & 7.60 $\pm$ 0.61 & 0.90 $\pm$ 0.11     & 0.12 $\pm$ 0.06    & $<$ 2.69                      & 2380 $\pm$ 121 \\
		J155725.27+260252.8 & 2.82 & 4.58 $\pm$ 0.43 & 5.10 $\pm$ 0.17 & 0.93 $\pm$ 0.10     & 0.18 $\pm$ 0.03    & $<$ 2.32                      & 1172 $\pm$ 85  \\
		J212055.57+065244.7 & 2.60 & 1.80 $\pm$ 0.32 & 5.57 $\pm$ 0.27 & 1.13 $\pm$ 0.05     & 0.20 $\pm$ 0.02    & $<$ 0.78                      & 625 $\pm$ 75   \\
		J163343.85+261026.2 & 3.07 & 2.52 $\pm$ 0.22 & 5.76 $\pm$ 0.08 & $<$ 0.83            & $<$ 0.14           & $>$ 1.35                      & 983 $\pm$ 52   \\
		J162025.94+170721.5 & 2.03 & 2.03 $\pm$ 0.18 & 6.77 $\pm$ 0.11 & $<$ 0.90            & $<$ 0.13           & $>$ 1.03                      & 1005 $\pm$ 74 \\
		\hline
		\multicolumn{8}{c}{Class 4}                                                                                                                                \\
		\hline
		J162327.66+312204.3 & 2.34 & 6.00 $\pm$ 0.50 & 5.11 $\pm$ 0.81 & 0.32 $\pm$ 0.10     & 0.06 $\pm$ 0.02    & --                            & 994 $\pm$ 90   \\
		J011506.65-015307.0 & 2.33 & 2.50 $\pm$ 0.79  & 8.37 $\pm$ 0.19  & --                  & --                 & --                            & 1901 $\pm$ 64  \\
		J022051.68-012403.3 & 2.60 & 6.03 $\pm$ 0.46  & 9.43 $\pm$ 0.20  & --                  & --                 & --                            & 2613 $\pm$ 74  \\
		J075119.09+130202.4 & 2.64 & 2.51 $\pm$ 0.13  & 4.76 $\pm$ 0.33  & --                  & --                 & --                            & 788 $\pm$ 88   \\
		J081950.96+111507.9 & 2.81 & 2.74 $\pm$ 0.05  & 7.82 $\pm$ 0.34  & --                  & --                 & --                            & 987 $\pm$ 49   \\
		J084005.00+344832.1 & 2.55 & 1.50 $\pm$ 0.13  & 9.12 $\pm$ 0.80  & --                  & --                 & --                            & 1300 $\pm$ 107 \\
		J100916.93+031128.9 & 2.69 & 8.55 $\pm$ 0.14  & 9.18 $\pm$ 0.26  & --                  & --                 & --                            & 1252 $\pm$ 39  \\
		J114542.07+401318.4 & 3.30 & 4.59 $\pm$ 0.65  & 8.03 $\pm$ 0.06  & --                  & --                 & --                            & 1155 $\pm$ 91  \\
		J125148.53+060027.4 & 2.27 & 2.97 $\pm$ 0.81  & 5.77 $\pm$ 0.09  & --                  & --                 & --                            & 1365 $\pm$ 53  \\
		J150549.73+074309.0 & 3.32 & 4.08 $\pm$ 0.08  & 5.79 $\pm$ 0.85  & --                  & --                 & --                            & 1354 $\pm$ 20  \\
		J153306.06+322649.6 & 2.71 & 9.21 $\pm$ 0.62  & 15.53 $\pm$ 0.41 & --                  & --                 & --                            & 1600 $\pm$ 25  \\
		J160900.01+190534.8 & 2.54 & 1.81 $\pm$ 0.17  & 8.88 $\pm$ 0.26  & --                  & --                 & --                            & 1051 $\pm$ 113 \\
		J162812.51+233734.8 & 2.39 & 2.51 $\pm$ 0.69  & 6.67 $\pm$ 0.35  & --                  & --                 & --                            & 1702 $\pm$ 99  \\
		J163343.85+261026.2 & 3.07 & 2.52 $\pm$ 0.52  & 5.76 $\pm$ 0.08  & --                  & --                 & --                            & 983 $\pm$ 52   \\
		J222946.61+005540.5 & 2.37 & 4.71 $\pm$ 0.63  & 9.00 $\pm$ 0.68  & --                  & --                 & --                            & 2036 $\pm$ 103 \\
		J004423.20+035715.5 & 2.22 & 1.58 $\pm$ 0.44  & 6.35 $\pm$ 0.19  & --                  & --                 & --                            & 931 $\pm$ 25   \\
		J004600.48+000543.6 & 2.46 & 4.65 $\pm$ 0.47  & 6.89 $\pm$ 0.27  & --                  & --                 & --                            & 1781 $\pm$ 29  \\
		J012552.08-015218.3 & 3.18 & 10.28 $\pm$ 0.44 & 12.96 $\pm$ 0.37 & --                  & --                 & --                            & 1799 $\pm$ 120 \\
		J014607.15+121119.9 & 3.14 & 6.82 $\pm$ 0.25  & 10.89 $\pm$ 0.24 & --                  & --                 & --                            & 947 $\pm$ 107  \\
		J015700.14+073116.0 & 3.05 & 6.16 $\pm$ 0.41  & 17.06 $\pm$ 0.78 & --                  & --                 & --                            & 1050 $\pm$ 102 \\
		J020245.82+000848.4 & 2.22 & --               & 7.19 $\pm$ 0.04  & --                  & --                 & --                            & 1029 $\pm$ 35  \\
		J020643.64+010403.3 & 2.66 & 3.39 $\pm$ 0.46  & 4.55 $\pm$ 0.12  & --                  & --                 & --                            & 511 $\pm$ 84   \\
		J020728.19+033833.5 & 2.76 & 5.39 $\pm$ 0.52  & 6.34 $\pm$ 0.22  & --                  & --                 & --                            & 1562 $\pm$ 35  \\
		J024525.95+025310.7 & 3.18 & 2.79 $\pm$ 0.26  & 4.63 $\pm$ 0.10  & --                  & --                 & --                            & 1472 $\pm$ 25  \\
		J025339.00+042154.5 & 2.85 & 7.53 $\pm$ 0.45  & 10.87 $\pm$ 0.63 & --                  & --                 & --                            & 1189 $\pm$ 49  \\
		J073637.54+225917.4 & 2.27 & 3.43 $\pm$ 0.70  & 7.92 $\pm$ 0.64  & --                  & --                 & --                            & 1000 $\pm$ 72  \\
		J073851.85+422010.9 & 2.19 & 1.80 $\pm$ 0.37  & 5.35 $\pm$ 0.80  & --                  & --                 & --                            & 1372 $\pm$ 46  \\
		J080428.80+111906.6 & 2.59 & 2.12 $\pm$ 0.32  & 4.90 $\pm$ 0.16  & --                  & --                 & --                            & 1240 $\pm$ 64  \\
		J081812.72+223755.3 & 2.46 & 3.74 $\pm$ 0.16  & 7.35 $\pm$ 0.21  & --                  & --                 & --                            & 726 $\pm$ 78   \\
		J082530.67+120658.4 & 2.35 & 3.31 $\pm$ 0.06  & 4.99 $\pm$ 0.06  & --                  & --                 & --                            & 733 $\pm$ 91   \\
		J082550.58+332543.1 & 2.87 & 3.99 $\pm$ 0.41  & 5.81 $\pm$ 0.56  & --                  & --                 & --                            & 694 $\pm$ 67   \\
		J083851.81+022907.1 & 2.87 & 1.36 $\pm$ 0.24  & 9.37 $\pm$ 0.05  & --                  & --                 & --                            & 1130 $\pm$ 45  \\
		J084949.57+044330.9 & 3.10 & 1.14 $\pm$ 0.12  & 6.41 $\pm$ 0.49  & --                  & --                 & --                            & 898 $\pm$ 47   \\
		J090612.64+030900.4 & 2.50 & 4.92 $\pm$ 0.46  & 8.61 $\pm$ 0.45  & --                  & --                 & --                            & 1285 $\pm$ 55  \\
		J091025.50+042944.4 & 3.78 & 4.43 $\pm$ 0.54  & 5.19 $\pm$ 0.16  & --                  & --                 & --                            & 1148 $\pm$ 52 \\
		\hline
		\hline
	\end{tabular}
\end{table*}

\begin{table*}
	\caption{continued.}
	\label{class_tab2}
	\begin{tabular}{cccccccc}
		\hline
		\hline
		SDSS name           & z    & NV/HeII          & CIV/HeII         & [NeIV]2422/HeII & [NeIV]2422/CIV & [NeIV]1602/[NeIV]2422 & FWHM$_{CIV}$ (km s$^{-1}$)    \\
		\hline
		\multicolumn{8}{c}{Class 4}                                                                                                                                  \\
		\hline
		J091357.87+005530.7 & 3.21 & 5.37 $\pm$ 0.49  & 9.60 $\pm$ 0.64  & --                  & --                 & --                            & 1559 $\pm$ 68  \\
		J094826.45+035834.5 & 2.27 & 2.16 $\pm$ 0.18  & 7.70 $\pm$ 0.03  & --                  & --                 & --                            & 1221 $\pm$ 74  \\
		J095819.35+013530.5 & 3.06 & 2.34 $\pm$ 0.19  & 6.98 $\pm$ 0.23  & --                  & --                 & --                            & 1865 $\pm$ 41  \\
		J100133.85+415107.9 & 2.36 & 3.22 $\pm$ 0.70  & 10.33 $\pm$ 0.20 & --                  & --                 & --                            & 1862 $\pm$ 97  \\
		J100210.53+381245.0 & 2.94 & 5.96 $\pm$ 0.36  & 12.01 $\pm$ 0.70 & --                  & --                 & --                            & 1347 $\pm$ 57  \\
		J104133.36+393325.4 & 2.43 & 3.88 $\pm$ 0.52  & 15.04 $\pm$ 0.72 & --                  & --                 & --                            & 2328 $\pm$ 57  \\
		J105324.11+445444.4 & 2.31 & 4.91 $\pm$ 0.18  & 6.46 $\pm$ 0.23  & --                  & --                 & --                            & 1768 $\pm$ 41  \\
		J105344.18+395402.3 & 3.23 & 3.49 $\pm$ 0.28  & 5.46 $\pm$ 0.59  & --                  & --                 & --                            & 939 $\pm$ 53   \\
		J114856.13+012731.6 & 2.33 & 3.51 $\pm$ 0.35  & 6.27 $\pm$ 0.29  & --                  & --                 & --                            & 778 $\pm$ 43   \\
		J115335.78+051221.8 & 3.07 & 0.72 $\pm$ 0.17  & 4.55 $\pm$ 0.38  & --                  & --                 & --                            & 1308 $\pm$ 113 \\
		J115411.95+040556.4 & 3.32 & 2.96 $\pm$ 0.13  & 5.56 $\pm$ 0.62  & --                  & --                 & --                            & 1084 $\pm$ 65  \\
		J115510.34+044455.4 & 2.43 & 5.24 $\pm$ 0.25  & 7.04 $\pm$ 0.17  & --                  & --                 & --                            & 1176 $\pm$ 119 \\
		J115947.86+351418.6 & 2.47 & 4.21 $\pm$ 0.49  & 6.98 $\pm$ 0.29  & --                  & --                 & --                            & 1089 $\pm$ 28  \\
		J133059.32+014610.3 & 2.34 & 5.15 $\pm$ 0.41  & 5.92 $\pm$ 0.15  & --                  & --                 & --                            & 683 $\pm$ 101  \\
		J133417.04+362753.0 & 2.29 & 3.16 $\pm$ 0.13  & 14.33 $\pm$ 0.25 & --                  & --                 & --                            & 1166 $\pm$ 109 \\
		J140625.75+042919.4 & 3.38 & 2.32 $\pm$ 0.17  & 6.27 $\pm$ 0.09  & --                  & --                 & --                            & 1457 $\pm$ 39  \\
		J144441.05-001343.4 & 2.55 & 5.27 $\pm$ 0.37  & 18.37 $\pm$ 0.36 & --                  & --                 & --                            & 1290 $\pm$ 95  \\
		J150145.46+360148.4 & 2.19 & 4.12 $\pm$ 0.15  & 5.59 $\pm$ 0.02  & --                  & --                 & --                            & 632 $\pm$ 99   \\
		J150451.51+202507.4 & 3.81 & 2.68 $\pm$ 0.29  & 7.80 $\pm$ 0.50  & --                  & --                 & --                            & 1288 $\pm$ 66  \\
		J151544.01+175753.0 & 2.40 & 3.43 $\pm$ 0.22  & 4.22 $\pm$ 0.20  & --                  & --                 & --                            & 908 $\pm$ 53   \\
		J151747.00+005550.7 & 2.66 & 1.12 $\pm$ 0.18  & 4.60 $\pm$ 0.13  & --                  & --                 & --                            & 874 $\pm$ 55   \\
		J151815.55+273618.0 & 3.34 & 8.45 $\pm$ 0.32  & 14.03 $\pm$ 0.19 & --                  & --                 & --                            & 1250 $\pm$ 73  \\
		J152105.83+182318.3 & 2.66 & 0.94 $\pm$ 0.10  & 4.80 $\pm$ 0.42  & --                  & --                 & --                            & 866 $\pm$ 37   \\
		J155200.53+175722.7 & 2.70 & --               & 5.34 $\pm$ 0.18  & --                  & --                 & --                            & 801 $\pm$ 96   \\
		J160103.85+195436.2 & 2.45 & 3.39 $\pm$ 0.18  & 6.30 $\pm$ 0.14  & --                  & --                 & --                            & 911 $\pm$ 31   \\
		J160158.53+260926.1 & 2.79 & 5.27 $\pm$ 0.42  & 9.35 $\pm$ 0.27  & --                  & --                 & --                            & 1186 $\pm$ 81  \\
		J161343.40+314524.8 & 3.21 & 1.56 $\pm$ 0.24  & 10.78 $\pm$ 0.40 & --                  & --                 & --                            & 1034 $\pm$ 110 \\
		J161447.97+354221.2 & 3.30 & 1.76 $\pm$ 0.26  & 4.27 $\pm$ 0.11  & --                  & --                 & --                            & 981 $\pm$ 92   \\
		J162025.94+170721.5 & 2.03 & 4.45 $\pm$ 0.48  & 6.77 $\pm$ 0.21  & --                  & --                 & --                            & 1005 $\pm$ 74  \\
		J162500.57+165705.8 & 2.20 & 2.65 $\pm$ 0.35  & 5.24 $\pm$ 0.31  & --                  & --                 & --                            & 1377 $\pm$ 76  \\
		J162651.76+293111.3 & 2.48 & 1.43 $\pm$ 0.19  & 5.79 $\pm$ 0.26  & --                  & --                 & --                            & 762 $\pm$ 72   \\
		J162806.01+331551.1 & 2.43 & 3.28 $\pm$ 0.25  & 5.44 $\pm$ 0.18  & --                  & --                 & --                            & 1189 $\pm$ 113 \\
		J165525.54+322936.9 & 2.72 & 2.56 $\pm$ 0.14  & 11.25 $\pm$ 0.38 & --                  & --                 & --                            & 1450 $\pm$ 65  \\
		J170110.12+301502.9 & 3.27 & 6.57 $\pm$ 0.25  & 6.57 $\pm$ 0.17  & --                  & --                 & --                            & 1441 $\pm$ 109 \\
		J170558.64+273624.8 & 2.45 & 5.58 $\pm$ 0.21  & 12.09 $\pm$ 0.29 & --                  & --                 & --                            & 1197 $\pm$ 47  \\
		J171908.90+310438.6 & 2.32 & 0.99 $\pm$ 0.16  & 6.13 $\pm$ 0.37  & --                  & --                 & --                            & 1281 $\pm$ 83  \\
		J213843.08-010133.6 & 2.80 & 1.44 $\pm$ 0.17  & 4.60 $\pm$ 0.26  & --                  & --                 & --                            & 1073 $\pm$ 47  \\
		J215341.33+041132.7 & 2.42 & 2.39 $\pm$ 0.22  & 4.85 $\pm$ 0.17  & --                  & --                 & --                            & 1032 $\pm$ 70  \\
		J225607.63+052520.2 & 3.15 & 4.62 $\pm$ 0.33  & 10.42 $\pm$ 0.30 & --                  & --                 & --                            & 962 $\pm$ 71   \\
		J230451.68+005135.1 & 4.17 & 12.09 $\pm$ 0.29 & 14.16 $\pm$ 0.34 & --                  & --                 & --                            & 1601 $\pm$ 66 \\
		\hline
		\multicolumn{8}{c}{Class 5}                                                                                                                                 \\
		\hline
		J163343.85+261026.2 & 3.07 & 2.52 $\pm$ 0.42 & 5.76 $\pm$ 0.08  & $<$ 0.83            & $<$ 0.14           & $>$ 1.35                      & 983 $\pm$ 52   \\
		J162025.94+170721.5 & 2.03 & 4.45 $\pm$ 0.48 & 6.77 $\pm$ 0.21  & $<$ 0.90            & $<$ 0.13           & $>$ 1.03                      & 1005 $\pm$ 74 \\
		\hline
		\hline
	\end{tabular}
\end{table*}


\bsp	
\label{lastpage}
\end{document}